\documentclass[a4paper,colorlinks,unicode,final]{cas-sc}

\usepackage[utf8]{inputenc}
\usepackage[english]{babel}
\usepackage{textcomp}
\usepackage{cancel}

\usepackage{graphicx}
\graphicspath{{assets/}{assets/tikz/}}

\usepackage{mathtools}
\usepackage{amsmath}
\usepackage{amssymb}
\usepackage[overload]{empheq}
\usepackage{amsthm}
\newtheorem{theorem}{Theorem}

\usepackage{bm}
\newcommand{\mat}[1]{{\ensuremath{\bm{#1}}}}
\renewcommand{\vec}[1]{{\ensuremath{\bm{#1}}}}
\newcommand{\Tau}{\ensuremath{\mathrm{T}}}
\newcommand{\Beta}{\ensuremath{\mathcal{B}}}

\usepackage{siunitx}
\usepackage[capitalise]{cleveref}

\usepackage{longtable}
\usepackage{tabularx}
\usepackage{booktabs}
\usepackage{subfigure}
\usepackage{caption}
\usepackage{subcaption}

\usepackage{tikz}
\DeclareRobustCommand{\tikzline}[1]{\raisebox{2pt}{\tikz{\draw[black,solid,line width=0.9pt,#1] (0,0) -- (5mm,0);}}}
\definecolor{refcolor}{HTML}{0072B2}
\definecolor{respcolor}{HTML}{D55E00}

\makeatletter
\def\@xfootnote[#1]{%
  \protected@xdef\@thefnmark{#1}%
  \@footnotemark\@footnotetext}
\makeatother

\usepackage[square,numbers]{natbib} %
\usepackage{bigints}

\begin{document}
\let\WriteBookmarks\relax
\def\floatpagepagefraction{1}
\def\textpagefraction{.001}

\shorttitle{Dynamic Inversion: An Incrementally Evolving Methodology for Flight Control Design}    

\shortauthors{Milz, Looye}

\title [mode = title]{Dynamic Inversion: An Incrementally Evolving Methodology for Flight Control Design\texorpdfstring{\footnote[\( \star \)]{This article is an homage to Dale Enns', Dan Bugajski's, Russ Hendrick's, and Gunter Stein's article ``Dynamic inversion: an evolving methodology for flight control design''~\cite{Enns1994}}}{}}

\author[1,2]{Daniel Milz}[orcid=0000-0001-9704-2036]
\fnmark[1]
\ead{Daniel.Milz@DLR.de}
\credit{Initiation, Conceptualization, Investigation, Methodology, Coordination, Visualization, Writing -- original draft, Writing -- review \& editing}

\author[1,3]{Gertjan Looye}[orcid=]
\cormark[1]
\fnmark[2]
\ead{Gertjan.Looye@DLR.de}
\credit{Conceptualization, Investigation, Methodology, Visualization Sketching, Writing -- original draft, Writing -- review \& editing} 

\cortext[cor1]{Corresponding author}
\fntext[fn1]{Research Associate, Department of Flight Control and Simulation}
\fntext[fn2]{Head of Department of Control of Aeroelastic Systems, Senior Expert at Department of Flight Control and Simulation}

\affiliation[1]{organization={Institute of Flight Systems, German Aerospace Center (DLR)}, street={M\"unchner Str. 20}, city={We\ss{}ling}, postcode={82234}, country={Germany}}
\affiliation[2]{organization={Technical University of Munich (TUM)}, street={Lise-Meitner-Str. 9}, city={Ottobrunn}, postcode={85521}, country={Germany}}
\affiliation[3]{organization={Institute of Aeroelasticity, German Aerospace Center (DLR)}, street={M\"unchner Str. 20}, city={We\ss{}ling}, postcode={82234}, country={Germany}}


\begin{abstract} %
	Nonlinear Dynamic Inversion (NDI) has become a standard methodology in flight control law design. It offers an intuitive approach to decouple commanded variable responses, handle system nonlinearities, and adapt to operating conditions. NDI also comes with a well-structured architecture that reduces design effort by addressing various functional aspects in separate components, and that allows straightforward integration of extended capabilities, such as envelope protection, control saturation handling, and compensating for faults or damage. A development that has resulted in considerable renewed attention is the use of (angular) acceleration sensors to partially replace inverse model equations. Known as incremental NDI, or INDI, this development offers reduced sensitivity to modeling errors and lower control law complexity. Incremental NDI, however, lacks useful design degrees of freedom in the previously present inverse model equations and underlying feedback signal synthesis, and comes with pitfalls in design aspects like control allocation, disturbance rejection, and inter-disciplinary couplings. This has given rise to recently developed hybrid and mathematically restructured approaches. The aim of this article is to give an up-to-date, structured overview of the various evolved forms of NDI from conceptual, historical, architectural, and mathematical perspectives. It hereby intends to provide useful considerations for future flight control law developments by reviewing its various forms and potentials from methodological, design, and application points of view.
\end{abstract}

\maketitle

\section*{Nomenclature}

{\begin{longtable*}{rl}%
\multicolumn{2}{l}{\textit{System description and signals}} \\
		\( \mathcal{X}, \mathcal{U}, \mathcal{Y}, \mathcal{D} \) & State, input, output, and disturbance space \\
		\( \vec{x} \in \mathcal{X} \subseteq \mathbb{R}^{n_x}\) & State vector of the model to be inverted \\
		\( \vec{u} \in \mathcal{U} \subseteq\mathbb{R}^{n_u}\) & Physical control input vector \\
		\( \vec{\mathfrak{u}} \) & Generalized control input (input to control allocation) \\
		\( \vec{y} \in \mathcal{Y} \subseteq\mathbb{R}^{n_y}\) & Output vector (commanded variables) \\
		\( \vec{d} \in \mathcal{D} \subseteq\mathbb{R}^{n_d}\) & Disturbance vector \\
		\( \vec{p} \in \mathbb{R}^{n_p} \) & Model parameter vector \\
		\( \vec{n} \in \mathbb{R}^{n_n} \) & Sensor noise vector \\
		\( \vec{\nu} \) & Pseudo control input \\
		\( \vec{e} = \vec{\nu} - \vec{y}^{(\rho)} \) & Inversion error \\
		\( \Sigma_g, \Sigma_G \) & Non-affine and control-affine form of the system \\
		\( \vec{\mathfrak{x}}, \vec{F}, \vec{H}, \vec{\mathfrak{y}} \) & State, dynamics, output function, and output of the full model \\
		\\
\multicolumn{2}{l}{\textit{Dynamics, inversion, and control allocation}} \\
		\( \vec{f}: \mathcal{X} \to \mathbb{R}^{n_x} \) & Drift (internal) dynamics function \\
		\( \vec{g}: \mathcal{X} \times \mathcal{U} \to \mathbb{R}^{n_x} \) & Non-affine input dynamics function \\
		\( \mat{G}: \mathcal{X} \to \mathbb{R}^{n_x \times n_u} \) & Control-affine input dynamics function \\
		\( \vec{h}: \mathcal{X} \to \mathcal{Y} \) & Output dynamics function \\
		\( \vec{\alpha}: \mathcal{X} \to \mathbb{R}^{n_y} \) & Transformed and allocated drift dynamics \\
		\( \vec{\beta}: \mathcal{X} \times \mathcal{U} \to \mathbb{R}^{n_y} \) & Transformed and allocated non-affine input dynamics \\
		\( \mat{\Beta}: \mathcal{X} \to \mathbb{R}^{n_y \times n_y} \) & Transformed and allocated control effectiveness matrix \\
		\( \mat{\Beta}_u: \mathcal{X} \to \mathbb{R}^{n_y \times n_u} \) & Control effectiveness w.r.t.\ the physical controls \\
		\( \vec{\delta} \) & Stacked disturbance term collecting the entries \( D_i \) \\
		\( \vec{f}_I, \mat{G}_I \) & Rows of \( \vec{f}, \mat{G} \) belonging to the internal states \\
		\( \mathcal{M}: \vec{\mathfrak{u}} \mapsto \vec{u} \) & Control allocation function \\
\\
\multicolumn{2}{l}{\textit{Geometric control and normal form}} \\
		\( \rho_i, \vec{\rho}, \rho \) & Relative degree of output \( i \), relative degree vector, total relative degree \\
		\( \bar{\rho} = \max_i \rho_i \) & Maximum relative degree \\
		\( \mathcal{L}_\vec{f} \vec{h} \) & Lie derivative of \( \vec{h} \) along \( \vec{f} \) \\
		\( \mathrm{ad}_\vec{f} \vec{g} \) & Adjoint operator (iterated Lie bracket) \\
		\( \Delta(\vec{x}) \subset T_\vec{x} \mathcal{X} \) & Distribution on the tangent space of \( \mathcal{X} \) \\
		\( T: \mathcal{X} \mapsto \mathcal{Z} \) & State transformation into normal (canonical control) form \\
		\( \vec{\xi} \) & External state vector \\
		\( \vec{\eta} = \vec{x}_I \) & Internal state vector \\
		\( \vec{\xi}_0 \) & External states held at their zero dynamics values \\
		\( \vec{\phi} \) & Zero dynamics coordinate functions, \( \vec{\eta} = \vec{\phi}(\vec{x}) \) \\
		\( \vec{z} = \left[ \vec{\xi} \ \vec{\eta} \right]^\Tau \) & Transformed state vector \\
\\
\multicolumn{2}{l}{\textit{Operators and notation}} \\
		\( \hat{(\cdot)} \) & Estimated, measured, or modeled quantity \\
		\( \tilde{(\cdot)} \) & Actual quantity, as opposed to the commanded one \\
		\( \Delta (\cdot) \) & Uncertain (residual) part of a quantity \\
		\( (\cdot)^{(\rho)} \) & \( \rho \)-th time derivative \\
		\( (\cdot)^\Tau \) & Transpose \\
		\( \mat{I}_n \) & \( n \times n \) identity matrix \\
		\( \mat{0}, \vec{0} \) & Zero matrix, zero vector \\
		\( \nabla \) & Nabla operator / gradient of function or vector field \\
		\( \mathcal{O}(\cdot) \) & Landau order symbol (higher-order terms) \\
		\( \left\Vert \cdot \right\Vert \) & Induced matrix norm \\
		\( s \) & Laplace complex frequency variable \\
		\( \mathcal{L}^{-1} \left\{ \cdot \right\} \) & Inverse Laplace transformation \\
\\
\multicolumn{2}{l}{\textit{Flight mechanics and control}} \\
		\( K, \mat{K} \) & Controller gains or controller \\
		\( \omega_n, \zeta \) & Second-order low pass natural frequency and damping \\
		\( \vec{r} \) & Position vector in NED \\
		\( \vec{v} \) & Velocity vector in body frame \\
		\( \vec{V}_{A_B} \), \( V_\mathrm{tas} \) & Aerodynamic velocity vector in body frame, true airspeed \\
		\( \vec{\omega} = \left[ p,\ q,\ r \right]^\Tau \) & Angular rates in body frame \\
		\( \vec{\Theta} = \left[ \phi, \ \theta, \ \psi \right]^\Tau \) & Euler angles attitude vector for roll, pitch, and yaw \\
		\( \mat{R}_\mathrm{EB} \) & Rotation matrix from B to E frame \\
		\( \alpha, \beta \) & Angle of attack, angle of sideslip \\
		\( \gamma, \chi, \mu \) & Flight path angle, course angle, flight-path bank angle \\
		\( \vec{f}_\mathrm{a}, \vec{f}_\mathrm{p}, \vec{m} \) & Aerodynamic force, propulsive force, and moment vector \\
		\( m \) & Aircraft mass \\
		\( \mat{I} \) & Moment of inertia matrix \\
		\( \bar{q} = \frac{1}{2} \rho V_\mathrm{tas}^2 \) & Dynamic pressure, with air density \( \rho \) \\
		\( S, c_\mathrm{ref} \) & Aerodynamic reference surface area and chord width \\
		\( C \) & Aerodynamic coefficient (e.g., \( C_{n_\beta} \)) \\
		\( \vec{\delta}_\mathrm{cs}, \delta_\mathrm{a}, \delta_\mathrm{e}, \delta_\mathrm{r} \) & Control surface (aileron, elevator, rudder) deflection \\
		\( n_y \) & Lateral load factor \\
		\( M_\mathrm{a} \) & Mach number \\
\end{longtable*} }%

	\section{Introduction}\label{sec:introduction}

	The design of flight control algorithms is both a highly rewarding, as well as a very challenging task in the development of a new aircraft or derivative thereof. Rewarding, because the algorithms equip the aircraft (or flight vehicle) with the functions that are necessary for its safe control, as well as with advanced ones that ensure operational efficiency, reduced crew workload, and/or that provide the capability of automated or autonomous operation. A major challenge of the design task is in the large number of design requirements that have to be taken into account, and eventually verified by means of analysis and test~\cite{Fielding2000}. Due to limited modeling accuracy and the occurrence of external disturbances, flight control algorithms nearly always close multiple feedback loops between available sensors and actuated control devices. Feedback control, in turn, makes seemingly independent systems, including the aircraft itself, intrinsically interdependent. The underlying engineering discipline comes with requirements and metrics that address this interdependence in terms of closed-loop stability margins, static and dynamic performance, robustness to uncertainties, etc. At the same time, each part of the loop contributes to the requirement specification in order to ensure its safe operation. For example, apart from the basic functional and safety requirements, control activity often must be limited from a systems point of view to maximize actuator lifetime, stable aeroelastic dynamics must be ensured and flight loads on the airframe kept within limits. Furthermore, computational load must be acceptable from an avionics point of view, and software development must adhere to strict development standards. A further major challenge lies in the fact that most aircraft operate in a large flight envelope under varying atmospheric conditions, with variable (fuel) loading, and in different configurations, and have to safely cope with failure conditions. It has to be ensured that the aircraft actually stays within safe operational bounds at all times. From a flight physical point of view, aircraft may have different, highly coupled axes to control, expose considerable nonlinear dynamic behavior, have limited but redundant control devices, and are always operated with two references: the inertial one in which basic dynamics are derived and that ensures operation along planned trajectories and arrival at the intended destination, as well as the air-mass referenced one to ensure safe flight at all times. In addressing the challenges above, the design of flight control algorithms combines multiple engineering disciplines, like feedback controls, flight physics, software development, and systems. 
	
    Flight control law design was originally based on a divide-and-conquer approach, applying Single-Input Single-Output (SISO) control techniques and metrics to a multitude of flight points, and combining the resulting parameters by means of gain schedules. Over the last decades, a considerable number of feedback control design methods have emerged that enable the control design team to handle the above (and many other) challenges more successfully and with less effort, especially in the aspects that the method is tailored towards. Comparisons can be found in~\cite{Robustcontrol1997},~\cite{Tischler2002}, and~\cite{Mubeena2025}.
	
	A method that has become favored for designing flight control laws in both research and industrial applications is \textit{Nonlinear Dynamic Inversion (NDI)}~\cite{Enns1994,Balas2003,Steinert2025a,Steinert2025b,Kim2026}.
	The underlying design methodology and architecture are particularly useful for addressing the aforementioned flight physical challenges related to non-linearity, coupled multivariable command responses, and varying operating conditions. To this end, \emph{dynamic} behavior of the aircraft is compensated and transformed into a linear, decoupled equivalent system by using model \emph{inversion} and coordinate transformation in an inner control law core -- hence the designation \emph{Dynamic Inversion}. Since the inverse model uses measured or estimated current state values, it is said that the methodology achieves linearization by means of feedback, or \emph{feedback linearization}. Although the latter is actually the objective, the methodology to achieve this is often also referred to as such. As a consequence, the actual control functions around the inverse model core can be designed relatively independently using any (linear) control methodology, avoiding the need for gain scheduling.

	In 1994, a key publication with a nearly identical title assessed NDI as a promising methodology for the design of nonlinear flight control systems. This paper by Dale Enns, Dan Bugajski, Russ Hendrick, and Gunter Stein provided an excellent summary of the current status of development, as well as a review of the method from a \emph{design} point of view: how to effectively use NDI to address handling quality requirements, robustness, and implementation aspects in the development of aircraft flight control laws. It also pointed at the potential for improving the underlying process -- avoiding the aforementioned divide-and-conquer approach. 
	In these respects, the paper complemented preceding publications that focused more on NDI from a \emph{methodological}, or \emph{application} point of view, either derived directly from physical model equations or starting from formal mathematical principles. Three decades later, NDI is not only controlling a wide variety of flight vehicles~\cite{Balas2003,Harris2018,Kim2023}, but has also become the basis for control architectures with advanced capabilities, like handling of highly nonlinear flight dynamics~\cite{Walker2002,Brinker2001,Goldthorpe2010,Smith2000,Kim2023,Gaessler2025,Milz2026jgcd,Bugajski1990,Snell1992,Steinhauser2004,Panish2023}, adaptation in (near) real time~\cite{Johnson2000,Holzapfel2004,Lombaerts2009,Snyder2018}, envelope protection ~\cite{Holzapfel2004,Lombaerts2012}, control allocation~\cite{Harris2018,Milz2026jgcd,Raab2024} and saturation handling schemes~\cite{Johnson2000,Holzapfel2004,Lombaerts2010}, etc.
	In recent years, so-called \textit{incremental} formulations (Incremental NDI, or INDI) have attracted considerable interest due to the simplicity of the control law and lower model dependency~\cite{Smith1998,Bacon2000,Chen2008,Sieberling2010,Vlaar2014,Smeur2016,Grondman2018}.
	
	This article provides a renewed, comprehensive overview of current NDI variants, highlighting their common basis while critically evaluating how their individual features address different design problems. The methodology is hereby examined from multiple perspectives --- from its theoretical foundations to its practical applications. First, NDI is reviewed from a conceptual and architectural perspective in~\cref{sec:conceptual,sec:architecture}, respectively, addressing its basic principles and relations to other inversion-based control architectures.
	The evolution of dynamic inversion can (retrospectively) be examined from two perspectives: The aforementioned architect's view (\cref{sec:architecture}) and a historian's view (\cref{sec:history}).
	These views elucidate \textit{how} and \textit{why} certain developments took place, and thus are essential to a holistic understanding of factors shaping NDI and its derivatives. %
	Recognizing that derivation approaches vary significantly in depth and rigor,~\cref{sec:mathematical} provides a mathematical perspective to establish a sound and common theoretical foundation for all NDI variants.
	
	Like the 1994 publication by Enns \textit{et al.}, this article specifically addresses \emph{design aspects} of NDI. \cref{sec:method} introduces and analyzes evolved NDI variants and relates them to the mathematical foundation from~\cref{sec:mathematical}.
	\cref{sec:design} provides a deeper review from a design perspective, discussing how to handle the challenges listed at the beginning of this section, including conflicting requirements, stability, control allocation, uncertainties, and failures, as well as engage with other disciplines, implement, verify, validate, and ultimately test in flight. %
	\cref{sec:application} demonstrates the practical application on the Cessna Citation II \textit{PH-LAB} experimental aircraft, covering all design stages, implementation, and flight experiments, and effectively implementing all NDI variants.

	\section{A Conceptual Perspective}\label{sec:conceptual}
	
\begin{figure*}[!htb]
		\centering
		\includegraphics[width=\textwidth]{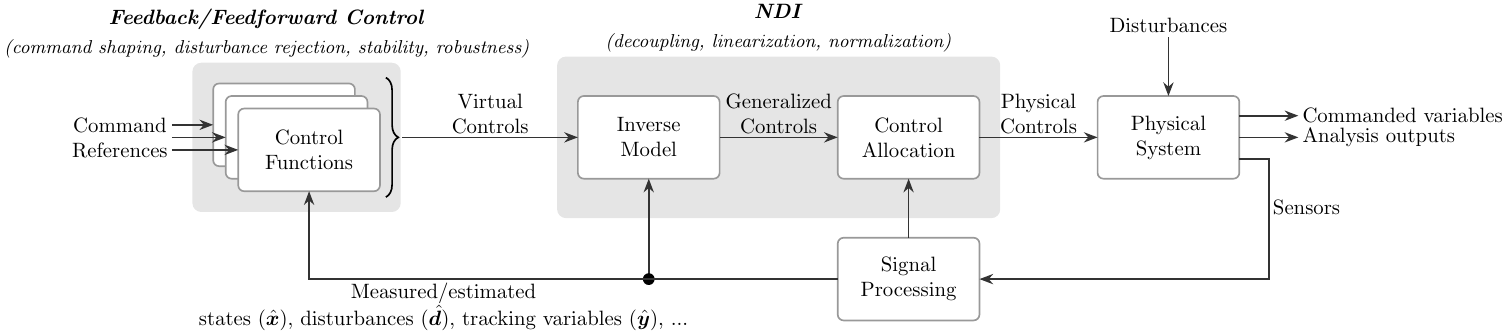}
		\caption{Typical NDI-based control law architecture.}
		\label{fig:NDItypicalArchitecture}
	\end{figure*}

	The main reasons for the wide applicability of nonlinear dynamic inversion are best explained by looking at a generic NDI-based control architecture in~\cref{fig:NDItypicalArchitecture}.
	
	The methodology comes with a structured control law architecture that allocates the various principal and supporting functions to individual components. The core of the architecture is an \textit{inverse model} representation of the physical system. It computes control commands to the physical system either directly, or, in the case of redundant devices, via generalized controls and a specific \textit{control allocation} (cf.~\cref{sec:method:mod:ca}).
	The complexity of the latter may range from virtual hard-wiring (ganging) to versatile algorithms with online optimization~\cite{Enns1998,Johansen2013,Steinert2025b}. The inverse model addresses time-varying, nonlinear dynamic behavior and couplings between commanded variables that are to be tracked. The combination of an inverse model, control allocation, and the physical system ideally yields a decoupled, linearized system between its \emph{virtual control inputs} and \emph{commanded variable outputs}. 
	The virtual controls are usually time derivatives of the latter (normalization). The principal \textit{control functions} are, in turn, arranged around the NDI core and designed to realize desired command response behavior, ensure stability, provide disturbance rejection, and achieve overall robustness to mismatches between the physical system and the model used for control design. Individual functions may be designed more or less independently and usually do not need gain scheduling. Apart from provisions for envelope protection and control allocation, aircraft dependency of the control laws is concentrated in the inverse model. This means that most components are reusable, and adaptation to changes in aircraft design or other updates in models ideally focus on adaption the latter. As pointed out in~\cite{Enns1994}, this has been a major factor in shifting from the traditional divide-and-conquer paradigm (typically involving manual tuning and scheduling of control law gains) towards a highly modular, well-structured, and reusable control law design. 
	
	NDI-based design does not come with method-specific metrics that serve as synthesis or analysis objectives, like induced systems norms or the structured singular value in robust control. Most publications use standard structures and metrics for the design of the \textit{Control Functions} in~\cref{fig:NDItypicalArchitecture} (detailed in \cref{sec:design:generalarch}). In fact, the architecture may be completed by any controller synthesis method of preference. NDI should therefore maybe not be considered as a stand-alone design method, but rather as an architecture that decouples, normalizes, and linearizes the physical system to considerably reduce design effort and complexity of the principal control functions. For example, gain scheduling is avoided or at least simplified, and measures for decoupling between control axes are (in principle) no longer needed. %
	Functions may even be individually developed using different control methods. The applied synthesis method effectively sees a decoupled linear system as rendered by the NDI architecture (to the right in~\cref{fig:NDItypicalArchitecture}). For Verification and Validation (V\&V) standard and advanced metrics, like gain and phase margins, system norm-based ones, or the structured singular value \( \mu \)~\cite{Hyde1995,Packard1993}, are well applicable to assess closed-loop system performance, stability, and robustness~\cite{Pollack2024,Hyde2001,Looye2001,Ito2002,Looye2008}. 

	The term \emph{inversion} unavoidably raises concerns about the robustness of the closed-loop system \cite{Snell1992b}. These concerns are valid, but have proven to be manageable for two reasons. First of all, state variables in the inverse model equations are obtained from measurement or estimation (cf. Signal Processing in~\cref{fig:NDItypicalArchitecture}), rather than from simulation of an internal model\footnote{In the linear case, this would be equivalent to an inverse transfer function (matrix).}. The latter may easily lead to deviations between the system and inverse model states due to model errors in the state equations, disturbances, or integration drift. With NDI, state variable deviations are limited to sensing or estimation errors or time delays. Second, the NDI core is only part of the overall design and does not even attempt to establish robustness by itself. The inverse model core  in~\cref{fig:NDItypicalArchitecture} is necessarily \emph{encapsulated} by external feedback control functions, as the inversion is just an open-loop mapping from virtual to generalized controls. These functions allow handling of uncertainties and achieving the required stability margins in the same way as with traditional control methods~\cite{Enns1994,Buffington1993,Reiner1996,Yuan2013}. A consideration may be, if the use of highly uncertain data in the inverse model core makes achieving robustness more difficult. For example, design of functions assumes sufficient decoupling by the NDI core; if this is not established due to uncertainties, their independent synthesis may not achieve desired performance and robustness levels. 
	
	\begin{figure*}[!htb]
		\centering
		\includegraphics{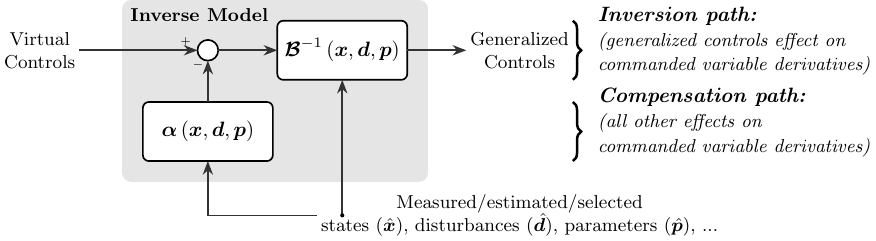}
		\caption{NDI inverse model structure.}
		\label{fig:NDIinverseModel}
	\end{figure*}
	
	The main characteristics of NDI, as well as its implementation variants, are best explained by having a closer look at the inverse model from~\cref{fig:NDItypicalArchitecture}, see~\cref{fig:NDIinverseModel}. The actual inversion is in the relation between generalized controls and the variables to be controlled. The latter are differentiated individually until a physically meaningful relation with one or more generalized control inputs emerges that can be inverted algebraically. These derivatives become the depicted virtual control inputs to the inverse control effectiveness matrix $\mat{\Beta}^{-1}(\vec{x},\vec{d},\vec{p})$. This matrix may, in turn, depend on measured or estimated states $\hat{\vec{x}}$, disturbances $\hat{\vec{d}}$ (or any other influences) acting on the physical system, as well as selectable parameters \(\hat{\vec{p}}\). The states and disturbances also directly influence the tracking variable derivatives, as modeled by the function $\vec{\alpha}(\vec{x},\vec{d},\vec{p})$. This influence is computed and explicitly compensated for by subtraction from the virtual controls, as depicted in~\cref{fig:NDIinverseModel}. 
	
	As stated before, the inversion thus aims at matching virtual controls and (derivatives of) variables to be tracked as accurately as possible. The heavy reliance on the availability of a sufficiently good model is usually not a limiting factor, as the verification and validation of any flight control system already impose demanding requirements on model availability and quality. In case the model functions $\vec{\alpha}(\vec{x},\vec{d},\vec{p})$ and $\mat{\Beta}(\vec{x},\vec{d},\vec{p})$ do not match the physical reality exactly, the intended compensation of nonlinearities and decoupling, as well as normalization, may be inadequate to some unknown extent, though.
	As a first main critique of NDI, the only available leverage from a methodological point of view is to maximize the accuracy of $\vec{\alpha}$ and $\mat{\Beta}$, which may easily result in disproportional modeling, online computational and software development effort, and sensor requirements. In practice, aspects like computational effort, common requirements to avoid unnecessary control activity, and avoidance of structural coupling result in the preferred use of simplified inverse models. NDI further assumes that system states and, to some extent, disturbances can be estimated from sensor measurements. Sensor failures must usually be accounted for by some form of redundancy in order to keep the functions in working order, which may be addressed in the Signal Processing function in~\cref{fig:NDItypicalArchitecture}. As mentioned before, handling any of the aforementioned inversion inadequacies is left to be addressed in the design of the outer control functions, unavoidably requiring performance to be traded in. 
	
	A second critique of NDI lies in the explicit compensation of non-controls-related influences, as modeled by  $\vec{\alpha}(\vec{x},\vec{d},\vec{p})$. This helps to achieve the desired response behavior of the commanded variables, but may also render some states in $\vec{x}$ invisible to the outer functions. Their dynamics are effectively left to themselves if not addressed in any of the other functions. These \emph{zero dynamics} have to be carefully attended to at the architectural level~\cite{Isidori1985,Slotine1991,Lin1994}, and ensured to be stable with a sufficient margin. This more often involves the use of deeper physical insight and engineering skills, rather than mathematical and generally applicable solutions approaches~\cite{Hauser1992,Devasia1996,Horn2019}. 
	
	A third critique, to some extent, is that controlled variables and generalized or physical controls are initially linked via time derivatives of the first (i.e. the virtual controls). In flight control design it is more common to establish this link from a quasi-static point of view, as this makes placement of integrators, which basically provide built-in trimming capability, more obvious. Good examples are air speed and flight path angle. In NDI, the derivatives naturally link to thrust and angle of attack respectively. Once a new equilibrium in a new flight condition is reached, speed effectively sets the angle of attack and a changed flight path angle is sustained by the thrust setting~\cite{Lambregts2013a,Soule1969}. In an NDI design this is resolved indirectly \cite{Lombaerts2012}, but may lead to problems in case physical control means (e.g. thrust) saturate. This also requires deeper physical insight in the selection of suitable commanded variables \cite{Rysdyk2002}. 
	
	Since the publication by Enns \textit{et al.}~\cite{Enns1994}, the methodology has considerably \textit{evolved}, as the title of the paper already foresaw. A considerable number of extensions have emerged, either further exploiting the discussed merits of NDI, extending capabilities and functionality, or addressing the aforementioned disadvantages of the methodology. First of all, NDI has been shown to go hand in hand with the method of Backstepping~\cite{Krstic1995}. Starting from a (to be chosen) Lyapunov function as a basic metric for ensuring closed-loop stability of the nonlinear system, NDI is often the first control law structure of choice towards fulfilling this stability metric. At the same time, the metric may provide a solid basis for experimenting with reducing the complexity of the $\vec{\alpha}(\vec{x},\vec{d},\vec{p})$ term, e.g., by not compensating for influences that actually help rather than hamper achieving the desired dynamic behavior. 
	Reference~\cite{Littleboy1998} introduces bifurcation analysis as a guiding metric for NDI-based design. The NDI architecture has also proven to be an excellent basis for adaptive control systems. Given that system dependencies are largely captured by the inverse model terms contained in $\vec{\alpha}(\vec{x},\vec{d},\vec{p})$ and $\mat{\Beta}(\vec{x},\vec{d},\vec{p})$, these functions may be adapted to online identified model data~\cite{Johnson2000,Holzapfel2004,Lombaerts2010b,Lombaerts2010,Lombaerts2012}. A general problem in the field of feedback control, particularly when applying adaptive control to damaged aircraft, is control saturation. The aforementioned references show that this can be elegantly solved in the NDI architecture by means of \emph{Pseudo-Control Hedging} (PCH).        
	
	In its basic form, NDI computes the $\vec{\alpha}(\vec{x},\vec{d},\vec{p})$ function using model equations and measured (estimated) states and disturbances. Recent developments have taken a different approach by computing this function from measured or estimated derivatives of tracking variables, subtracting the part induced by current control deflections~\cite{Smith1998,Bacon2000}. This obviously reduces sensitivity to model uncertainties and, in its evolved form, results in a remarkably simple control law structure with excellent robustness properties and disturbance-rejection capabilities~\cite{Chen2008,Sieberling2010}. This variant, better known as \emph{incremental NDI} or \emph{INDI}, as well as evolved hybrid developments~\cite{Kumtepe2022,Milz2024e,Milz2026jgcd}, retain the basic structure depicted in~\cref{fig:NDIinverseModel} and will be further detailed in~\cref{sec:method}.

	\section{An Architectural Perspective}\label{sec:architecture}
	
	The use of inverse models in control laws is quite common and not limited to the NDI architecture as discussed in the previous section. %
	Explicit use of model knowledge, especially in some inverted form, has long since been an intuitive approach in control system design to achieve accurate, decoupled tracking performance. 
	The \textit{good regulator theorem} already stated that ``every good regulator of a system must be a model of that system''~\cite{Conant1970}. The \textit{internal model principle}~\cite{Francis1976} continues on this, showing that asymptotic regulation requires a model of the exogenous reference and disturbance dynamics inside the loop. %
	The broader idea of carrying a model of the system within the controller later motivated forward and, subsequently, inverse models for control.
	Among others, Brockett~\cite{Brockett1965}, Silverman~\cite{Silverman1968,Silverman1969}, and Hirschorn~\cite{Hirschorn1979} laid the groundwork by formalizing inversion of linear and nonlinear systems. The use of inverse system models as part of the control system design was proposed in later research, e.g.,~\cite{Francis1976,Devasia1996}. Interestingly, the principle of internal and inverse models is even assumed to be present in cerebellar motor control~\cite{Ito1970,Kawato1999}.
	
	Inverse system models can be used in various ways in a control architecture, both in feed-forward and feedback paths~\cite{Fliess1987,Looye2005}. In all approaches, the inverse system is driven by a desired output trajectory and yields the control inputs necessary to achieve that output. Evidently, inversion requires a direct causality and an algebraic relation between outputs and inputs. As detailed later in~\cref{sec:mathematical:reldeg}, this can be accomplished by differentiating each output \(y_i\) a sufficient number of times (denoted as the relative degree\( \rho_i \)) prior to the actual inversion, assuming a well-defined system (see~\cref{sec:mathematical} for the assumptions on the system). %
	
	In the following, five common methodologies will be described and compared: inverse feed-forward --, (explicit) model-following --, trajectory-referenced --, internal model --, and dynamic inversion control. Although their development did not happen independently or sequentially, these methodologies can be regarded as an evolution from an architectural point of view. %
	
	\subsection{Inverse Feed-Forward Control}\label{sec:inversearchitecture:ff}
	
\begin{figure}[!htb]
		\centering
		\includegraphics{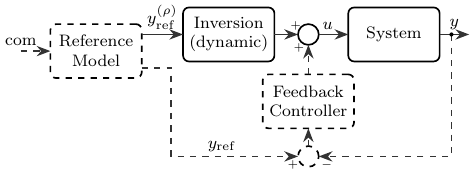}
		\caption{Inverse feed-forward control using a dynamical ``self-breathing'' inverse system with optional feedback control and reference model (\tikzline{dashed}).}
		\label{fig:inversion_feed-forward}
	\end{figure}

	Pure feed-forward control can achieve high tracking precision if the inverted model is accurate and the disturbances acting on the system can be neglected. In practice, this principle is always part of a two-degree-of-freedom control architecture~\cite{Horowitz1963}, adding feedback control loops to address model uncertainty and reject external disturbances (see~\cref{fig:inversion_feed-forward} and references~\cite{Brockett1963,Devasia1996}).
	In this setup, the inverse model does not affect closed-loop stability but must be stable itself (requiring stable zero dynamics). Note that a reference model might be used to shape the command appropriately as shown in~\cref{fig:inversion_feed-forward}. The applications presented in the references \cite{Thuemmel2001} and \cite{Reiner2011} use the structure for compensation of elasticity in robot joints.
	
	Inverse feed-forward control realizes the design principle Rudolf Brockhaus put forward in his book on flight control (translated from German): \emph{Compensate known influences using feed-forward control, unknown influences with the help of feedback control}~\cite[p. 581]{Brockhaus2011}. In this way, faster responses are achieved in comparison with command following being sorted out via feedback only.
	
	Apart from use for commanded variable tracking, a similar structure may also be used for disturbance rejection. The effect of estimated disturbances on, for example, external forces and moments, may be compensated by inversely computed control deflections. This approach is adopted by~\cite{Koenig1995,Hecker2007} in a quasi-static form. A key aspect of attention is the accurate timing of control deflections.

	From \cref{fig:inversion_feed-forward} two further observations can be made:
	First, the feedback controller may need to handle nonlinearities and coupled reference command responses, and adapt to different operating conditions. This is commonly addressed using gain scheduling. In the feed-forward path, the inverse model automatically takes care of these aspects.
	Second, the inverse model is a dynamic system, containing and integrating its own state equations (i.e., ``self-breathing''). These states may increasingly differ or even diverge from the plant's states due to model uncertainty and external disturbances. Therefore, states resulting from open integration should be avoided or provided externally from measurements.
	Inverse feed-forward models thus have three sources of uncertainty: the model equations themselves, external disturbances, and the current value of the self-integrated states.
	This raises the question: ``Should model-based inverse inputs be used as feed-forward under plant uncertainty?''~\cite{Devasia2002}. Devasia argues that, under sufficiently small model uncertainties, using a two-degrees-of-freedom control approach and adding an inversion-based feed-forward control can lead to performance improvements, and proposes using the feed-forward model selectively at lower-frequency bands with lower uncertainties~\cite{Devasia2002}.
	This can also be argued on a more abstract level: feed-forward does not affect closed-loop stability but might degrade robustness if model errors drive the plant into non-operational regions. Therefore, an architecture must include ``robustifying'' feedback and possibly limited feed-forward.
	Still, current developments often deploy this architecture, for instance, using physics-guided neural networks to learn the (self-breathing) inverse dynamics for feed-forward control~\cite{Bolderman2024}.
	Another aspect that needs to be covered when looking at inverse systems is potential non-minimum phase behavior, which causes instability when inverting directly~\cite{Martin1996,Hauser1992}. Bounded feed-forward solutions can still be obtained by stable-inversion techniques that accept non-causality and exploit preview of the reference trajectory~\cite{Devasia1996}, or by inverting a minimum phase approximation of the system and leaving the residual to feedback~\cite{Hauser1992}.
	
	A variation of this architecture is to invert the closed-loop system with the Feedback Controller integrated, commanding references to the latter, instead of controls of the system. In \cite{Kavaja2023} it is proven that this variant is equivalent to the one in~\cref{fig:inversion_feed-forward}.
	
	\subsection{Trajectory-referenced control}
	
\begin{figure}[!htb]
		\centering
		\includegraphics{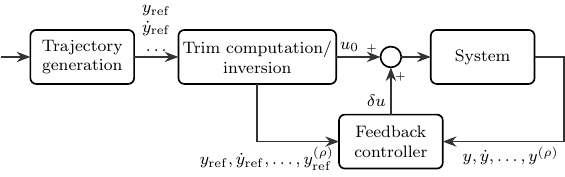}
		\caption{Trajectory-referenced control}
		\label{fig:trajectory_referenced}
	\end{figure}

	A very similar control structure is also used in the field of optimal trajectory generation and tracking. The inverse model and its commanded input in~\cref{fig:inversion_feed-forward} may just as well be a pre-computed or on-line generated trajectory that is differentiable a number of times that at least equals the relative degrees (will defined in \ref{sec:mathematical}) w.r.t. the physical controls , see~\cref{fig:trajectory_referenced}. This is often used in exploiting so-called differential flatness of a system~\cite{Fliess1995,Murray1995,VanNieuwstadt1997}, where the \textit{Output/state references} become \textit{Output/output derivative references}. Trajectory generation includes computation of (smooth) trajectories of relevant outputs, as well as of system states and required inputs by means of trim computations or pre-computed trim maps~\cite{Kaminer1998}. The latter is effectively a model inversion problem that is solved by means of numerical nonlinear equation solvers. Based on the trim solutions, the system is linearized around the state trajectories and the feedback controller is designed using linear methods and gain scheduling. Alternatively, the trim problem may be formulated in the form of an explicit model inversion, see for example~\cite{Boyarko2011}.	
	
	The trajectory-referenced control architecture is very similar to the inverse feed-forward one in the previous section. The feature that makes it stand out is the fact that the commanded variable trajectory and derivatives references are obtained by defining a feasible trajectory for the system in the first place, and dynamically trimming it along the trajectory in the second. The reference model is thus the system on the commanded trajectory, with no state equations involved. The feedback controller may be designed linearly as proposed in the listed references and scheduled if necessary. Alternatively, a reference model may be used in design as shown in~\cref{fig:inversion_feed-forward}, replacing it with a trajectory generation algorithm in the implementation~\cite{Looye1997}.  
	
	\subsection{Internal Model Control}\label{sec:inversearchitecture:imc}
	
\begin{figure}[!htb]
		\centering
		\includegraphics{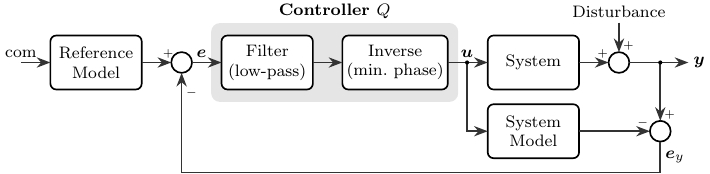}
		\caption{Internal Model Control architecture~\cite{Garcia1982,Economou1986}}
		\label{fig:imc}
	\end{figure}

	Internal Model Control (IMC)~\cite{Garcia1982,Morari1983,Rivera1986,Economou1986} embeds a model of the plant in the controller, continuing on the internal model principle~\cite{Francis1976} by using a forward rather than an inverse model.
	Architecturally, IMC is characterized by a structure in which a parallel model of the plant is embedded within the control loop, as shown in~\cref{fig:imc}. The difference between the measured and predicted outputs is fed back, isolating the effects of unmeasured disturbances and model mismatches~\cite{Garcia1982}. It can be seen as an architecture that uses an estimate of the system to design the controller, typically involving the inverse of the estimated system model combined with a filter~\cite{Garcia1982}. IMC architectures are applied across various industrial applications, including extensions for unstable processes~\cite{Morari1989,Ranjan2023}.
	
	For open-loop stable plants, this structure provides a practical realization of the Youla-Kucera (or $Q$-) parameterization~\cite{Youla1976a,Youla1976b,Kucera1975,Mahtout2020}. In the Youla framework, all stabilizing controllers can be parameterized by a stable transfer function, often denoted as \( Q \). In IMC, the Q-parameter corresponds directly to the product of the internal model's inverse and the filter~\cite{Ranjan2023}. Unstable plants require a coprime factorization and hence architectural modifications of the plain IMC structure~\cite{Ranjan2023}.
	
	To ensure physical realizability and safe reference tracking, IMC systematically partitions the plant model into invertible (minimum-phase) and non-invertible (e.g., time delays or right-half-plane zeros) components~\cite{Garcia1982}. The controller is formed by inverting the invertible portion, augmented with a low-pass filter~\cite{Garcia1982}. This filter ensures the controller is causal and proper, especially when the plant model contains non-invertible dynamics or time delays, and additionally serves as the primary tuning parameter that shapes the closed-loop response~\cite{Morari1989}.
	
	When this design procedure is applied to systems with significant transport delays, it natively yields an architecture equivalent to the Smith Predictor~\cite{Smith1957,Garcia1982}. Designed specifically to address the challenges of systems with significant dead time (time delays), the Smith Predictor utilizes a model of the plant without the delay to provide a direct feedback signal to the controller. By effectively removing the delay from the feedback loop, the Smith Predictor allows for higher controller gains and faster response times than would be possible with standard feedback. While the Smith Predictor focuses primarily on delay compensation, IMC generalizes this principle to encompass a broader range of plant dynamics and uncertainty handling.
	
	Nonlinear generalizations exist, of which one prominent example is combined with dynamic inversion for quadcopter control~\cite{Bouzid2016,Nascimento2019}.
	Additionally, IMC has a similar architecture to \( \mathcal{L}1 \) adaptive control~\cite{Hovakimyan2011}.

	\subsection{Model-Following Control}\label{sec:inversearchitecture:mf}
	
\begin{figure}[!htb]
		\centering
		\includegraphics{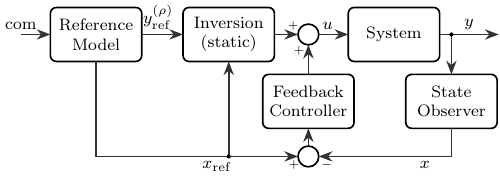}
		\caption{Explicit model-following control architecture.}
		\label{fig:inversion_modelfollowing}
	\end{figure}

	(Explicit) Model-following control, as depicted in~\cref{fig:inversion_modelfollowing}, is a control strategy designed to enforce a desired closed-loop behavior by making the system outputs or states follow a specified reference model~\cite{Tyler1964}.
	This is achieved using an (optimal) reference model, often in combination with an inverse model~\cite{Godbole1972} in the feed-forward path, effectively shaping the command response to impose these dynamics on the system.
	This control approach originated from the field of optimal control for aerospace systems and rose in the 1960s to 1970s~\cite{Kalman1962,Rynaski1964,Tyler1964,Asseo1970}.
	The key difference with the inverse feed-forward approach discussed previously is that the inverse model does not integrate its own states; instead, it uses those of the reference model (potentially in a transformed way). A detailed example can be found in~\cite{Duda1997a,Duda1997b}.
	A comparable approach is presented in~\cite{Devasia1996}.
	Early control designs already included adaptive elements~\cite{Kalman1962,Landau1972,Young1978}.
	
	To ensure robustness and disturbance rejection, a feedback controller is included. Compared to the previous approach, this controller ensures that the system states follow the reference states as closely as possible. Evidently, the states of the reference model need to match the system's states both from a physical and a numerical point of view. As this is typically the case in applications like in-flight simulation~\cite{Motyka1972,Robustcontrol1997}, the method has become a standard in this field. With the feedback control ensuring that the reference and actual system states match, self-integrated states are largely removed as a source of uncertainty in the inverse model. However, the feedback controller still needs to account for the potentially nonlinear and coupled command response behavior of the system, usually requiring some form of gains scheduling. 
	
	\subsection{Dynamic Inversion Control}\label{sec:inversearchitecture:di}
\begin{figure}[!htb]
		\centering
		\includegraphics{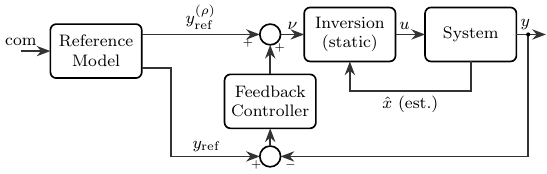}
		\caption{Dynamic inversion or feedback linearization control architecture.}
		\label{fig:inversion_dynamicinversion}
	\end{figure}

	The third approach, and the focus of this article, is dynamic inversion, depicted in~\cref{fig:inversion_dynamicinversion}. 
	The key difference to the previous architectures is that the inverse model becomes part of the closed-loop system. This way, as seen from controller components in surrounding loops, it effectively linearizes the system through state feedback. Placing the feedback controller before the inversion reduces the complexity as the feedback control functions do not need to handle the nonlinearities and couplings, as shown in~\cite{Meyer1981}.
	This is a fundamental difference compared to previous approaches and implies that outer-loop feedback functions can be largely linear, as nonlinear and coupled command response behavior, as well as the effects of varying operating conditions, are readily accounted for by the inverse model. For this reason, the approach in its most general form has become known as \emph{feedback linearization}~\cite{Isidori1985}. A second fundamental difference is that the inverse model does not bring its own differential equations, but uses the measured states from the system instead. This effectively reduces sensitivity to errors in the inverse model equations~\cite{Morgan1964,Freund1971}. Valuable comparisons with inverse feed-forward and model-following control in this respect are made in \cite{Ireland2017} and \cite{Saetti2018,Kocurek1997} respectively. As already pointed out in~\cref{sec:conceptual}, the encapsulation of the inverse model in the feedback loop and the use of state feedback make uncertainties in the inverse model well manageable in control law design practice. An integrated approach between NDI and model following control is discussed in \cite{Autenrieb2022,Pei2023}.

	\section{A Historical Perspective}\label{sec:history}
	
	While the first encounter with feedback linearization was probably already in the 1930s~\cite{Black1934}, the first structured approach to an NDI-like control synthesis was carried out in~\cite{Boksenbom1950} in the late 1940s~\cite{Guardabassi2001,Liu2019}. 
	In the latter, Boksenbom and Hood published an algebraic method for controlling gas-turbine engines with multiple interacting variables~\cite{Boksenbom1950}. They solved the linear multivariable decoupling problem by imposing algebraic conditions so that the closed‐loop system splits into independent single-input single-output (SISO) channels. This early work essentially obtained the ideal decoupling matrix for a linear system, predating modern state-space theory.
	
	In the 1960s and 1970s, the groundwork for NDI was laid. At first, in the form of necessary conditions for the decoupling and inversion~\cite{Falb1967}, and later by structured synthesis techniques for multivariable linear systems~\cite{Morgan1964,Brockett1965,Silverman1968}. This was extended by parallel developments in nonlinear systems, where geometric nonlinear control helped develop rigorous theory for the feedback linearization of nonlinear systems~\cite{Brockett1978,Hirschorn1979,Jakubczyk1980a,Isidori1985}. 
	The developments leading to NDI will subsequently be discussed in~\cref{sec:history:early}.
	In practice, most systems of interest are multivariable, characterized by multiple inputs and outputs, couplings, and interactions, and are often nonlinear in their nature. The first serious applications came up in the 1970s~\cite{Meyer1975} and 1980s, with aerospace systems as a major driver of developments towards NDI. However, it was not until the 1990s that flight control computers allowed the routine computational realization of NDI in production programs~\cite{Wacker2001}. In the 1980s to 1990s, the robustness of dynamic inversion needed for practical implementation became focus of attention~\cite{Enns1994,Snell1992b}.
	The current state of the art will be presented in~Section~\ref{sec:history:ndi}. 
	In the 21st century, endeavors aimed to overcome limitations of NDI, particularly its reliance on the on-board plant model, and to integrate advanced capabilities for handling of failures and damages, envelope protection, control allocation, etc.
	Incremental developments in the theory and engineering of NDI have led to the partial replacement of model dependencies with sensor measurements, manifested in the \textit{sensory derivatives} like incremental NDI (INDI)~\cite{Smith1998,Chen2008,Sieberling2010}, as explained in~\cref{sec:history:indi}.
	Current efforts aim to bridge the gap between model information and sensor measurements, yielding new \textit{hybrid} methods~\cite{Kumtepe2022,Pollack2024,Milz2024i,Milz2026jgcd} that will be explained in~\cref{sec:history:hndi}.

	\subsection{Early Developments}\label{sec:history:early}
	
	Following the work of~\cite{Boksenbom1950}, several publications throughout the 1950s developed synthesis methods for achieving non-interacting control of multivariable systems, e.g.,~\cite{Freeman1957,Kavanagh1958}.
	Non-interacting control decouples the system so that each command input affects only one output, yielding a set of single-variable systems that can be handled using existing control design techniques.
	One of the first systematic design approaches to non-interacting control for transfer function matrices was published in the early 1960s by Chen \textit{et al.}~\cite{Chen1962}, which addressed issues such as physical realizability and stability by leveraging Bode diagrams to simplify these considerations. 
	Synthesis of decoupling compensators, which transform a matrix of transfer functions into a diagonal one, was, however, a ``formidable manual work''~\cite{Morgan1964} known as \emph{Morgan's problem}~\cite{Morgan1964,Warren1975}. 
	During the 1960s, the advent of state-space methods enabled extending non-interacting control approaches to multivariable state-space systems, where state feedback was now used.
	This led to a significant reduction in design complexity and allowed for a more formalized design, as well as the utilization of early digital computers. Furthermore, issues arising from overlooked zero dynamics due to transfer function cancellation could be addressed~\cite{Gilbert1969}.
	An important step was the introduction of synthesis techniques~\cite{Morgan1964}.
	Similar findings have been proposed by Brockett in~\cite{Brockett1965}, where he proposed using the inverse of the system for an optimal control problem and characterizes the effects of state feedback on the pole-zero configuration (cf.~\cref{sec:inversearchitecture:ff}).
	
	Parallel developments investigated the determination of inverse systems. Brockett introduced the inverse of a single-input single-output system initially in 1963~\cite{Brockett1963,Brockett1965}. In the following years, the concept of the inverse system was extended to time-variant single-input single-output systems~\cite{Silverman1968} and to multivariable linear systems~\cite{Falb1967,Silverman1969}.
	
	Necessary and sufficient conditions for decoupling multivariable linear systems with constant coefficients by state feedback have been introduced in~\cite{Falb1967}, while giving a synthesis technique for a desired closed-loop pole configuration.
	Decoupling those systems by state feedback in a canonical form was introduced in~\cite{Gilbert1969}, which significantly reduced design complexity and generalized the decoupling of multivariable systems by state feedback, even enabling an algorithmic implementation~\cite{Gilbert1969Computer}.
	Finally, Wonham and Morse~\cite{Wonham1970} provided an approach based on geometric control theory.
	Freund proposed using the decoupling approach through inverse systems as a design method for time-variant multivariable systems~\cite{Freund1971}. This can be regarded as the inception of feedback linearization as a general synthesis method.

	All that laid the foundation for applying this principle to nonlinear systems. 	In the early 1970s, Porter~\cite{Porter1970} derived sufficient conditions for the diagonalization and inversion of general nonlinear systems, followed by Nazar and Rekasius~\cite{Nazar1971}, who decoupled a particular class of nonlinear systems with additive nonlinearities and distinguished between coupled and decoupled systems.
	Singh and Rugh~\cite{Singh1972} further developed this approach by extending it to control-affine nonlinear systems in general.	An early application to a flight control design problem was published by Asseo \cite{Asseo1973}, demonstrating achieved decoupling between responses around pitch and roll axes.
	
\begin{figure}[!htb]
        \centering
        \hfill
        \subfigure[NASA DHC-6 Twin Otter (NASA 720), used in early nonlinear inverse control flight tests. Source: NASA / Chuck Ritchie, Photo ID AC77‑0845‑015]{\label{fig:dhc6_nasa_color}\includegraphics[width=.46\textwidth]{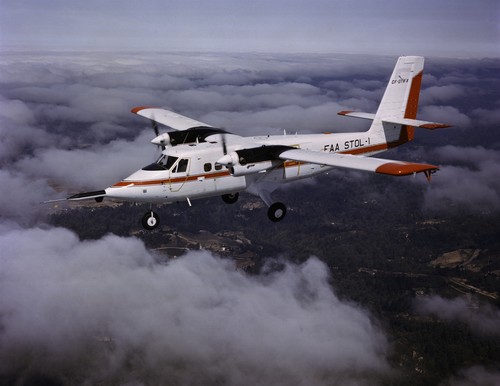}}
        \hfill
        \subfigure[de Havilland Canada C-8A experimental aircraft \copyright{} In the collection of Robert Thomas]{\label{fig:DHC8A}\includegraphics[width=.515\textwidth]{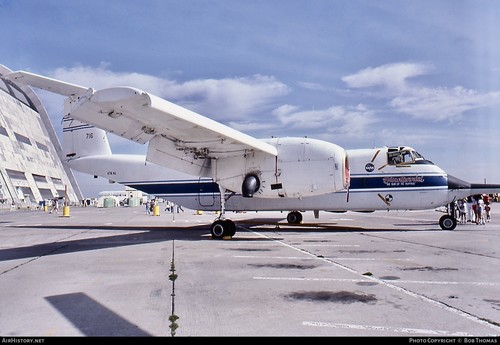}}
        \hfill
        \vspace{-0.5cm}
        \caption{First NDI flight test platforms}
        \label{fig:DHC}
    \end{figure}

	A major development that likely took place in parallel (as the above references are not cited in reports) was definition of the flight control laws for automated Short Take-Off and Landing (STOL) operations. The system, originally called \emph{Total Automatic Flight Control System (TAFCOS)}, was initially flight tested on a \emph{de Havilland Canada DHC-6} (\cref{fig:dhc6_nasa_color})~\cite{Wehrend1980} in preparation for implementation and use on the heavily modified \emph{de Havilland Canada C-8A Buffalo} (civil designation \emph{DHC-5}), shown in~\cref{fig:DHC8A}. The experimental aircraft was developed in the frame of the NASA Augmentor Wing Jet STOL Research Aircraft project~\cite{Ashleman1972}.

	In the underlying architecture, Meyer and Cicolani take the step from feed-forward to feedback controls (cf.~\cref{sec:inversearchitecture:di}) and developed a structure for automatic flight control laws~\cite{Meyer1975} using inverse trim maps for computation of attitude and control surface and thrust (nozzle) commands. They even use the principle of time scale separation between translational and rotational dynamics, which later became known as partial nonlinear dynamic inversion~\cite{Elgersma1988}. Although not yet called as such, these authors are probably the first to apply NDI to (fixed-wing) aircraft and perform successful flight tests. This is quite a remarkable achievement, given the complexity of the STOL-capable aircraft flight dynamics and limited capabilities of the Flight Control Computer (FCC). An inversion-based control law structure for flight path and speed tracking was flight tested on a further modified version of the aircraft type by Franklin \textit{et al.}~\cite{Franklin1986}.
	
	In the 1970s, a new approach to nonlinear systems was established that used geometric control and Lie algebras, promising a structured approach to nonlinear control. Essential work was done by Krener, who used approaches from geometric control and Lie algebra to show the equivalence between nonlinear systems and their linearized counterparts~\cite{Krener1973}, i.e., exact linearization.
	Brockett stated the condition for linearizability of affine nonlinear systems~\cite{Brockett1978} and introduced the Lie algebra formulation of NDI.
	In~\cite{Brockett1978}, Brockett also characterized \textit{feedback invariants}, i.e., properties of a system's nonlinearity that cannot be removed by feedback, thereby defining fundamental limits of linearization.
	Both~\cite{Brockett1978} and~\cite{Jakubczyk1980} show local equivalence of feedback-linearized systems and the nonlinear system.
	Towards the 1980s, the feedback linearization problem for nonlinear systems was considered solved, and necessary and sufficient conditions for transforming a nonlinear system into a linear one were provided~\cite{Brockett1978,Jakubczyk1980a,Hunt1983,Isidori1985}.
	Isidori formalized the feedback linearization theory in his book~\cite{Isidori1985} using the differential geometric approach, which has become the canonical reference work.

	\subsection{State of the Nonlinear Dynamic Inversion}\label{sec:history:ndi}
	
	In 1994, Enns \textit{et al.}~\cite{Enns1994} describe NDI as a methodology that shifts control design towards a modular approach, which simplifies the design process in that regard, as it increases scalability and provides flexibility to change or add new functions.
	Modern (flight) control laws are designed to meet multiple criteria simultaneously. NDI allows multiple design aspects to be addressed individually, including performance, disturbance rejection, command shaping, anti-windup, and handling qualities, enabling a straightforward, requirement-driven design and implementation of flight control systems. Additionally, the adoption of NDI facilitates the reuse of these modules by integrating plant-specific dependencies into the inversion, leaving the outer loop functions mission-specific.
	
\begin{figure}[!htb]
        \centering
        \subfigure[F-18 HARV. Source: NASA, Photo ID EC94-42645-9 ]{\label{fig:harv}\includegraphics[width=.36\textwidth]{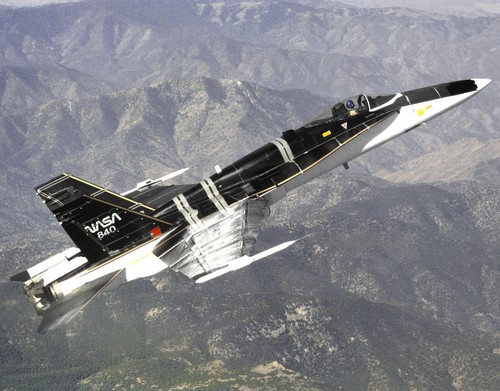}}
        \hspace{1cm}
        \subfigure[X-38 prototype. Source: NASA / Carla Thomas, Photo ID EC99-45080-21]{\label{fig:x38}\includegraphics[width=.4\textwidth]{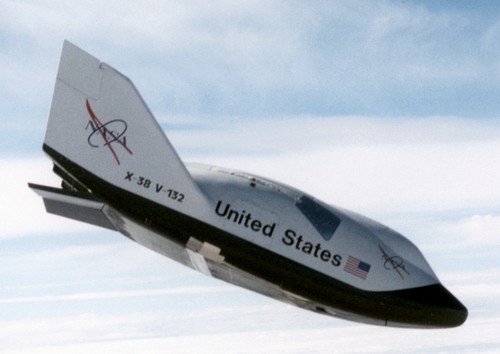}}
        \caption{NASA flight test platforms that played a key-role in early development and application of NDI}
        \label{fig:harvx38}
    \end{figure}

	As computational performance (on flight control computers) rose, NDI could be practically realized. These aforementioned findings and developments were incorporated in Honeywell's NDI-based \emph{MACH (Multi-Application Control Honeywell)} approach~\cite{Wacker2001}, which saw first applications to the F-18 High Angle-of-attack Research Vehicle (HARV), and its first flight on the X-38 prototype in 1998 (\cref{fig:harv} and \cref{fig:x38}). %

\begin{figure}[!htb]
        \centering
        \subfigure[DLR's VFW-614 ATTAS during an automatic landing with NDI in operation in September 2000. \copyright DLR]{\label{fig:ATTAS}\includegraphics[width=.48\textwidth]{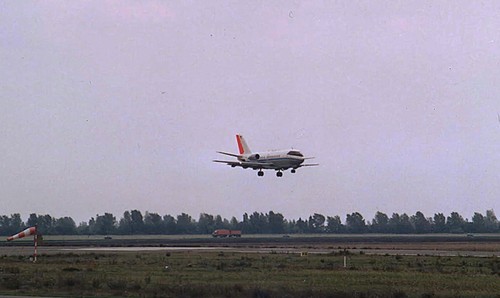}}
       \hspace{0.5cm}
       \subfigure[Conventional (yellow -- 1x), steep (blue -- 2x), and helical (red -- 3x) approach trajectories to Braunschweig runway 26 (blue marker) under 20kt cross wind conditions, flown with NDI. Ground tracks are black, at red markers noise is measured. Source: \cite{Bertsch2010}.]{\label{fig:HeNAP}\includegraphics[width=.44\textwidth]{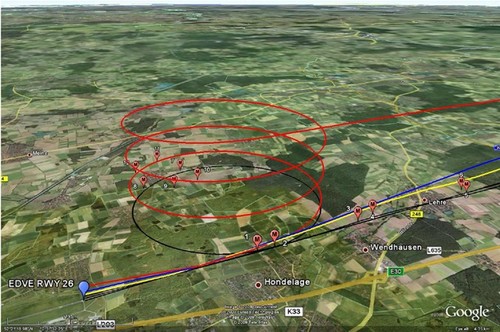}}
        \caption{Flight testing of control laws for automatic landing and helical approach paths}
        \label{fig:ATTAStimes2}
    \end{figure}

	However,~\cite{Enns1994,Wacker2001} highlight three main concerns in a practical NDI application: robustness to model uncertainties, stabilization of the zero dynamics, and computational complexity.
	Uncertainties, mainly from aerodynamics but also weight and balance, may cause the inverse model core not to deliver the expected level of decoupling and linearization. Like in any control system design, they may also cause reduced stability margins, or even instabilities~\cite{Balas1992,Papageorgiou2004}. Thus, different means to achieve robustness
	have been explored, including the use of robust control for designing the feedback controllers~\cite{Balas1992,Reiner1995,Reiner1996,Bennani1998,Buffington1993}, adaptive model updating for reconfiguration and slowly varying parameters~\cite{Lombaerts2009}, or robust offline optimization of the inverted model parameters~\cite{Looye2001}. The latter involved application to an automatic landing system that was successfully flight tested~\cite{Looye2001} on DLR's VFW-614 test aircraft ATTAS (Advanced Technologies Testing Aircraft System), see~\cref{fig:ATTAS}. Advanced developments were flight tested in 2009, tracking a so-called Helical Noise Abatement Procedure (HeNAP, see \cref{fig:HeNAP}) \cite{Bertsch2011,Looye2011}.
	
	Dynamic inversion has multiple overlaps with other, similar, nonlinear control approaches. 
	Differential flatness and flat systems represent a ``nonlinear extension of Kalman's controllability''~\cite{Fliess1995} that originated from feedback linearization~\cite{Fliess1995,Fliess1999}. With respect to a flat output, they have no zero dynamics.
	Backstepping, introduced by Kokotovic and others~\cite{Kokotovic1992,Krstic1995}, is a recursive design methodology for nonlinear systems that guarantees stability via Lyapunov functions. Thus, multiple control laws can be suitable candidates for Backstepping, including NDI, which is often a suitable and self-evident candidate.
	However, Backstepping often leads to an explosion in complexity due to its step-by-step, recursive nature and, like NDI, its model dependency. 
	So-called Block Backstepping has been shown to offer superior global stability guarantees over NDI~\cite{Knoos2012}.
	
	NDI has shown its practical applicability in several flight tests~\cite{Balas2003}, for instance, on the ATTAS~\cite{Looye2006}, the PH-LAB~\cite{Grondman2018,Weiser2024}, the X-35~\cite{Walker2002}, X-36~\cite{Brinker2001}, and X-38~\cite{Wacker2001}, X-48B~\cite{Goldthorpe2010}, and similarly in the EF-2000~\cite{Osterhuber2004}.
	Today, NDI is an established methodology that is widely applied to aerospace systems, especially to decouple complex system behavior along and around the aircraft axes, e.g. see~\cite{Milz2026jgcd}.
	
	\subsection{Incremental Developments}\label{sec:history:indi}
	
\begin{figure}[!htb]
		\centering
		\subfigure[VAAC Harrier \copyright{} Tony Hisgett, CC-BY-2.0]{\label{fig:VAAC}\includegraphics[width=0.56\textwidth]{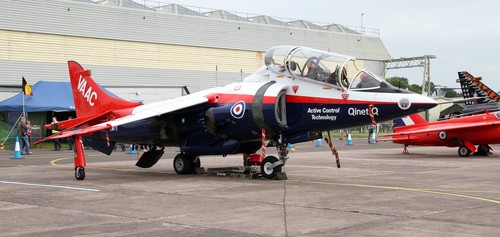}}
		\subfigure[Lockheed-Martin F-35A (Photo by Sr. Airman Julius Delos Reyes, U.S. Air Force Photo, Public Domain)]{\label{fig:F35}\includegraphics[width=0.395\textwidth]{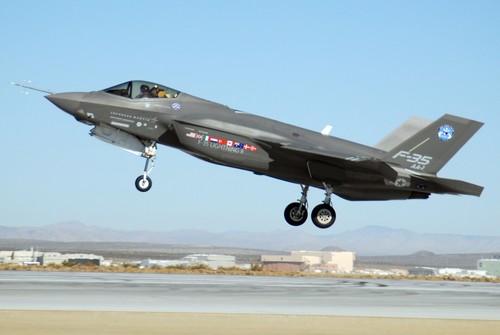}}
		\caption{Application of incremental NDI in military aircraft.}\label{fig:INDIappl}
	\end{figure}

	In 1991, the first approaches to flight control that used ``measurement feedback to eliminate the need for detailed aircraft
	models in outer-loop control applications''~\cite{Antoniewicz1991} were implemented. Although not equal to current concepts, they can be seen as the first step towards incorporating measurements to (partly) eliminate the on-board plant model in NDI.
	In 1998, Smith introduced ``a simplified approach to nonlinear dynamic inversion-based flight control,'' using rotational accelerometer measurements to simplify the inversion of the rotational equations of motion~\cite{Smith1998}. This approach, demonstrated in simulation, was the first complete use of sensor measurements in NDI control to reduce dependence on the system model, particularly the aerodynamic model. The resulting control law requires only information on the control effectiveness. The approach has subsequently been flight-tested on the VAAC Harrier~\cite{Smith2000}, see \cref{fig:VAAC}. 
	
	Bacon and Ostroff extended this approach to an online design of a ``reconfigurable flight control using nonlinear dynamic inversion with a special accelerometer implementation.''~\cite{Bacon2000} They use the Taylor series approximation of the nonlinear system to handle non-affine control inputs and apply a weighted pseudo-inverse to the linearized control effectiveness matrix.
	In~\cite{Chen2008} and~\cite{Sieberling2010}, the authors introduced the concept of incremental nonlinear dynamic inversion (INDI), subsequently referred to as \textit{incremental NDI}. This concept was based on a new derivation and resulted in a more simple formulation of the control algorithm.
	Sieberling \textit{et al.} considered the robustness or insensitivity to uncertainties in the model as a main motivation, albeit a sensitivity to sensor measurements~\cite{Sieberling2010}.
	
\begin{figure}[!htb]
		\centering
		\subfigure[FASER UAV during preparations for first incremental NDI flight that applied synchronization principle. \copyright DLR]{ \label{fig:FASER}\includegraphics[width=0.405\textwidth]{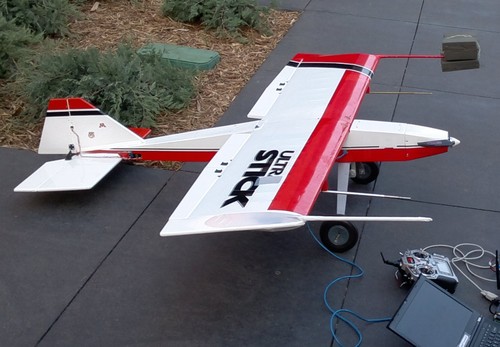}}
		\subfigure[Cessna Citation II (Model 550) Research Aircraft \textit{PH-LAB}\ \copyright DLR]{\label{fig:phlab2}\includegraphics[width=0.5\textwidth]{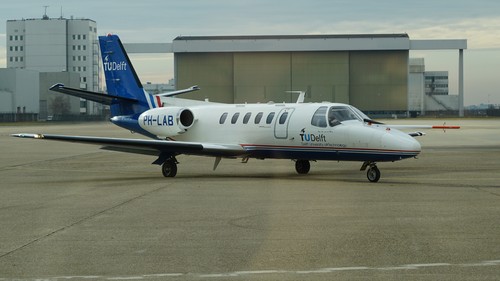}}
		\caption{Research platforms used for flight testing NDI-based and back-stepping control laws}\label{fig:INDIappl2}
	\end{figure}

	In~\cite{Chen2008}, Chen \textit{et al.} used a weighted pseudo-inverse of the control effectiveness matrix to distribute control commands across three aerodynamic surfaces and a 2-D thrust-vector system to track angular rates.
	The first implementation and flight test of Incremental NDI in its evolved form was done on a UAV in 2013~\cite{Vlaar2014}, see~\cref{fig:FASER}. It was during the preparation for this work that the principle of \textit{synchronization} between angular acceleration and control deflection measurement or estimation was found to be key to successful implementation with reasonably low sampling times. A formal analysis was published~\cite{Smeur2016}, with first application to, and implementation on a quadrocopter.
	
	The first successful implementation and test on a passenger aircraft took place in early 2017~\cite{Grondman2018} using the Cessna Citation PH-LAB (\cref{fig:phlab2}, \cite{Scholten2020}). In this campaign, actual control surface measurement, and later angular acceleration sensors were introduced~\cite{Keijzer2019}, replacing estimates of those. Both the FASER and Citation aircraft were used to test and compare back-stepping and LPV control laws as well \cite{Ekeren2018,Weiser2018,Keijzer2019,Weiser2020}. The production Lockheed Martin F-35 JSF~\cite{Harris2018} (\cref{fig:F35}) employs a similar control concept to incremental NDI for its control laws, enabling maneuvers up to 50 degrees Angle-of-Attack without necessarily requiring supplementary effectors like Thrust Vectoring Control, thus exploiting the full aerodynamic performance envelope~\cite{Kim2023}.
	
	Incremental NDI also has many similarities with other control concepts. %
	Noteworthy are parallel developments in robotics that led to a method called Time Delay Control (TDC) in the late 1980s~\cite{YoucefToumi1987,Hsia1989}. This method is applicable to nonlinear plants with unknown dynamics and unexpected disturbances. As shown in~\cite{Acquatella2017}, TDC is largely similar to incremental NDI.
	The interplay between NDI and Backstepping is also evident in developments such as Incremental Backstepping, proposed by Acquatella \textit{et al.}~\cite{Acquatella2013}, which blends the recursive nature of Backstepping with incremental control concepts to mitigate model dependency, albeit relying on time-scale separation for its derivation. 
	The evolution towards incremental methods for both NDI and Backstepping highlights a shared pursuit of robustness and reduced model dependence in the face of complex, uncertain nonlinear dynamics.
	
	Incremental NDI uses direct sensor feedback of the matching virtual controls (usually (angular) accelerations) and current control deflections to partially replace model computations, thereby reducing dependency and sensitivity to uncertainties in the latter. However it also raises issues with sensor faults, noise, delays, and disturbances. Despite its robustness to model uncertainties, performance and control activity depend heavily on the quality of sensor measurements, including delays~\cite{Pollack2022}, noise, and accuracy. Because disturbances are directly fed back through sensor measurements, structural loads can become critical when in gust scenarios~\cite{Kier2020}. %
	
	Main topics in current incremental NDI research include control allocation and synchronization. As current trends in aerospace introduce new configurations with multiple distributed actuators, and incremental NDI can be used to achieve general moment (and sometimes force) commands, it is crucial to employ a control allocation that distributes the required moments across the available effectors. For incremental NDI, incremental nonlinear control allocation (INCA)~\cite{Matamoros2018} is a widely used method. Current research, for example, focuses on applying findings from robust control to incremental NDI~\cite{Pollack2023,Pollack2024,Encarnacao2026}. Additionally, synchronization of measurements - especially actuator and aircraft sensor measurements - is an ongoing field of research~\cite{Pollack2022,Steffensen2022,Steffensen2023,Raab2019}.
	Significant efforts are directed toward extending NDI capabilities with learning-based and adaptive concepts~\cite{Smeur2016,Johnson2000b,Wise1999,Brinker2001,Calise2000,Holzapfel2004}. Since dynamic inversion is a model-based control method that - also in incremental NDI - requires accurate knowledge of certain model components, data-driven adaptation is a logical extension. 
	Common approaches combine dynamic inversion with \( \mathcal{L}1 \)-adaptive control~\cite{Snyder2022,Li2022,Cheng2023}, where the inverted control command is augmented with the adaptive control command, and neural network-based approaches~\cite{Johnson2000b,Wise1999,Zhang2026,Kim2026,Gu2020}. Notably, the dynamic inversion architecture seems to remain advantageous in these contexts, as it is frequently integrated into such concepts. For example, the X-36 program employed an adaptive element with an inversion core to handle the complex and unstable flight dynamics~\cite{Brinker2001,Calise2000}.
	
	Milz \textit{et al.}~\cite{Milz2022} published a sensory NDI formulation, an evolutionary step between NDI and incremental NDI, that calculates absolute actuator position commands and traces back to sensor measurements to reduce reliance on an explicit aircraft model, as in~\cite{Smith1998}. The method is termed \textit{sensory} instead of \textit{sensor-based}, as all NDI variants rely on sensor measurements, e.g., for state reconstruction, but sensory NDI additionally replaces the compensated model dynamics with direct sensor measurements.
	
	A thorough overview of the developments and the state-of-the-art in incremental NDI and its derivatives is provided by Steinert \textit{et al.} in~\cite{Steinert2025a,Steinert2025b}.
	
	\subsection{Hybrid Nonlinear Dynamic Inversion}\label{sec:history:hndi}
	
	Hybrid Nonlinear Dynamic Inversion, also known as Hybrid Incremental Nonlinear Dynamic Inversion~\cite{Kumtepe2022}, has evolved into a robust flight control methodology to address the limitations inherent in the previously mentioned approaches. Conventional NDI exhibits susceptibility to model uncertainties, unmodeled dynamics, and system failures due to its reliance on an approximated on-board plant model. Sensory and incremental NDI emerged as more model-independent alternatives, deriving (incremental) control commands from measured changes in states and inputs, thereby enhancing robustness against such uncertainties. However, pure sensory and incremental NDI face challenges related to sensor noise, measurement delays, and the accurate estimation of angular accelerations through differentiation. The ``hybrid'' concept within NDI was developed to mitigate these conflicting requirements by combining model-based and sensor-based information. This approach employs a complementary filter to fuse the higher-frequency components of smooth estimates from an on-board plant model with the lower-frequency components from on-board sensor measurements~\cite{Jiali2016}. Since then, it has been used in multiple publications, including~\cite{Steffensen2023,Kumtepe2022,Akkinapalli2018,Ji2021,Kim2021,Milz2024e,Surmann2024,Autenrieb2025}.
	Previous works had selected the hybrid approach primarily to handle measurement noise in the angular acceleration feedback signal. It is, however, not just the noise that causes trouble, but also input disturbances fed into the dynamics.
	Pollack~\cite{Pollack2024} identified three hybrid architectures distinct through their compensation gain and (low-pass) filter function configurations. 
	The \textit{complementary filter scheme}~\cite{Kumtepe2022} fixes the gain while optimizing the filter, while the \textit{scaling gain scheme}~\cite{Kim2021} constrains the filter and optimizes the compensation gain.
	The \textit{scaled complementary filter scheme}, allowing both parameters to be free design variables, has been demonstrated to be an effective solution for control design~\cite{Pollack2024} by introducing lead-lag compensation that provides an additional design degree of freedom.
	Despite requiring an on-board model and additional sensor measurements, this method significantly reduces sensitivity to uncertainties and disturbances. Furthermore, in the case of an air data failure, degradation can be mitigatedDynamic  in certain configurations by disabling the air data-dependent model and reducing feedback to inertial measurements only~\cite{Milz2024e}.
	
	Although initially labeled as hybrid INDI, it represents a hybrid between incremental or sensory NDI and NDI. We will use the term \textit{hybrid NDI} in the remainder of this paper, since the incremental form of NDI will not be used.

	\begin{figure}[!htb]
		\centering
		\subfigure[Joby Aviation S4 experimental eVTOL aircraft \copyright{} Harlan Huntington - US Air Force photo ID 230920-F-CC248-1015]{ \label{fig:joby}\includegraphics[width=0.455\textwidth]{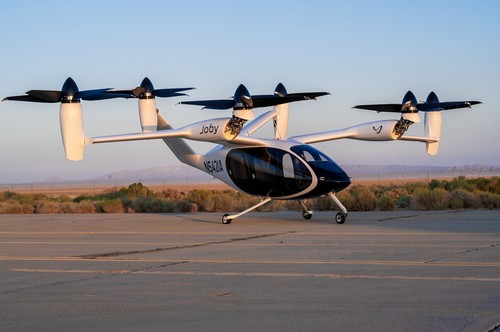}} \hspace{0.5cm}
		\subfigure[Vertical Aerospace VX4 at the 2026 Farnborough Airshow \copyright{} Oren Rozen CC BY-SA 4.0]{\label{fig:vertical}\includegraphics[width=0.458\textwidth]{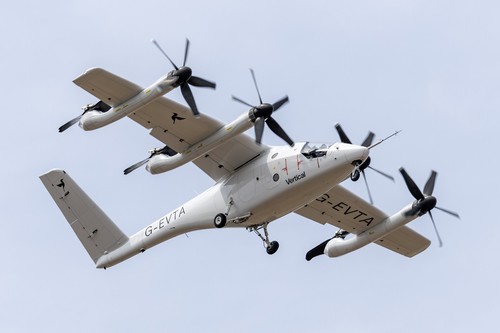}}
		\caption{eVTOL applications of NDI under certification}\label{fig:eVTOL}
	\end{figure}	
	
	\subsection{Applications in Aerospace}
	The historical development outlined above is mirrored by a steadily broadening range of applications that made it into flight test or operation. The first flight applications were realized on fixed-wing research aircraft, starting with the NASA/FAA DHC-6 and NASA C-8A Buffalo in the 1970s~\cite{Meyer1975,Franklin1986}, followed by military experimental programs like the X-35~\cite{Walker2002}, X-36~\cite{Brinker2001}, and X-38~\cite{Wacker2001}, and the X-48B blended wing body~\cite{Goldthorpe2010}. Military industrial programs like the production Lockheed Martin F-35~\cite{Harris2018} use NDI, as well as, in a similar form, the Eurofighter Typhoon~\cite{Osterhuber2004}. Honeywell's MACH program demonstrated the reusability of the architecture across applications~\cite{Wacker2001} and the technology has been transferred to several civil aircraft. On the transport-category side, new (at the time) NDI variants have been flight tested on DLR's VFW-614 ATTAS~\cite{Looye2006} and the TU-Delft/NLR Cessna Citation II PH-LAB~\cite{Grondman2018,Weiser2024}, as well as on the VAAC Harrier for V/STOL research~\cite{Smith2000}. Beyond fixed-wing aircraft, dynamic inversion methods have been applied to helicopters~\cite{Simplicio2013}, (micro) drones and multirotors~\cite{Hussein2022,Smeur2016}, and, more recently, eVTOL configurations~\cite{Raab2018,Milz2026jgcd}. Young companies like Fusion Engineering offer flight control computers with standard control laws based on incremental NDI~\cite{Hussein2022}. NDI even found its way into airship \cite{Paiva2025} and underwater and aerial/underwater robotic applications~\cite{Wang2026,Chen2020}. %
	
	Two commercial eVTOLs where NDI is planned to be certified with the production models are under development by Joby Aviation~\cite{Joby2026}, see \cref{fig:joby}, and Vertical Aviation~\cite{MacMillen2024} (\cref{fig:vertical}). %

	\section{A Mathematical Perspective}\label{sec:mathematical}
	
	The objective of this section is to lay the mathematical groundwork that Nonlinear Dynamic Inversion is built on. This will be the basis for a deeper discussion on methodological aspects of the various variants.
	
	Let \( \vec{x} \in \mathcal{X} \) denote the state vector, \( \vec{u} \in \mathcal{U} \) the input vector, \( \vec{y} \in \mathcal{Y} \) the output vector, and \( \vec{d} \in \mathcal{D} \) an exogenous disturbance. These influence the state derivatives and the output within the sets of reachable states \( \mathcal{X} \subseteq \mathbb{R}^{n_x} \), realizable inputs \( \mathcal{U} \subseteq \mathbb{R}^{n_u} \), achievable outputs \( \mathcal{Y} \subseteq \mathbb{R}^{n_y} \), and disturbances \( \mathcal{D} \subseteq \mathbb{R}^{n_d} \).
	Furthermore, let \( F: \mathcal{X} \times \mathcal{U} \times \mathcal{D} \to W_F \subseteq \mathbb{R}^{n_x} \) and \( H: \mathcal{X} \times \mathcal{U} \times \mathcal{D} \to W_H \subseteq \mathbb{R}^{n_y} \) be smooth and differentiable vector fields comprising the state and output dynamics.
	Then, a generic nonlinear state-space system \( \Sigma \) can be stated as
	\begin{subequations} \label{eq:gennlss1}
		\begin{alignat}{1}[left=\Sigma : \empheqlbrace]
			\dot{\vec{x}} &= F(\vec{x},\vec{u},\vec{d}) \\
			\vec{y} &= H(\vec{x},\vec{u},\vec{d}) \label{eq:gennlss}
		\end{alignat}
	\end{subequations}
	For dynamic inversion, only a subset of the output, consisting of the commanded variables \( \vec{y} \) on which the inversion is performed and which are to be tracked, is considered (cf.~\cref{sec:method:mod:ca}).
	
	In the following, assume that the output is solely state-dependent and has no direct feed-through, as this is a prerequisite for feedback linearization.
	For now, we will neglect the disturbance \( \vec{d} \) (see~\cref{sec:method} for a discussion with disturbances).
	For the derivation of certain dynamic inversion functions, it is handy to separate the dynamics into a state- and input-dependent part.
	Thus, let \( \vec{f}: \mathcal{X} \to W_f \subseteq \mathbb{R}^{n_x} \), \( \vec{g}: \mathcal{X} \times \mathcal{U} \to W_g \subseteq \mathbb{R}^{n_x} \), and \( \vec{h}: \mathcal{X} \to \mathcal{Y} \) be smooth vector fields comprising the \textit{drift}, also called \textit{internal}, \textit{input}, and \textit{output} dynamics.
	The non-affine system \( \Sigma_g \) can be written as
	\begin{subequations}\label{eq:nonaffine_system}
		\begin{alignat}{1}[left=\Sigma_g : \empheqlbrace]
			\dot{\vec{x}}        & = \vec{f}(\vec{x}) + \vec{g}(\vec{x}, \vec{u}) \\
			\vec{y} &= \vec{h}(\vec{x})
		\end{alignat}
	\end{subequations}
	
	In most cases, the input function \vec{g} is (assumed to be) input- or control-affine, or is approximated as such in the sense of approximate feedback linearization~\cite{Krener1984,Hauser1992,Hauser1992vstol}, i.e., 
	\begin{equation}
		\vec{g}(\vec{x}, \vec{u}) = \mat{G}(\vec{x}) \, \vec{u} = \sum\limits_{i=1}^{n_u} \vec{g}_i(\vec{x}) u_i
	\end{equation}
	with the control input matrix \( \mat{G} (\vec{x}): \mathcal{X} \to \mathbb{R}^{n_x\, \times\, n_u}\), yielding the control-affine nonlinear state-space system
	\begin{subequations}\label{eq:affine_system}
		\begin{alignat}{1}[left=\Sigma_G : \empheqlbrace]
			\dot{\vec{x}} & = \vec{f}(\vec{x}) + \mat{G}(\vec{x}) \, \vec{u} \\
			\vec{y}       & = \vec{h}(\vec{x})
		\end{alignat}
	\end{subequations}
	
	In the following, the control-affine system is mostly considered, as the theoretical framework is more rigorous in that case. However, there is also literature considering the non-affine case~\cite{Henson1990,Nijmeijer1990}.
	Given the systems, different approaches to dynamic inversion can be made:
	\begin{enumerate}
		\item The practical and control design approach, where the inverse of a system is sought (cf.~\cref{sec:inversearchitecture:di}), as shown in, e.g.,~\cite{Steinert2025a,Steinert2025b} and in~\cref{sec:method}. This corresponds to the input-output linearization approach shown in~\cref{sec:mathematical:lie}.
		\item The mathematical or system theoretic approach - the \textit{geometric control} view (cf.~\cref{sec:mathematical:lie}) - where the system is transformed into an equivalent linear representation using the differential geometry theory, as shown in, e.g.,~\cite{Krener1973,Brockett1978,Jakubczyk1980a,Hunt1983}. In this approach, either exact input-state or full state linearization, or input-output linearization (cf.~\cref{sec:mathematical:lie}) can be performed.
	\end{enumerate}
	A thorough and rigorous description of the theory introduced subsequently is provided by~\cite{Isidori1985,Khalil2002,Slotine1991,Lin1994}.

	\subsection{Geometric Control Theory}\label{sec:mathematical:lie}
	
	The core idea of geometric control is to shift the perspective from analyzing equations to analyzing geometry using differential geometry. This allows a transfer of the concepts of controllability, reachability, and observability to nonlinear systems.
	Feedback linearization can then be employed to transform the nonlinear system into an equivalent linear one, yielding:
	\begin{enumerate}
		\item \textbf{Input-state}, \textbf{full state}, or \textbf{exact feedback linearization}: The entire dynamics and the full state vector are transformed to evolve according to linear dynamics. This is a less common method for practical design, as the prerequisites are strict and hardly apply to practically relevant systems.
		\item \textbf{Input-output linearization} uses specified outputs to achieve a direct relation between input and output, without directly considering the states (cf.~\cref{sec:mathematical:reldeg}f.). However, this approach can leave out part of the system dynamics, resulting in internal or zero dynamics (cf.~\cref{sec:mathematical:normal}).
	\end{enumerate}

	A crucial step in geometric control theory is the coordinate-invariant representation of the system and control concepts.
	The state space \( \mathcal{X} \) is an \( n_x \)-dimensional differentiable manifold, i.e., locally Euclidean.
	At each point \( \vec{x}_0 \in \mathcal{X} \), there exists a tangent space \( T_{\vec{x}_0} \mathcal{X} \), which is obtained by differentiating the manifold at \( \vec{x}_0 \) and represents all possible velocity vector directions at \( \vec{x}_0 \). It can also be seen as the local linearized system.
	The collection of all tangent spaces on \( \mathcal{X} \) is the tangent bundle \( T \! \mathcal{X} \).
	The (time-dependent) state trajectory \( \vec{x}(t) \in \mathcal{X} \) is an integral curve on \( \mathcal{X} \), i.e., its tangent vector \( \dot{\vec{x}}(t) \in T \! \mathcal{X} \) is given by smooth vector fields that assign every point \( \vec{x}_0 \in \mathcal{X} \) a tangent vector in \( T_{\vec{x}_0} \mathcal{X} \) consisting of:
	\begin{itemize}
		\setlength{\itemsep}{0pt}
		\item The drift vector field \( \vec{f}(x) \) describes the system dynamics in the absence of control.
		\item The control vector field \( \mat{G}(\vec{x}) = \left[ \vec{g}_1, \ldots, \vec{g}_{n_u} \right] \) defines the direction in which the control input \( \vec{u} \) influences the state's velocity.
	\end{itemize}
	
\begin{figure}[!htb]
		\centering
		\includegraphics{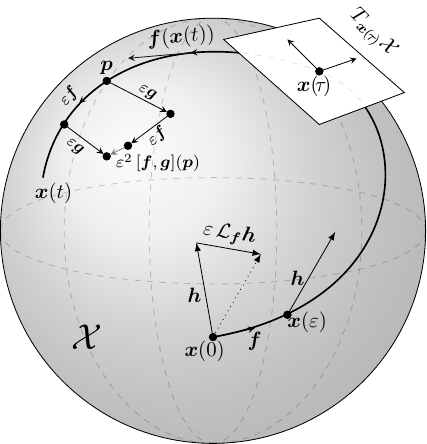}
		\caption{Graphical sketch of geometric control concepts on a manifold \( \mathcal{X} \) defined by the system \( \dot{\vec{x}}(t) = \vec{f}(\vec{x}(t)) + \vec{g}(\vec{x}(t)) \vec{u} \) and the observer function \( \vec{h} \) including the Lie derivative \( \mathcal{L}_\vec{f} \vec{h} \), the tangent space \( T_x \mathcal{X} \), and the Lie bracket \( \left[ \vec{f}, \vec{g} \right] \) at \( \vec{p} \) for an infinitesimal \( \varepsilon \).}
		\label{fig:lie_manifold}
	\end{figure}

	Lie derivatives generalize the concept of differentiation to be coordinate-invariant and applicable along vector fields. They allow analyzing how a smooth (output) vector field function \( \vec{h}(\vec{x}) \) changes along a trajectory defined by a smooth vector field \( \vec{f}(\vec{x}) \) as
	\begin{align}
		\mathcal{L}_\vec{f} \vec{h}(\vec{x}) &= \frac{\partial \vec{h}}{\partial \vec{x}} \vec{f}(\vec{x}) = \sum\limits_i \frac{\partial \vec{h}}{\partial x_i} f_i(\vec{x})
	\end{align}
	where a concatenation of Lie derivatives is denoted as, e.g., \( \mathcal{L}_\vec{f}^2 \vec{h}(\vec{x}) = \mathcal{L}_\vec{f} \left( \mathcal{L}_\vec{f} \vec{h}(\vec{x}) \right) \).
	Geometrically, \( \mathcal{L}_\vec{f} \vec{h}(\vec{x}) \) is the rate of change of \( \vec{h} \) for an observer moving along the integral curve of the vector field \( \vec{f} \). This concept is graphically illustrated in~\cref{fig:lie_manifold}.
	
	The Lie bracket of two vector fields \( \vec{f} \) and \( \vec{g}_i \) is a third vector field, denoted \( \left[ \vec{f}, \vec{g}_i \right] \), that captures the non-commutativity of the nonlinear dynamics and the interactions between the flows, interpreted as ``virtual control direction'', and is defined as:
	\begin{equation}
		\left[ \vec{f}, \vec{g}_i \right] (\vec{x}) := \frac{\partial \vec{g}_i}{\partial \vec{x}} \vec{f} (\vec{x}) - \frac{\partial \vec{f}}{\partial \vec{x}} \vec{g}_i (\vec{x})
	\end{equation}
	The new vector field \( \left[ \vec{f}, \vec{g}_i \right] \) points in the direction of the change created by moving along \( \vec{f} \) and \( \vec{g}_i \) combined.
	The geometric interpretation is the residuum of an infinitesimal motion sequence along \( \vec{f} \) then \( \vec{g}_i \) versus \( \vec{g}_i \) then \( \vec{f} \), which is zero exactly when the flows of the two vector fields commute, but is in general non-zero. Even for a linear system (\(  \dot{\vec{x}} = \mat{A} \vec{x} + \mat{B} \vec{u} \)), the bracket \( \left[ \vec{f}, \vec{g} \right] = - \mat{A} \mat{B} \) does not vanish but its repeated application precisely produces the direction of Kalman's controllability matrix.
	Thus, a non-zero Lie bracket means that the system's natural drift \( \vec{f} \) can drag the control direction \( \vec{g}_i \) to create motion in a new direction that was not directly accessible through \( \vec{f} \) or \( \vec{g} \) alone. This new direction is a virtual control vector field, which can be employed by the controller.
	The Lie bracket can be nested as the drift acting on the new direction can generate yet another direction, i.e., \( \left[ \vec{f}, \left[ \vec{f}, \vec{g} \right] \right] \). Repeated Lie brackets generate an entire set of virtual control directions and are defined as
	\begin{equation}
		\mathrm{ad}_\vec{f} \vec{g} := \left[ \vec{f}, \vec{g} \right], \ \mathrm{ad}_\vec{f}^2 \vec{g} = \left[ \vec{f}, \left[ \vec{f}, \vec{g} \right] \right], \ \ldots
	\end{equation}
	
	A distribution \( \Delta (\vec{x}) \subset T_\vec{x} \mathcal{X} \) on a manifold \( \mathcal{X} \) is a smooth assignment of a subspace to the tangent space \( T \! \mathcal{X} \). For a control system, distributions are used to represent the subspaces of directions in which the system can evolve, e.g.,
	\begin{equation}
		\Delta_\vec{g} (\vec{x}) = \mathrm{span} \left\{ \vec{g}_1 (\vec{x}), \ldots, \vec{g}_{n_u} (\vec{x}) \right\}
	\end{equation}
	represents the directions accessible through the control input.
	A distribution is \textit{involutive} if the Lie bracket of any two vector fields within it remains within the distribution itself, i.e.,  for any two smooth vector fields \( \vec{f}, \vec{g} \),
	\begin{equation}
		\vec{f}(\vec{x}), \vec{g}(\vec{x}) \in \Delta (\vec{x}) \implies \left[ \vec{f},\vec{g} \right](\vec{x}) \in \Delta (\vec{x}) \quad \forall \vec{x} \in \mathcal{X}
	\end{equation}
	
	Involutivity is a closure property under the Lie bracket operation. Thus, if the distribution \( \Delta_\vec{g} \) is involutive, it describes all possible directions in which a system with absent drift dynamics can evolve through a control input.
	A distribution \( \Delta \) is said to be \textit{integrable} if the manifold \( \mathcal{X} \) can be partitioned into an submanifold \( \mathcal{X}_1 \subset \mathcal{X} \), called an integral manifold of \( \Delta \), such that the tangent space of \( \mathcal{X}_1 \) is precisely \( \Delta (\vec{x}) \):
	\begin{equation}
		\exists \mathcal{X}_1 \subset \mathcal{X}:\ T_\vec{x} \mathcal{X}_1 = \Delta (\vec{x}) \quad \forall \vec{x} \in \mathcal{X}_1
	\end{equation}
	Loosely speaking, the distribution (representing the state's velocity) is integrable if its integral is part of the manifold (representing the state space). The Frobenius theorem combines both concepts: a smooth nonsingular (constant-dimensional) distribution \( \Delta \) on a manifold \( \mathcal{X} \) is integrable if and only if it is involutive.
	
	With these definitions, we can state the necessary and sufficient conditions for a system to be exactly state-linearizable~\cite{Jakubczyk1980a}:
	
	\begin{theorem}{(Existence of an exact feedback linearization~\cite{Jakubczyk1980a,Su1982,Hunt1983})}
		Consider the control-affine system \( \Sigma_G \) in~\cref{eq:affine_system} with a single input (\( n_u = 1,\ \mat{G} = \vec{g} \)). \( \Sigma_G \) is \emph{exactly state feedback linearizable} in a neighborhood of \( \vec{x}_0 \in \mathcal{X} \), i.e., there exists a diffeomorphism \( \vec{z} = T(\vec{x}) \) and a regular static feedback \( u = ... \) that transform \( \Sigma_G \) into a controllable linear system, if and only if:
		\begin{enumerate}
			\setlength{\itemsep}{0pt}
			\item Rank (controllability) condition: The set of vector fields \( \left\{ \vec{g}, \mathrm{ad}_\vec{f} \vec{g}, \ldots, \mathrm{ad}_\vec{f}^{n_x-1} \vec{g} \right\} \) must be linearly independent, which is equivalent to the ``nonlinear controllability matrix'' \( \left[ \vec{g}(\vec{x}_0) \ \ \mathrm{ad}_\vec{f} \vec{g}(\vec{x}_0)\ \cdots \ \mathrm{ad}_\vec{f}^{n_x-1} \vec{g}(\vec{x}_0) \right] \) having rank \( n_x \).
			\item Involutivity condition: The distribution spanned by the vector fields \( \left\{ \vec{g}, \mathrm{ad}_\vec{f} \vec{g}, \ldots, \mathrm{ad}_\vec{f}^{n_x-2} \vec{g} \right\} \) is involutive in a neighborhood of \( \vec{x}_0 \).
		\end{enumerate}
	\end{theorem}
	Proofs are given in~\cite{Jakubczyk1980a}, \cite[Thm.~4.2.3]{Isidori1995}, \cite[Thm.~13.2]{Khalil2002}, and~\cite[Prop.~6.16]{Nijmeijer1990}.
	The system is \emph{locally} exact state-linearizable if the conditions hold in a neighborhood of \( \vec{x}_0 \in \mathcal{X} \). If the two conditions hold at every \( \vec{x} \in \mathcal{X} \), the system is linearizable around every point. However, a single \emph{global} transformation exists only under additional assumptions~\cite{Hunt1983}. %
	For the multi-input case, analogous rank and involutivity conditions are stated on a nested sequence of distributions spanned by the columns of \( \mat{G} \) and their repeated Lie brackets with \( \vec{f} \)~\cite{Isidori1985,Hunt1983}.
    The conditions generalize to the nested distributions \( \mathcal{G}_i = \mathrm{span}\{ \mathrm{ad}_\vec{f}^k \vec{g}_j : 0 \le k \le i,\ 1 \le j \le n_u \} \): each \( \mathcal{G}_i \), \( i \le n_x - 2 \), must be involutive and of constant dimension, with \( \dim \mathcal{G}_{n_x-1} = n_x \)~\cite{Jakubczyk1980a,Hunt1983,Isidori1995}.
	
	The rank condition is the counterpart to the linear control concept of controllability, where the set of vector fields is similar to Kalman's controllability matrix. The involutivity condition ensures that all states can be fully linearized by state feedback. Frobenius' Theorem guarantees the existence of a coordinate system in which these directions are aligned with the coordinate axes, allowing us to construct the final linearizing transformation.
	
	For the broader case of input-output feedback linearization, the necessary and sufficient conditions are looser and mainly depend on the relative degree (cf.~\cref{sec:mathematical:reldeg}). The main theorem can be stated as~\cite{Kravaris1987,Hirschorn1979}:
	
	\begin{theorem}{(Existence of an input-output feedback linearization~\cite{Isidori1995,Falb1967,Kravaris1987,Henson1990})}\label{thm:iofbl}
		The control-affine system \( \Sigma_G \) from~\cref{eq:affine_system} is, in a neighborhood of \( \vec{x}_0 \in\mathcal{X} \), input-output feedback linearizable by a regular static state feedback if and only if it possesses a finite positive integer vector \( \vec{\rho} = [ \rho_1, \ldots, \rho_{n_y} ] \) (the relative degree vector) such that
		\begin{equation}
			\left( \frac{\partial \vec{h}_i}{\partial \vec{x}} \right)^T \ \mathrm{ad}_\vec{f}^{\rho_i - 1} \vec{g}_j (\vec{x}) = {(-1)}^{\rho_i - 1} \mathcal{L}_{\vec{g}_j} \mathcal{L}_\vec{f}^{\rho_i - 1} \vec{h}_i (\vec{x}) \neq 0 \ \ \forall j,\ 1 \leq i \leq n_y
		\end{equation}
		where \( \rho_i \) is the smallest positive integer fulfilling the equation.
		Furthermore, a system is input-output linearizable if and only if it is left-invertible~\cite{Henson1990}.
	\end{theorem}
	A vector relative degree at \( \vec{x}_0 \) (with \( n_u = n_y \)) implies local left-invertibility: the input is reconstructed from \( \vec{x}_0 \) and the output derivatives via the inverse system~\cite{Hirschorn1979b,Silverman1969}.
	For multivariable systems, left-invertibility does not necessarily imply static input-output linearizability. Square, analytic, left-invertible systems can, however, be input-output	linearized and decoupled by \emph{dynamic} state feedback~\cite{Singh1981,Descusse1985,NijmeijerRespondek1988} (cf.~\cref{sec:mathematical:nonaffine}).
	
	A detailed derivation of input-output feedback linearization is provided subsequently (cf.~\cref{sec:mathematical:reldeg}f.), as this is the main approach applied in control engineering.
	The geometric framework reveals that exact state linearization is not a separate theory but rather a special case of input-output linearization. Both scenarios try to find a coordinate system that aligns with the structure of the system dynamics. Input-output linearization achieves this for a portion of the state space tied to a given output, potentially leaving behind internal dynamics. Exact state linearization can only be applied if the system's structure is so regular that an output can be defined for which the linearized portion encompasses the entire state space, resulting in no zero dynamics.

	\subsection{Relative Degree}\label{sec:mathematical:reldeg}
	
	To invert a given system, we evidently need an algebraic relation between the sought input \( \vec{u} \) and the desired output \( \vec{y} \). As this relation is in general not given a priori, we differentiate the output until this relation occurs, i.e., effectively seeking the immediate output (derivative) response to an input, simultaneously ensuring a proper transfer function.
	The number of times an output needs to be differentiated to establish this relation is called the \textit{relative degree}.
	
	Let \( \rho_i \) denote the relative degree of the output \( \vec{y}_i \) and be defined as the smallest integer such that~\cite{Isidori1985,Henson1990}
	\begin{subequations}\label{eq:reldeg}
		\begin{alignat}{3}
			\Sigma_g: \quad & \nabla_\vec{u} \mathcal{L}_\vec{g} \mathcal{L}_\vec{f}^{\rho_i-1} \vec{h}_i \, (\vec{x},\vec{u}) && \neq \mat{0} \\
			\Sigma_G: \quad & \mathcal{L}_\vec{G} \mathcal{L}_\vec{f}^{\rho_i-1} \vec{h}_i \, (\vec{x}) && \neq \mat{0}
		\end{alignat}
	\end{subequations}
	This means that the \( \rho_i \)-{th} derivative of the \( i \)-{th} output has a direct, algebraic, same-order relation to the input.
	For multiple inputs and outputs, a well-defined \textit{vector relative degree} additionally requires the decoupling matrix \( \left[ \mathcal{L}_{\vec{g}_j} \mathcal{L}_\vec{f}^{\rho_i-1} \vec{h}_i \, (\vec{x}) \right]_{ij} \) to be nonsingular~\cite{Isidori1985}, which corresponds to the invertibility of \( \mat{\Beta}(\vec{x}) \) required later on (cf.~\cref{sec:mathematical:affine,sec:method:mod:ca}).
	
	Consequently, the relative degree vector of the system \( \vec{\rho} = \left[ \rho_1, \ldots, \rho_{n_y} \right] \) and the system's total relative degree \( \rho = \sum_i \rho_i \leq n_x \) can be defined.
	If \( \rho < n_x \), zero dynamics with internal states of dimension \( n_x - \rho \) occur~\cite{Enns1994}.
	If \( \rho = n_x \), no internal dynamics remain and the system is \textit{full-state linearizable}~\cite{Hauser1992} by static state feedback, with \( \vec{y} \) constituting a flat output. The converse does not hold: \textit{differential flatness}~\cite{Fliess1995} is the weaker property of being linearizable by dynamic feedback, so a flat system need not admit a static-feedback full-state linearization.
	Thus, selecting an appropriate {(pseudo-)}output map \( \vec{h} : \vec{x} \mapsto \vec{y} \) is an important design decision (cf.~\cref{sec:method:mod:ca}).
	The relative degree vector allows us to determine the ``responsive time derivative''~\cite{Steinert2025a} \( \vec{y}^{(\rho)} \) for a system, which has a direct link to the input \( \vec{u} \).

	\subsection{Normal or Canonical Control Form, and Zero Dynamics}\label{sec:mathematical:normal}
	Given the relative degree vector \( \vec{\rho} \) of a system, we can systematically find a way to transform the system into an equivalent system in Brunovsk\'{y} normal form~\cite{Brunovsky1970} or canonical control form~\cite{Su1982,Hunt1983}, i.e., \( \vec{y}^{(\rho)} = \vec{\nu} \) with the pseudo control input \( \vec{\nu} \) of the transformed system and assuming a controllable system. In the presence of internal dynamics (\( \rho < n_x \)), the resulting representation is also known as the Byrnes-Isidori normal form~\cite{Isidori1985}.
	
	Let \( \mathcal{Z} \simeq \mathcal{X} \) be a diffeomorphic manifold of \( \mathcal{X} \). Given a relative degree vector of the system \( \vec{\rho} \) with \( \rho \leq n_x \), then a transformation diffeomorphism \( T: \mathcal{X} \mapsto \mathcal{Z} \) exists that puts the system into normal form with a new state vector \( \vec{z} = T(\vec{x}) \in \mathcal{Z} \). 
	The new state vector \( \vec{z} = \left[ \vec{\xi}, \ \vec{\eta} \right]^T \) consists of \textit{external} (observable) states \( \vec{\xi}, \dim\vec{\xi} = \rho \), and \textit{internal} (non-observable) states \( \vec{\eta}, \dim\vec{\eta} = n_x - \rho \).
	
	The external states can be defined using the first \( \rho_i - 1 \) derivatives of the corresponding output: %
	\begin{equation}
		\vec{\xi}_i = \begin{bmatrix} \vec{y}_i \\ \dot{\vec{y}}_i \\ \vdots \\ \vec{y}_i^{(\rho_i-1)} \end{bmatrix} = \begin{bmatrix} \vec{h}_i (\vec{x}) \\ \mathcal{L}_\vec{f} \vec{h}_i(\vec{x}) \\ \vdots \\ \mathcal{L}^{\rho_i-1}_\vec{f} \vec{h}_i(\vec{x}) \end{bmatrix}
	\end{equation}
	so that \( \dim \vec{\xi}_i = \rho_i \), while the derivative of the last entry,
	\begin{equation}
	    \vec{y}_i^{(\rho_i)} = \mathcal{L}^{\rho_i}_\vec{f} \vec{h}_i(\vec{x}) + \mathcal{L}_\vec{g} \mathcal{L}^{\rho_i-1}_\vec{f} \vec{h}_i (\vec{x},\vec{u})
	\end{equation}
	is the first one to be directly influenced by the input, since \( \mathcal{L}_\vec{g} \mathcal{L}^{k}_\vec{f} \vec{h}_i (\vec{x},\vec{u}) = 0 \) for \( k < \rho_i-1 \) due to \cref{eq:reldeg}. For the non-affine system \( \Sigma_g \) the corresponding condition is \( \nabla_\vec{u} \mathcal{L}_\vec{g} \mathcal{L}^{k}_\vec{f} \vec{h}_i = 0 \) for \( k < \rho_i-1 \); assuming, without loss of generality, \( \vec{g}(\vec{x},\vec{0}) = \vec{0} \) (any input-independent part of \( \vec{g} \) is absorbed into \( \vec{f} \)), these terms then vanish identically as well.
	For a better representation, the external state vector may be reordered such that all \( j \)-th derivatives of the outputs are grouped, i.e., \( \vec{\xi} = \left[ \vec{y}_1 \ \ldots \ \vec{y}_{n_y} \ \dot{\vec{y}}_1 \ \ldots \ \dot{\vec{y}}_{n_y} \ \vec{y}_1^{(\rho_{n_1}-1)} \ \ldots \ \vec{y}_{n_y}^{(\rho_{n_y}-1)} \right] \), where, in the case of differing relative degrees, the \( j \)-th derivative of output \( i \) is only present if \( j \leq \rho_i - 1 \).

	For \( \rho < n_x \), internal dynamics or \textit{zero dynamics} unobservable from the output \( \vec{y} \) exist and represent the remaining dynamics.
	The internal states are described by \( n_x - \rho \) smooth functions \( \vec{\eta} = \vec{\phi}(\vec{x}) \) that complete the transformation \( T \) to a (local) diffeomorphism, i.e., that render its Jacobian nonsingular. If the input distribution \( \Delta_\vec{g} \) is involutive, the Frobenius theorem additionally guarantees that the functions \( \vec{\phi} \) can be chosen such that
	\begin{align}
	    \mathcal{L}_{\vec{g}_j} \phi_i (\vec{x}) = \frac{\partial \phi_i}{\partial \vec{x}} \vec{g}_j(\vec{x}) = 0 \ \forall i,j \label{eq:zdstates}
	\end{align}
	so that the internal dynamics are not directly driven by the input~\cite{Isidori1985}.
	The internal states can be thought of as being independent from the output \( \vec{y} \), i.e., not observable according to classical means. They depend solely on the external states, not directly on the input.
    According to Enns \textit{et al.}, ``conceptually, zero dynamics are nothing more than the remaining motions permitted [...] when the [controlled variables] are constrained to be constant or prescribed.''~\cite{Enns1994}
	It is important to ensure their stability~\cite{Enns1994}.
	A more thorough discussion of internal or zero dynamics is given, e.g., in~\cite{Enns1994,Isidori1985,Khalil2002,Slotine1991,Lin1994}.

\begin{figure*}[!htb]
		\centering
		\includegraphics[]{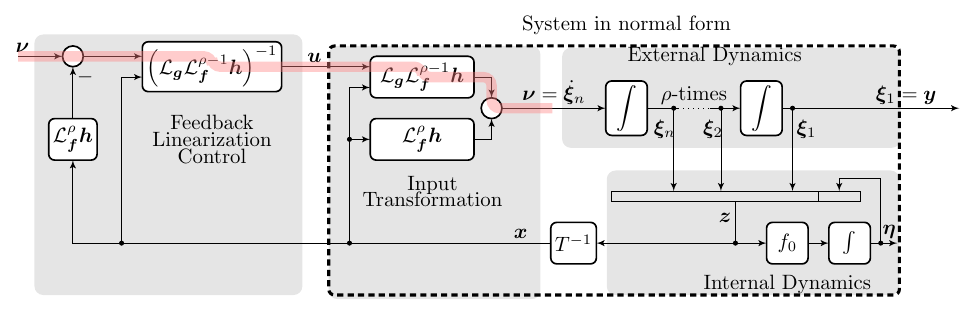}
		\caption{Block diagram schematics of the transformed system in Brunovsk\'{y} normal form or canonical control form.}
		\label{fig:normal_form}
	\end{figure*}

	As the transformation \( T \) is a diffeomorphism, we can use the inverse mapping \( \vec{x} = T^{-1} (\vec{z}) \) to transform the system from~\cref{eq:nonaffine_system} into
	\begin{subequations}\label{eq:canonicalcontrolform}
		\begin{align}
			\dot{\vec{\xi}}  & = \begin{bmatrix} \mat{0} & \mat{I}_{\rho-1} \\ \mat{0} & \mat{0} \end{bmatrix} \vec{\xi} + \begin{bmatrix} \mat{0} \\ \mat{I}_{n_y} \end{bmatrix} \left( \mathcal{L}^{\rho}_\vec{f} \vec{h}(T^{-1}(\vec{z})) + \mathcal{L}_\vec{g} \mathcal{L}^{\rho-1}_\vec{f} \vec{h}(T^{-1}(\vec{z}), \vec{u}) \right) \notag \\
			&=  \mat{A}_\mathrm{c}  \xi + \mat{B}_\mathrm{c} \nu \\
			\dot{\vec{\eta}} & = \mathcal{L}_\vec{f} \vec{\phi} (T^{-1}(\vec{z})) = \vec{f}_0 (\vec{\eta}, \vec{\xi}) \\
			\vec{y} &= \begin{bmatrix} \mat{I}_{n_y} & \mat{0} \end{bmatrix} \vec{\xi} = \mat{C}_\mathrm{c} \vec{\xi}
		\end{align}
	\end{subequations}
	where \( \mat{A}_\mathrm{c} \) and \( \mat{B}_\mathrm{c} \) represent the controllable canonical form of the observable part. The output is \( \vec{y} = \vec{z}_1 \).
	This form separates the system into an input-output linear part around the external dynamics (\( \vec{\xi} \)) and, optionally, the internal dynamics (\( \vec{\eta} \)).
	\cref{fig:normal_form} sketches the transformed system in normal form. The dynamic inversion law can be achieved by inverting the ``feedback linearization'' block, i.e., commanding a pseudo control input \( \vec{\nu} \) while providing a state \( \vec{x} \) to get the desired input \( \vec{u} \).
	
	\subsection{Non-Affine Systems}\label{sec:mathematical:nonaffine}
	For the case of a non-input-affine system \( \Sigma_g \) as defined in \cref{eq:nonaffine_system}, the normal form becomes
	\begin{align}\label{eq:nonaffine_normal}
		\vec{y}^{(\rho)} &= \left[ \mathcal{L}_\vec{f}^{\rho_i} \vec{h}_i \, (\vec{x}) + \mathcal{L}_\vec{g} \mathcal{L}_\vec{f}^{\rho_i-1} \vec{h}_i \, (\vec{x}, \vec{u}) \right]_{i=1 \ldots n_y} \notag \\
		&= \vec{\alpha} (\vec{x}) + \vec{\beta}(\vec{x},\vec{u})
	\end{align}
	
	To be suitable for dynamic inversion, \vec{\beta} must be sufficiently smooth and bijective (or at least surjective), which implies the existence of its (non-)unique (right-)inverse \( \vec{\beta}^{-1} (\vec{x}, \vec{\upsilon} ):  \mathcal{X} \times W_\beta \to \mathcal{U} \) satisfying~\cite{Nijmeijer1990}
	\begin{equation}\label{eq:nonaffine_control_normal}
		\vec{\upsilon} = \vec{\beta} \left( \vec{x} , \vec{\beta}^{-1} (\vec{x}, \vec{\upsilon} ) \right) %
	\end{equation}
	for all reachable states \( \vec{x} \in \mathcal{X} \) and outputs \( \vec{\upsilon} \) describing the effect of the control input on the system.
	Practical implementations of \( \vec{\beta}^{-1} \) can be realized, for instance, with analytic inversions using exact feedback linearization, look-up tables~\cite{Steinhauser2004}, implicit solvers~\cite{Looye2007}, employing time-scale separation~\cite{Hovakimyan2008}, or local linearization and inversion via the Taylor series approximation of \vec{\beta}~\cite{Milz2026jgcd}. See~\cite{Steinert2025a,Nijmeijer1990} for a thorough discussion on non-affine system control using dynamic inversion. %
	
	An alternative approach is to dynamically extend the non-affine system by treating the input as a state and introducing its derivative as a new control input \( \vec{u}^\prime \), thereby converting the system \( \Sigma_g \) from  \cref{eq:nonaffine_system} into an augmented control-affine system~\cite{Zhan1991,Isidori1995,vandeVen2000}
	\begin{equation}
		\begin{bmatrix} \dot{\vec{x}} \\ \dot{\vec{u}} \end{bmatrix} = \begin{bmatrix} \vec{f}(\vec{x}) + \vec{g}(\vec{x}, \vec{u}) \\ \vec{0} \end{bmatrix} + \begin{bmatrix} \mat{0} \\ \mat{I}_{n_u} \end{bmatrix} \vec{u}^\prime
	\end{equation}
	with a higher relative degree, but with affine input dynamics on the input derivatives. The same concept can be applied to systems that lack a properly defined relative degree~\cite{Isidori1995}.

	\subsection{Input-Affine or Affine-in-Control Systems}\label{sec:mathematical:affine}
	
	Most literature on feedback linearization considers only input-affine systems, as the theory for this class is more thorough and rigorous.
	The normal form of this system can be achieved analogously as
	\begin{equation}
		\vec{y}^{(\rho)} = \mathcal{L}_\vec{f}^{\rho_i} \vec{h}_i \, (\vec{x}) + \mathcal{L}_\vec{G} \mathcal{L}_\vec{f}^{\rho_i-1} \vec{h}_i \, (\vec{x}) \cdot \vec{u} = \vec{\alpha} (\vec{x}) + \mat{\Beta}_u(\vec{x}) \, \vec{u} \label{eq:affine_normal}
	\end{equation}
	where \( \mat{\Beta}_u (\vec{x}): \mathcal{X} \to \mathbb{R}^{n_y\, \times\, n_u} \) has to be (pseudo-)invertible for all reachable states \vec{x} in order to guarantee the existence of its (pseudo-)inverse \( {\mat{\Beta}_u^\dagger(\vec{x})} \).
	Input-affinity is typically used because 1) feedback linearization has rigorous theoretical backup and much experience and literature exist for this case, 2) the design complexity reduces drastically, and 3) the assumption holds sufficiently well in many aerospace systems.
	However, current developments bring up novel vehicle configurations with inherently non-affine control inputs, e.g., active flow control, thrust vectoring, and transformational vehicles~\cite{Milz2026jgcd}.

	\section{A Methodological Perspective}\label{sec:method}
	
	In the previous section it has been shown how a system model can be transformed in a normal form relative to its outputs, and how it can be effectively linearized and decoupled by means of feedback and appropriate selection of corresponding virtual control inputs. These are properties that are of great use as a guiding principle for control law design for multi-variable and dynamically nonlinear physical systems, and that build the very core idea behind NDI (\cref{sec:conceptual}). 
	
	The principal methodological aspect is in the appropriate configuration and derivation of an inverse model core within the control law architecture. At this point it is assumed that the overall architecture, including selection of commanded variables and control effectors to be utilized, have been decided upon. These are overarching aspects that are not necessarily NDI-specific and will be addressed from an overall design perspective in~\cref{sec:design}. 
	\vskip 1ex
	
	\noindent Starting point is the nonlinear state-space system in~\cref{eq:gennlss1}:
	\begin{subequations}
		\begin{alignat}{1}[left=\Sigma : \empheqlbrace] %
			\dot{\vec{x}} &= F(\vec{x},\vec{u},\vec{d}) \notag \\
			\vec{y} &= H(\vec{x},\vec{u},\vec{d}) \notag
		\end{alignat}
	\end{subequations}
	The objective is to arrive at~\cref{eq:nonaffine_normal} derived in the previous section:
	\begin{align}
		\vec{y}^{(\rho)} &= \vec{\alpha} (\vec{x}) + \vec{\beta}(\vec{x},\vec{u}) \tag{\ref{eq:nonaffine_normal}}
	\end{align}
	or its affine form:
	\begin{align}
		\vec{y}^{(\rho)} &= \vec{\alpha} (\vec{x}) + \mat{\Beta}(\vec{x}) \vec{u} \tag{\ref{eq:affine_normal}}
	\end{align}
	that can be practically inverted and implemented as a core in the NDI control law. 
	
	The particular relevance of this equation is in that it contains all aspects and influences that the NDI core will be capable of compensating for. It may be decided that specific influences are neglected and left to be sorted out by feedback loops in outer functions. Even though the subsequent inversion is mathematically unambiguous, the realization may be done in different ways, giving rise to the different variants of NDI introduced in~\cref{sec:history}. 
	
	As discussed in~\cref{sec:architecture}, the NDI model core normally does not come with any state equations. The state variables that the equations depend on are measured, computed, or estimated from sensors on the physical system. This means that the sensor system and signal processing algorithms (\cref{fig:NDItypicalArchitecture}) need to be capable to do so. A state transformation on beforehand, e.g. by using the normal form in~\cref{eq:canonicalcontrolform}, may reduce computational effort at this point. 
	
	From a methodological point of view, configuration and derivation of the inverse model core for a given design can be divided into three steps:
	\begin{itemize}
		\item Selection of suitable generalized controls, enabling actual inversion of~\cref{eq:nonaffine_normal} or~(\ref{eq:affine_normal}) in an unambiguous way. Generalized controls are in turn realized by a separately designed control allocation function.
		\item Model Configuration: Simplification of the model equations and computation of commanded variable outputs in~\cref{eq:nonaffine_normal} or~(\ref{eq:affine_normal}) to an appropriate level. This implies balancing between effects that the NDI core is explicitly tasked with to compensate for, and reducing complexity of the model equations, as these will eventually become part of the control law software;
		\item Inverse Model Realization: Derivation of the NDI core in a form that best suits the specific application.
	\end{itemize}
	The three key aspects will be detailed in the following sub-sections.
	
	\subsection{Selection of Generalized Controls}\label{sec:method:mod:ca}
	
	A key point in NDI is invertibility of \(\mat{\Beta}(\vec{x})\), or of \(\vec{\beta}(\vec{x},\vec{u})\) with respect to \(\vec{u}\), in the above equation. This requires the system to have at least as many inputs as commanded variables, i.e., \( n_u \ge n_y \) --- in fact, the inversion is only unambiguous for \( n_u = n_y \). From a physical point of view, this is achieved by appropriate selection of commanded variables \(\vec{y}\) on the one hand, and the availability of controls that actually allow these outputs to be effectively influenced on the other. In the case of equality (and a physically sensible selection), this \textit{full-order inversion} ensures that the commanded variables can be decoupled~\cite{Snell1991}, which is one of the fundamental objectives of NDI.
	
	Some systems pose an over-determined inversion problem, where \( n_u < n_y \). In this case, full decoupling will be compromised, and a pseudo-inverse may be used to find a solution that balances tracking accuracy across multiple commanded variables. Over-determined systems are usually not a good place to start from, but may occur in the case of system failures or specific (flight) conditions where control surfaces may be (locally) ineffective~\cite{Gaessler2025}.
	
	In flight control the under-determined (``over-actuated'') case is the more common one, i.e. \( n_u > n_y \). The reason is that 
	controls redundancy is the common approach to reduce the probability of loss of control to the low levels as demanded by certification requirements. Coordination of the redundant controls to provide required controllability around or along the given aircraft axes under all circumstances is the task of the control allocation function. Controls are assigned as a function of flight condition, aircraft configuration, and possible failure cases, in a way that is optimal relative to control power and other variables of interest (e.g., minimum drag). The development of the control allocation and the flight control law functions are therefore mostly approached as separate, but closely interacting design problems. Optimization of the allocation may be done both offline, as well as online~\cite{Durham2017,Johansen2013,Enns1998}. Note that in older non-fly-by-wire aircraft, the allocation is effectively mechanical, for example, ensuring differential aileron deflection for roll control and symmetrical deflection of both elevators~\cite{Grondman2018}. %
	
	As indicated in~\cref{fig:NDItypicalArchitecture}, the interface with the flight control laws consists of {\em Generalized Controls}, denoted by \( \vec{\mathfrak{u}} \) or \( \tau_\mathrm{c} \)~\cite{Johansen2013}). Their exact definition is a key design decision. From a methodological point of view, defining \( \vec{\mathfrak{u}} \) as (generalized) forces and moments along and around the aircraft axes can considerably reduce the complexity of the inversion, especially for non-affine systems, and realize the aforementioned {\em full-order inversion}. Control allocation algorithms can then be used to make the inversion affine, taking over the burden of distributing the virtual control input across the non-affine control inputs. 
	
	In the following, a control allocation algorithm will be mathematically described by the function \( \mathcal{M} \):
	\begin{align}
		\vec{u} &= \mathcal{M}(\vec{\mathfrak{u}}, ...)
	\end{align}
	where \( \vec{u} \in \mathcal{U} \). \(\mathcal{M}\) may depend on other variables, like current flight condition, detected failures, etc.
	The simplest form of control allocation is \textit{ganging}, where a fixed mapping between the quantities is used:
	\begin{align} %
		\vec{u} = \mat{M} \vec{\mathfrak{u}} 
	\end{align}
	The inverse mapping of the control allocator \( \mathcal{M}^{-1}: \vec{u} \mapsto \vec{\mathfrak{u}} \) represents the \textit{effects} of the inputs \( \vec{u} \) in terms of the generalized input \( \vec{\mathfrak{u}} \).

	\subsection{Model Configuration}\label{sec:method:mod}
	
	Starting point for an NDI design is usually a detailed system model in the mathematical form of~\cref{eq:gennlss1}. A direct use for inversion of this model is in most cases impractical and may result in unnecessarily complicated control algorithms. It is therefore common practice to reduce complexity to an appropriate level that includes all aspects and influences that the NDI core is actually supposed to compensate for. It must hereby be kept in mind that this can only be achieved if these aspects can actually be computed or estimated from the available system sensor signals. In literature this model, in its inverted form, is frequently referred to as \textit{OBAC} (On-Board AirCraft model) \cite{Enns2006}.  
	
	\subsubsection{Reduction of Model Complexity}\label{sec:modelcomplexity}
	The objective of this section is to discuss key design decisions in the derivation of~\cref{eq:nonaffine_normal} or~(\ref{eq:affine_normal}), starting from the \emph{physical system} model:
	\begin{subequations}\label{eq:full_system}
		\begin{align} \label{eq:systemmodel}
			\dot{\vec{\mathfrak{x}}} & = \vec{F}(\vec{\mathfrak{x}},\vec{u},\vec{d},\vec{p}) \\
			\vec{\mathfrak{y}}       & = \vec{H}(\vec{\mathfrak{x}},\vec{u},\vec{d},\vec{p},\vec{n}) 
		\end{align}
	\end{subequations}
	Apart from the outputs that are eventually to be commanded ($\vec{y}_C$), the outputs \(\vec{\mathfrak{y}}\) of this system model also include signals from all installed sensors ($\vec{y}_S$), as well as outputs that are needed for model analyses ($\vec{y}_{M}$):
	\begin{equation*}
		\vec{\mathfrak{y}} = \left[\vec{y}_S,\vec{y}_M,\vec{y}_C \right]^T
	\end{equation*}
	Sensor signals usually contain noise that is modeled and collected in the form of an additional model input vector \( \vec{n}  \in \mathbb{R}^{n_n}\). For practical and analysis purposes it is also helpful to collect relevant model parameters in an explicit vector \( \vec{p} \in \mathbb{R}^{n_p}\). On an aircraft, \( \vec{p}\) may include configuration settings, settings of control devices that are not used by the control laws, assumed tolerances on model data, variables that depend on aircraft loading, like mass and center of gravity location, or environmental variables that, once implemented, are obtained from the airdata system instead of standard environment models (e.g. static pressure, temperature). For now, no distinction between known and unknown parameters will be made. 
	
	The state vector $\vec{\mathfrak{x}}$ in~\cref{eq:systemmodel} contains all continuous states of the system. For an aircraft, this vector may usually be partitioned as follows:
	\begin{equation*}
		\vec{\mathfrak{x}} = \left[\vec{x}_B,\vec{x}_F,\vec{x}_A,\vec{x}_S \right]^T
	\end{equation*}
	where $\vec{x}_B$ describes the flight dynamical states of the aircraft center of gravity and mean body axes~\cite{Waszak1988} (in general the 6-DoF flight dynamics states), $\vec{x}_F$ contains all states related to dynamic deformation of the aircraft airframe relative to the mean body axes, $\vec{x}_A$ and $\vec{x}_S$ include all states related to the actuator and sensor systems. The flight dynamical states \(\vec{x}_B\) can in general be well determined from the sensors $\vec{y}_S$. For all other states, this is usually not the case.
	
	An accordingly structured model can for example be found in~\cite{Kier2024}. It must be conceded that, in most design applications, a set of discipline-specific models is used rather than a single integrated one~\cite{Kier2009,Looye2008}. In general, flight dynamics models are derived and can be found in for example \cite{Brockhaus2011,Stevens2016,Etkin1995,Stengel2022,Schmidt2023}. The equations have been summarized in Appendix~\ref{sec:eqm} for later reference in this work.
	
	\subsubsection*{Aeroelastic Dynamics} 
	
	State equations related to airframe states $\vec{x}_F$ in~\cref{eq:full_system} may be written as:
	\begin{equation}
		\dot{\vec{x}}_{F}  = \vec{F}_{\mathrm{F}}(\vec{\mathfrak{x}},\vec{d},\vec{p},\vec{u})
	\end{equation}
	The state vector $\vec{x}_{F}$ may contain mode shape multipliers, their rates, and unsteady aerodynamic states~\cite{Kier2009}. A common simplification is to set $\vec{x}_{F} = 0$. In case of larger deformations of the airframe, static contributions of mode shape multipliers should be taken into account by setting $\vec{x}_{F} = \vec{x}_{F_0}$, where $\vec{x}_{F_0}$ is obtained from (continuously) solving the above set of equations for $\dot{\vec{x}}_{F} = 0$~\cite{DiasMartins2026}. Another common approach is the use so-called flex-factors as a function of flight condition and vertical load factor instead~\cite{Seywald2014,Looye2008}.
	As long as the lowest airframe structural eigenfrequencies are well separated from the rigid ones, the aforementioned assumption can be and has been safely made in most flight control applications. Unlike most publications,~\cite{Kim2021} addresses coupling with aeroelastic deformation by means of signal processing and careful feedback design in the outer control functions. 
	
	If the structural and flight mechanical eigenfrequencies are close, explicit consideration in the inversion may become necessary. Although successful enhancements are presented in~\cite{Gregory2005,Wang2019b}, estimating (a subset of) $\vec{x}_{F}$ is challenging and adds complexity to the NDI core. The actuator bandwidth must be sufficient to compensate or improve structural dynamics. At the point that efforts and design complexity grow out of hand, the question may be raised if NDI is the most efficient approach for the flight control problem at hand.
	
	\subsubsection*{Actuator Dynamics} 
	
	A common simplification is to neglect dynamics of the actuators:
	\begin{equation}
		\dot{\vec{x}}_{A}  = \vec{F}_{\mathrm{A}}(\vec{\mathfrak{x}},\vec{d},\vec{p},\vec{u})
	\end{equation}
	The dynamics may be quite complex due to inherent nonlinear behaviors, including saturation effects, friction, operating temperature, aerodynamic loading, and electric or hydraulic system characteristics~\cite{Ravenscroft2000} -- hence the dependency on the full state vector $\vec{\mathfrak{x}}$.
	It is common to use simplified models that represent delayed response, approximate bandwidth, and rate and position limitations, rather than detailed physical behavior. Some publications even include such linear actuator dynamics models in system inversion~\cite{Raab2019,Steffensen2022,Steinert2025a,Steinert2025b,Raab2025}, e.g., when different controls have considerably different bandwidths~\cite{DePonti2026}. However, this has to be done carefully. Most NDI designs assume that actuator dynamics are fast enough to be ignored and that the resulting control deflections accurately match the commanded values $\vec{u}$. Rate and position limits then need to be addressed in control allocation and control saturation handling~\cite{Johnson2000}.
	
	\subsubsection*{Aerodynamic and Propulsion Model Data}
	
	A significant portion of aircraft model complexity is in the data used to compute aerodynamic and propulsive forces and moments. The relation between motion states and aerodynamic flow may be highly nonlinear, especially in transonic flow regions and at higher incidence angles. Tabulated data tables and matching application rules may consequently be of high complexity. Depending on the adopted form of NDI, these data become part of the flight control algorithms. Therefore, there are good reasons to critically analyze those for discontinuities or other artifacts, gradient reversals in control effectiveness (that may cause inversion problems, see for example~\cite{Steinhauser2004,Gaessler2025}), cause unwanted non-minimum phase behavior between control inputs and commanded variables, as well as to find a proper balance between model complexity and inversion accuracy.  
	
	It must always be kept in mind that, however detailed the model data is, a certain level of uncertainty relative to the physical system remains. Any adopted simplification will obviously add up to this uncertainty and is to be carefully analyzed in control law verification (cf.~\cref{sec:design:robustness} and~\cite{Looye2001}).
	
	\subsubsection*{The physical relative degree of (generalized) controls}
	Control devices are always designed with the system, based on requirements involving controllability of specific states or state combinations. These devices obviously will also affect other states and this may be relevant for NDI. In configuring the model, it is therefore important to ensure that the inverse model works along the principles that the physical control devices have been designed for at a systems level. A notorious example is vertical load factor control via elevator. The intended path is to pitch and adapt angle of attack of the aircraft by means of this device and, at least in the short term, adjust load factor by means of the angle of attack. Elevator deflection also causes a direct small change in lift that mathematically makes load factor controllable directly - with zero relative order. In case of elevators, this would immediately cause instability, as the initial response goes the wrong way (i.e. non-minimum phase behavior). This is avoided in the case of canards, but still results in excessive control deflections, as these are primarily intended to generate pitching moments as well. The usual solution is to remove such secondary effects from the model during preparation for inversion. In \cite{Bharadwaj1998} the effects are retained and a so-called \textit{global approximate vector relative degree} is introduced, where the effect of the controls are to be ``significant'' in determining the relative orders of the outputs.  Other approaches effectively avoid cancellation of the right half plane transmission zero~\cite{Snell2002}. Reference \cite{Buffington1998} addresses this by deriving sufficient conditions for a specific class of control allocation functions.
	
	\subsubsection*{Environment Models}
	Environment models are key elements in the model representation of~\cref{eq:systemmodel}. In the case of aircraft, these are often based on standards like the International Standard Atmosphere (ISA)~\cite{ISA1975}, or demanded for performing specific types of design analyses, like 1-cos gust models, or filters that realize atmospheric disturbances exhibiting standard spectra as proposed by Dryden or von K\'{a}rm\'{a}n~\cite{Dryden1980}.
	
	External disturbances $\vec{d}$ are therefore oftentimes exogenous effects depending on the state of the overall environment (e.g. wind fields) and on the state of the system, primarily to represent the mode of action. Locally, they can be approximated as a function of aircraft state and time:
	\begin{equation}
		d(x_B,t) \approx d_x(x_B) + d_t(t)
	\end{equation}
	The state dependent component $d_x(x_B)$ arises for example in modeled wind fields that are fixed relative to the Earth's surface. Local wind velocities can be determined as a function of aircraft position and height. The disturbance component can be considered as part of, and integrated into the state equations. Time dependency ($d_t(t)$) arises in the case of dynamically changing wind fields, or for example in the aforementioned turbulence models. 
	
	In configuring a model for inversion, it may be preferable to consider all external disturbances just as time dependent, i.e. $d(x_B,t)=d_t(t)$. Any detailed wind model will highly unlikely be able to match the actual weather conditions. If needed, atmospheric disturbances can be directly estimated from airdata and inertial measurements \cite{Looye2001b}. Atmospheric conditions, like static pressure and temperature, are usually available from the airdata system and may be represented in the model as entries in the parameter vector \(\vec{p}\). 
	
	\subsubsection{Model Representation for Inversion}\label{sec:modelsimplification}
	To facilitate analytical comparison of the different NDI variants, the physical system representation in~\cref{eq:full_system} is written in the form of~\cref{eq:nonaffine_system}, which, apart from the variables $\vec{d}$ and $\vec{p}$, is the one most commonly used in literature: 
	\begin{subequations}\label{eq:abnormal_nonaffine_system0}
		\begin{alignat}{1}%
			\dot{\vec{x}} & = \vec{{f}}(\vec{x},\vec{d},\vec{p})   + \vec{{g}}(\vec{x},\vec{d},\vec{p},\vec{u})  \\
			\vec{y}       & = \vec{h}(\vec{x},\vec{d},\vec{p})
		\end{alignat}
	\end{subequations}
	Outputs related to sensors and model analyses ($\vec{y}_S$, $\vec{y}_M$) are omitted for now, as these are irrelevant for actual inversion, i.e. $\vec{y} = \vec{y}_C$. The state vector \(\vec{\mathfrak{x}}\) has been reduced to \(\vec{x}\), in the case of aircraft usually containing flight dynamical states \(\vec{x}_B\) only. 
	
	Based on the considerations above a model representation of (further) reduced complexity may be formulated that is actually suitable for mathematical derivation of the inverse model equations in an NDI architecture. The main characteristics and limitations are:
	\begin{itemize}
		\item The state vector can be computed or estimated with sufficient accuracy from the system sensors \(\vec{y}_S\);
		\item Model data and equations can be evaluated with reasonable computational effort on the flight control computer (FCC);
		\item Environmental parameters are obtained from measurement, rather than internal models (included in \(\vec{p}\));
		\item The adopted simplifications can reasonably be expected to be sorted out by feedback control laws in outer functions;
		\item The model equations are formulated as a function of generalized controls \(\vec{\mathfrak{u}}\), allowing for exact inversion. 
	\end{itemize}
	The latter may be done by directly including an imposed allocation function \(\mathcal{M} (\vec{\mathfrak{u}}, ...)\), or by assuming that \(\vec{\mathfrak{u}}\) will be the effect of advanced allocation algorithms designed in a separate task (cf.~\cref{sec:method:mod:ca}). In case the number of physical controls equals the number of controlled outputs, \(\vec{\mathfrak{u}} = \vec{u}\) holds.
	The resulting model equations used for inversion are written in the following form:
	\begin{subequations}\label{eq:abnormal_nonaffine_system_hat}
		\begin{alignat}{1}%
			\dot{\vec{x}} & = \vec{\hat{f}}(\vec{x},\vec{d},\vec{p})   + \vec{\hat{g}}(\vec{x},\vec{d},\vec{p},\vec{\mathfrak{u}}) \\
			\vec{y}       & = \vec{h}(\vec{x},\vec{d},\vec{p})
		\end{alignat}
	\end{subequations}
	The modeled representations of the functions \(\vec{{f}}\) and \(\vec{{g}}\) have been marked with a hat and may thus differ from the physical representations in~\cref{eq:abnormal_nonaffine_system0}. In order to reduce complexity of subsequent derivations, it is assumed that (1) \(\vec{{h}}\) can be represented exactly, and that (2) the state vector definitions are identical. It is for this reason that this vector had been reduced beforehand in~\cref{eq:abnormal_nonaffine_system0}.
	
	\subsubsection*{Disturbances}
	Note that disturbances \( \vec{d} \) in~\cref{eq:abnormal_nonaffine_system0} are commonly disregarded in inverse models for NDI. Their effect is (partially) compensated for by using airmass-referenced state estimates in the resulting equations~\cite{Looye2001b} (see also \cref{sec:design:dist}). Examples of explicit consideration have been published in~\cite{Kim2024,Reiner2017}. In this paper, the disturbance vector \( \vec{d} \) will be retained in the derivation to facilitate a more detailed comparison between the various NDI implementations. Whether explicitly considered or left to outer feedback control functions, it has to be ensured that the overall design shows sufficient disturbance rejection capability (cf.~\cref{sec:design:dist}). %
	
	\subsubsection*{Model Uncertainty}
	Explicitly used model equations will unavoidably not exactly match physical reality. In order to compare accuracy of the various forms of NDI, it is assumed that~\cref{eq:abnormal_nonaffine_system0} and~\cref{eq:abnormal_nonaffine_system_hat} can be related by means of explicit uncertain terms denoted by \( \Delta .. \):
	\begin{subequations}\label{eq:abnormal_nonaffine_system}
		\begin{alignat}{1}[left=\Sigma^\prime_g : \empheqlbrace]
			\dot{\vec{x}} & = \underbrace{\left( \hat{\vec{f}}(\vec{x},\vec{d},\vec{p}) + \Delta {\vec{f}}(\vec{x},\vec{d},\vec{p}) \right)}_{= \vec{f}(\vec{x},\vec{d},\vec{p}) }  + \underbrace{\left( \hat{\vec{g}}(\vec{x},\vec{d},\vec{p}, \vec{\mathfrak{u}}) + \Delta {\vec{g}}(\vec{x},\vec{d},\vec{p},\vec{u}) \right)}_{=\vec{g}(\vec{x},\vec{d},\vec{p},\vec{\mathfrak{u}})} \\ %
			\vec{y}       & = \vec{h}(\vec{x},\vec{d},\vec{p})
		\end{alignat}
	\end{subequations}
	The control-affine version is written as:
	\begin{subequations}\label{eq:abnormal_affine_system}
		\begin{alignat}{1}[left=\Sigma_G^\prime : \empheqlbrace]
			\dot{\vec{x}} & = \underbrace{ \left( \hat{\vec{f}}(\vec{x},\vec{d},\vec{p}) + \Delta\! {\vec{f}}(\vec{x},\vec{d},\vec{p}) \right)}_{= \vec{f}(\vec{x},\vec{d},\vec{p}) } + \underbrace{ \left( \hat{\mat{G}}(\vec{x},\vec{d},\vec{p}) + \Delta {\mat{G}}(\vec{x},\vec{d},\vec{p}) \right)}_{= \mat{G}(\vec{x},\vec{d},\vec{p})} \,  \vec{\mathfrak{u}} \\ %
			\vec{y} & = \vec{h}(\vec{x},\vec{d},\vec{p})
		\end{alignat}
	\end{subequations}
	The generalized controls \(\vec{\mathfrak{u}}\) are realized by physical control deflections \(\vec{u}\) and therefore can be interpreted as demanded values to be realized by the allocation function \(\mathcal{M}(\vec{\mathfrak{u}}, ...)\). The additional uncertainty due to mismatches between demanded and realized generalized controls is considered part of the uncertainty in the terms \(\Delta {\vec{g}}(\cdot)\) and  \(\Delta {\vec{G}}(\cdot)\). 
	
	\subsubsection{Differentiation of Controlled Variables}\label{sec:differentiatemodel} %

Starting from~\cref{eq:abnormal_nonaffine_system0} the next step is differentiation of the output vector until an algebraic, invertible and physically meaningful relation with at least one entry of $\vec{\mathfrak{u}}$ is found. Consequently, the first derivative of $y_i$ becomes: %
\begin{align}
	\dot{y}_i  &= \frac{\partial \vec{h}_i(\vec{x},\vec{d},\vec{p})}{\partial \vec{x}}\dot{\vec{x}} + \frac{\partial \vec{h}_i(\vec{x},\vec{d},\vec{p})}{\partial \vec{d}}\dot{\vec{d}} \\ 
	&= \frac{\partial \vec{h}_i(\vec{x},\vec{d},\vec{p})}{\partial \vec{x}} f(\vec{x},\vec{d},\vec{p}) + 
	\frac{\partial\vec{h}_i(\vec{x},\vec{d},\vec{p})}{\partial \vec{x}}
	g(\vec{x},\vec{d},\vec{p},\vec{\mathfrak{u}}) + \frac{\partial \vec{h}_i(\vec{x},\vec{d},\vec{p})}{\partial \vec{d}}\dot{\vec{d}} \\ %
	&=\mathcal{L}_\vec{f} \vec{h}_i(\vec{x},\vec{d},\vec{p}) + \mathcal{L}_\vec{g} \vec{h}_i(\vec{x},\vec{d},\vec{\mathfrak{u}},\vec{p})  + \nabla_\vec{d} \vec{h}_i(\vec{x},\vec{d},\vec{p}) \dot{\vec{d}}\label{eq:dy_dt}
\end{align}
In the case $ \mathcal{L}_\vec{g} \vec{h}_i(\vec{x},\vec{d},\vec{\mathfrak{u}},\vec{p}) = 0$, differentiation proceeds. Since the last term in the above equation depends on $\dot{\vec{d}}$, the second derivative additionally comprises the term \( \frac{\partial \dot{y}_i}{\partial \dot{\vec{d}}} \, \ddot{\vec{d}} \):
\begin{align}
	\ddot{y}_i &= \frac{\partial\, \dot{y}_i}{\partial \vec{x}}\dot{\vec{x}} + \frac{\partial\, \dot{y}_i}{\partial \vec{d}}\dot{\vec{d}}  + \frac{\partial\, \dot{y}_i}{\partial \dot{\vec{d}}} \ddot{\vec{d}} %
\end{align}
where:
\begin{subequations}
	\begin{align*}
		\frac{\partial \dot{y}_i}{\partial \vec{x}}\ \dot{\vec{x}} &= \frac{\partial}{\partial \vec{x}}\left(\mathcal{L}_\vec{f} \vec{h}_i(\vec{x},\vec{d},\vec{p}) + \nabla_\vec{d} \vec{h}_i(\vec{x},\vec{d},\vec{p}) \dot{\vec{d}} \right)\dot{\vec{x}} \\
		&= \mathcal{L}_\vec{f}^2 \vec{h}_i(\vec{x},\vec{d},\vec{p}) 
		+ \left( \mathcal{L}_\vec{f} \nabla_\vec{d} \vec{h}_i \right)\, (\vec{x},\vec{d},\vec{p})\ \dot{\vec{d}}
		+ \mathcal{L}_\vec{g} \mathcal{L}_\vec{f} \vec{h}_i \, (\vec{x},\vec{d},\vec{\mathfrak{u}},\vec{p}) 
		+ \left(  \mathcal{L}_\vec{g} \nabla_\vec{d} \vec{h}_i \right)\,  (\vec{x},\vec{d},\vec{\mathfrak{u}},\vec{p})\ \dot{\vec{d}} \\
		\frac{\partial \dot{y}_i}{\partial \vec{d}}\ \dot{\vec{d}}
		&= \nabla_\vec{d}\left(\nabla_\vec{d} \vec{h}_i(\vec{x},\vec{d},\vec{p})\dot{\vec{d}}\right)\ \dot{\vec{d}} 
		+ \nabla_\vec{d}\left(\mathcal{L}_\vec{f} \vec{h}_i(\vec{x},\vec{d},\vec{p})\right)\ \dot{\vec{d}} \\
		\frac{\partial \dot{y}_i}{\partial \dot{\vec{d}}}\ \ddot{\vec{d}}
		&= \nabla_\vec{d} \vec{h}_i(\vec{x},\vec{d},\vec{p})\ddot{\vec{d}}
	\end{align*}
\end{subequations}
The differentiation process can be performed recursively:
\begin{align}\label{eq:ddy_dt2}
	y_i^{(\rho_i)} &= \frac{\partial y_i^{(\rho_i-1)}}{\partial \vec{x}} \ \left( \vec{f}(\vec{x},\vec{d},\vec{p}) + \vec{g}(\vec{x},\vec{d},\vec{\mathfrak{u}},\vec{p}) \right) + \sum\limits_{j=0}^{\rho_i - 1} \frac{\partial y_i^{(\rho_i-1)}}{\partial \vec{d}^{(j)}} \, \vec{d}^{(j+1)} \\
	&= \mathcal{L}_\vec{f}^{\rho_i} \vec{h}_i\, (\vec{x},\vec{d},\vec{p}) + \mathcal{L}_\vec{g}\mathcal{L}_\vec{f}^{\rho_i-1} \vec{h}_i\, (\vec{x},\vec{d},\vec{\mathfrak{u}},\vec{p}) + D_i ( \vec{x}, \vec{\mathfrak{u}}, \vec{d}, \dot{\vec{d}}, \ldots, \vec{d}^{(\rho_i)}, \vec{p} ) \notag
\end{align}
until the term $\mathcal{L}_\vec{g}\mathcal{L}_\vec{f}^{(\rho_i-1)} \vec{h}_i(\vec{x},\vec{d},\vec{\mathfrak{u}},\vec{p}) \neq 0$. The function \( D_i(\cdot) \) hereby collects all remaining terms that involve time derivatives of the disturbance, including the mixed terms that arise recursively from differentiating the products of partial derivatives of \( \vec{h}_i \) and \( \dot{\vec{d}} \), so that the first line and the second line of \cref{eq:ddy_dt2} are equivalent.
The formulation in \cref{eq:ddy_dt2} will be important in the following. Its first two terms are commonly known from literature and disturbances are actually considered in state estimation, as will be detailed in the following section~\cite{Looye2001b}. The additional disturbance terms are usually neglected, although references like~\cite{Chen2003,Chen2014,Kim2023b} describe approaches for their explicit consideration.

The individual entries of $y_i^{(\rho_i)}$ are collected into a single set of nonlinear equations, resulting in the form of~\cref{eq:nonaffine_normal}:
\begin{align}
	\vec{y}^{(\rho)} &= \left[ \mathcal{L}_\vec{f}^{\rho_i} \vec{h}_i\, (\vec{x},\vec{d},\vec{p}) + \mathcal{L}_\vec{g}\mathcal{L}_\vec{f}^{\rho_i-1} \vec{h}_i\, (\vec{x},\vec{d},\vec{p},\vec{\mathfrak{u}})  + D_i (\vec{x}, \vec{\mathfrak{u}}, \vec{d}, \vec{\dot{d}}, \ldots, \vec{d}^{(\rho_i)}, \vec{p} ) \right]_{i=1 \ldots n_y} \notag \\
	&= \vec{\alpha} (\vec{x},\vec{d},\vec{p}) + \vec{\beta}(\vec{x},\vec{d},\vec{p},\vec{\mathfrak{u}}) + \vec{\delta} (\vec{x}, \vec{p}, \vec{\mathfrak{u}}, \vec{d}, \vec{\dot{d}}, \ldots, \vec{d}^{(\bar{\rho})} ) \label{eq:nonaffine_normal2} \\
	\vec{y}^{(\rho)} &= \vec{\alpha} (\vec{x},\vec{d},\vec{p}) + \mat{\Beta}(\vec{x},\vec{d},\vec{p}) \vec{\mathfrak{u}} + \vec{\delta} (\vec{x}, \vec{p}, \vec{\mathfrak{u}}, \vec{d}, \vec{\dot{d}}, \ldots, \vec{d}^{(\bar{\rho})} ) \label{eq:affine_normal2}
\end{align}
with the allocated monomorphism \( \vec{\beta} \), i.e., an injective or left-invertible function, and the allocated regular or invertible matrix \( \mat{\Beta}(\vec{x}) \), which is assumed to have an affine control allocation for a simpler presentation in this section. Thus the corresponding (left-)inverse mappings \( \vec{\beta}^{-1}: ( \vec{x}, \vec{e}_\vec{\nu} ) \mapsto \vec{\mathfrak{u}}  \) (cf.~\cref{eq:nonaffine_control_normal}) and \( \mat{\Beta}^{-1} (\vec{x}) = \left( \mat{\Beta} (\vec{x}) \right)^{-1} \) exist. The stacked disturbance term \( \vec{\delta}(\cdot) \) collects the entries \( D_i(\cdot) \) and consequently depends on disturbance derivatives up to order \( \bar{\rho} = \max_i \rho_i \).

The above procedure can also be applied to the individual terms of the uncertain system form in~\cref{eq:abnormal_nonaffine_system} and~\cref{eq:abnormal_affine_system} respectively (cf.~\cref{eq:ddy_dt2}), yielding: 
\begin{align}
	\vec{y}^{(\rho)} &= 
	\left[ 
	\mathcal{L}_{\hat{\vec{f}}}^{\rho_i} \vec{h} \, (\vec{x},\vec{d},\vec{p}) + \sum\limits_{i=1}^{\rho_i} \mathcal{L}_{\hat{\vec{f}}}^{i-1} \mathcal{L}_{\Delta {\vec{f}}} \mathcal{L}_{\hat{\vec{f}} + \Delta {\vec{f}}}^{\rho_i-i} \vec{h} \, (\vec{x},\vec{d},\vec{p}) + \mathcal{L}_{\hat{\vec{g}}} \mathcal{L}_{\hat{\vec{f}}}^{\rho_i-1} \vec{h} \, (\vec{x},\vec{d},\vec{p},\vec{u}) \right. \notag \\
	& + \left( \mathcal{L}_{\Delta {\vec{g}}} \mathcal{L}_{\hat{\vec{f}} + \Delta \vec{f}}^{\rho_i-1} \vec{h} \, (\vec{x},\vec{d},\vec{p},\vec{u}) +  \sum\limits_{i=1}^{\rho_i-1} \mathcal{L}_{\hat{\vec{g}}} \mathcal{L}_{\hat{\vec{f}}}^{i-1} \mathcal{L}_{\Delta {\vec{f}}} \mathcal{L}_{\hat{\vec{f}} + \Delta {\vec{f}}}^{\rho_i-1-i} \vec{h} \, (\vec{x},\vec{d},\vec{p},\vec{u}) \right) \notag \\
	& \left. + \sum\limits_{i=1}^{\rho_i} \mathcal{L}_{\hat{\vec{f}}}^{i-1} \mathcal{L}_{\Delta {\vec{f}}} \mathcal{L}_{\hat{\vec{f}} + \Delta {\vec{f}}}^{\rho_i-i} \vec{h} \, (\vec{x},\vec{d},\vec{p}) +   \sum\limits_{i=1}^{\rho_i} \mathcal{L}_{\hat{\vec{g}} + \Delta \vec{g}} \mathcal{L}_{\hat{\vec{f}}}^{i-1} \mathcal{L}_{\Delta {\vec{f}}} \mathcal{L}_{\hat{\vec{f}} + \Delta {\vec{f}}}^{\rho_i-i} \vec{h} \, (\vec{x},\vec{d},\vec{p},\vec{u})
	\right]_{i=1 \ldots n_y} \notag \\ %
	&= \hat{\vec{\alpha}}(\vec{x},\vec{d},\vec{p}) + \Delta \vec{\alpha}(\vec{x},\vec{d},\vec{p}) + \hat{\vec{\beta}}(\vec{x},\vec{d},\vec{p},\vec{\mathfrak{u}}) + \Delta \vec{\beta}(\vec{x},\vec{d},\vec{p},\vec{\mathfrak{u}}) + \vec{\delta}(\vec{x},\vec{p},\vec{\mathfrak{u}},\vec{d}, \ldots,\vec{d}^{(\bar{\rho})}) \label{eq:yrho_unc_nonaffine} \\ 
	\vec{y}^{(\rho)} 
	&= \hat{\vec{\alpha}}(\vec{x},\vec{d},\vec{p}) + \Delta \vec{\alpha}(\vec{x},\vec{d},\vec{p}) + \left( \hat{\mat{\Beta}}(\vec{x},\vec{d},\vec{p}) + \Delta \mat{\Beta}(\vec{x},\vec{d},\vec{p}) \right) \vec{\mathfrak{u}} + \vec{\delta} (\vec{x},\vec{p},\vec{\mathfrak{u}},\vec{d}, \ldots,\vec{d}^{(\bar{\rho})}) \label{eq:yrho_unc_affine}
\end{align}
with the disturbance vector \( \delta \) comprising all disturbances and disturbance derivative terms (cf.~\cref{eq:ddy_dt2}). Unless indicated explicitly, the disturbance terms \( \vec{d} \) and \( \vec{\delta} \) are neglected subsequently for clarity.

\subsection{Realization of the inverse model} \label{sec:invmodelrealization}

In~\cref{sec:history} the historical development of different forms of NDI has been reviewed. These forms are equivalent from a mathematical point of view, but differ considerably from a design implementation perspective. The objective of this section is to give an overview of the most common forms, and to highlight advantages and disadvantages of their realizations in NDI.

\subsubsection{Mathematical model inversion}
Starting from the differentiated commanded variable output vector in~\cref{eq:nonaffine_normal2}:
\begin{align}\label{eq:nonaffine_normal3}
	\vec{y}^{(\rho)} &= \vec{\alpha} (\vec{x},\vec{d},\vec{p}) + \vec{\beta}(\vec{x},\vec{d},\vec{p},\vec{\mathfrak{u}}) + \vec{\delta} (\vec{x}, \vec{p}, \vec{\mathfrak{u}}, \vec{d}, \vec{\dot{d}}, \ldots, \vec{d}^{(\bar{\rho})} )  \mathrm{\ \ or:} \notag \\
	\vec{y}^{(\rho)} &= \vec{\alpha} (\vec{x},\vec{d},\vec{p}) + \mat{\Beta}(\vec{x},\vec{d},\vec{p}) \vec{\mathfrak{u}} + \vec{\delta} (\vec{x}, \vec{p}, \vec{\mathfrak{u}}, \vec{d}, \vec{\dot{d}}, \ldots, \vec{d}^{(\bar{\rho})} )
\end{align}
the perfect mathematical inversion of both forms would result in: 
\begin{align}
	\vec{\mathfrak{u}} &= {\vec{\beta}}^{-1} \left({\vec{x},\vec{d},\vec{p}}, \ \vec{\nu} - {\vec{\alpha}} ({\vec{x},\vec{d},\vec{p}})-\vec{\delta} (\vec{x}, \vec{p}, \vec{\mathfrak{u}}, \vec{d}, \vec{\dot{d}}, \ldots, \vec{d}^{(\bar{\rho})} ) \right)  \mathrm{\ \ resp.:} \notag \\
	\vec{\mathfrak{u}} &= {\mat{\Beta}}^{-1}\left({\vec{x},\vec{d},\vec{p}}\right) \left( \vec{\nu} - {\vec{\alpha}} ( {\vec{x},\vec{d},\vec{p}} )-\vec{\delta} (\vec{x}, \vec{p}, \vec{\mathfrak{u}}, \vec{d}, \vec{\dot{d}}, \ldots, \vec{d}^{(\bar{\rho})} ) \right) \label{eq:exactU}
\end{align}
where \(\vec{\nu} \equiv \vec{y}^{(\rho)}\) are the so-called virtual control inputs. Note that, since \( \vec{\delta}(\cdot) \) itself depends on \( \vec{\mathfrak{u}} \), these relations would be implicit in \( \vec{\mathfrak{u}} \) if \( \vec{\delta} \) would not be neglected in the control design.
Substitution in~\cref{eq:nonaffine_normal3} results in: 
\begin{equation} \label{exactNDInu}
	\vec{y}^{(\rho)} = \vec{\nu} 
\end{equation}
This is a remarkable simplification as compared with the original form of \cref{eq:nonaffine_normal3}, and has the potential to massively simplify design of of the control system. Obviously, reality is very different. The perfect inversion will not be attainable due to uncertainties, simplifications, disturbances, sensor offsets and noise, sensing limitations, neglected dynamics, etc. Nor will the perfect inversion be desirable, as physical controls have limited capability in dynamics and power, and zero or unmodeled dynamics may be excited unfavorably. Still, the objective in deriving an NDI control law is to retain as much of the essence as reasonably possible.

\subsubsection{Zero dynamics}
In most applications the total relative degree of the controlled variables will be lower than the dimension of \(\vec{x}\), so that \(\rho < n_x\). As a consequence, additional dynamics is hidden by \cref{exactNDInu} that has to be taken into account in overall analysis of stability and margins, as well regarding fulfillment of other control system requirements. In~\cref{sec:mathematical:normal} the state equations are rewritten based on a state transformation, splitting the state variables into \textit{external} states \(\vec{\xi}\), constructed from output variables \(\vec{y}\) and derivatives, and \textit{internal} states \(\vec{\eta}\). The condition for finding the transformation that results in the states \(\vec{\eta}\) is given in \cref{eq:zdstates}. The related state equations describe the internal dynamics. In practice it is often pretty clear which entries in the state vector \(\vec{x}\) need to be retained in order to complete the normal form of order \(\rho\) to retain rank \(n_x\), see for example the discussion on the topic in \cite{Lin1994}. These entries are collected in the vector \(\vec{x}_I\). As a consequence:
\begin{align}
	\vec{\eta} &= \vec{x}_I
\end{align}
With the inverse state transformation as defined in~\cref{sec:mathematical:normal}: \(\vec{x} = T^{-1}(\vec{\xi},\vec{\eta})\), the related state equations are (taking the affine form for now): 
\begin{align}
	\dot{\vec{x}}_I & = \vec{{f_I}}(\vec{x},\vec{d},\vec{p})+
	\vec{{G_I}}(\vec{x},\vec{d},\vec{p})\vec{\mathfrak{u}} \notag\\
	&= \vec{{f_I}}(T^{-1}(\vec{\xi},\vec{\eta}),\vec{d},\vec{p})   + \vec{{G_I}}(T^{-1}(\vec{\xi},\vec{\eta}),\vec{d},\vec{p})\vec{\mathfrak{u}} \notag\\
	&= \vec{{f_I}}(T^{-1}(\vec{\xi},\vec{x_I}),\vec{d},\vec{p})   + \vec{{G_I}}(T^{-1}(\vec{\xi},\vec{x_I}),\vec{d},\vec{p})\vec{\mathfrak{u}}
\end{align}
where the index \textit{I} just indicates the concerning rows of the state equations in~\cref{eq:abnormal_nonaffine_system0}. The zero dynamics can equivalently be found by substitution of \cref{eq:exactU} and enforcing \(\vec{y}=y_0\) by setting \(\vec{\nu}=0\), \(\vec{d}=0\) and setting the states related to \(\vec{y}\) and derivatives accordingly (i.e. \(\xi = \xi_0\) as in~\cref{eq:canonicalcontrolform}):
\begin{align}
	\dot{\vec{x}}_I & = \vec{{f_I}}(T^{-1}(\vec{\xi_0},\vec{\eta}),\vec{p})   - \vec{{G_I}}(T^{-1}(\vec{\xi_0},\vec{x_I}),\vec{p}){\mat{\Beta}}^{-1}\left({T^{-1}(\vec{\xi_0},\vec{x_I}),\vec{p}}\right){\vec{\alpha}} ( {T^{-1}(\vec{\xi_0},\vec{x_I}),\vec{p}} ) \\
	&= \vec{{f_I^*}}(\vec{\xi_0},\vec{x_I},\vec{p})
\end{align}
These state equations obviously need to be stable. A similar analysis is to be performed on the original model as described by~\cref{eq:systemmodel}, which may considerably extend \(\vec{x}_I\), with those that have been neglected or residualized in the process as described in \cref{sec:modelcomplexity}.

In control design practice, the above equations are usually not worked out in full detail, especially as neglected dynamics may make matters quite complex. Verification of design requirements, including internal stability and margins, will have to be done numerically on the full modeled closed-loop system. The derivation in its simplified form may still be very helpful in detecting physical implications of applying a specific NDI method and related architecture to the selected controlled variables \(\vec{y}\). This will be briefly revisited in~\cref{sec:design}.

\subsubsection{Comparative Metrics and Criteria for NDI Variants}
In the following subsections, the four most common forms of NDI will be derived based on the equations above. In order to compare these variants from methodological and design perspectives, the following metrics or criteria will be used:
\begin{itemize}
	\item \emph{Inversion Error}. To this end, the following error definition will be used:
	\begin{equation}
		\vec{e} = \vec{\nu} - \vec{y}^{(\rho)} \label{eq:inverror}
	\end{equation}
	This error can be mathematically derived using the explicit uncertainty form in~\cref{eq:yrho_unc_nonaffine} or ~(\ref{eq:yrho_unc_affine}). %
	\item \emph{Algorithmic Complexity}. Standard NDI relies on computation of model equations, incremental or sensory variants use measurement or direct estimation of the current values of \(\vec{y}^{(\rho)}\) instead. Although algorithmic complexity is difficult to capture without a concrete application, NDI variants can be compared to the fully model-based implementation that will be derived first.
	\item \emph{Disturbance Compensation}. Replacement of model equations with measurement of \(\vec{y}^{(\rho)}\) reduces algorithmic complexity. The latter will be closer to physical reality as opposed to the model-based variant, as effects of any disturbances are intrinsically captured. This is great from an accuracy perspective, but may not be desirable from a control activity or zero-dynamics perspective. Although difficult to capture as the one before, a comparison with model-based NDI can be made. 
	\item \emph{Controls Failure Handling}. The inversion stands or falls with operability of the physical control devices in \(\vec{u}\). As capabilities of the NDI variants differ significantly in this respect, this is considered a relevant aspect to attend to.
	
	\item \emph{Internal Dynamics}. Zero dynamics are invariant in the case of exact inversion \cite{Slotine1991}. Accuracy of compensating for internal dynamics differs between variants of NDI, especially when looking at the full system model. It is difficult to give a metric, but an analytical comparison can be made.
\end{itemize}

A recent publication by Yilmaz took an experimental approach in comparing NDI variants, see \cite{Yilmaz2026}.

\subsubsection{Nonlinear Dynamic Inversion}\label{sec:method:ndi}

The NDI control laws result from inversion of the non-affine (\cref{eq:yrho_unc_nonaffine}) and affine systems (\cref{eq:yrho_unc_affine}) with the terms \(\Delta \vec{\alpha}(\cdot)\), \( \Delta \vec{\beta}(\cdot) \), and \( \delta(\cdot) \) set to zero:
\begin{align}
	\vec{y}^{(\rho)} &= \hat{\vec{\alpha}}(\vec{x},\vec{d},\vec{p}) + \hat{\vec{\beta}}(\vec{x},\vec{d},\vec{p},\vec{\mathfrak{u}})  \label{eq:yrho_unc_nonaffine0} \\ 
	\vec{y}^{(\rho)} &= \hat{\vec{\alpha}}(\vec{x},\vec{d},\vec{p}) + \hat{\mat{\Beta}}(\vec{x},\vec{d},\vec{p}) \vec{\mathfrak{u}}  \label{eq:yrho_unc_affine0}
\end{align}
assigning \(\vec{y}^{(\rho)}\) as the new virtual control input \( \vec{\nu} \), inversion results in:
\begin{align}
	\vec{\mathfrak{u}} &= \hat{\vec{\beta}}^{-1} \left( (\hat{\vec{x}},\hat{\vec{d}},\hat{\vec{p}}), \ \vec{\nu} - \hat{\vec{\alpha}} (\hat{\vec{x}},\hat{\vec{d}},\hat{\vec{p}}) \right) \label{eq:ndi_nonaffine} \\
	\vec{\mathfrak{u}} &= \hat{\mat{\Beta}}^{-1}\left(\hat{\vec{x}},\hat{\vec{d}},\hat{\vec{p}}\right) \left[ \vec{\nu} - \hat{\vec{\alpha}} ( \hat{\vec{x}},\hat{\vec{d}},\hat{\vec{p}} ) \right] \label{eq:ndi_affine}
\end{align}
The realizations are sometimes referred to as model-based NDI to distinguish itself from other NDI variants~\cite{Encarnacao2026,Pollack2024,Pavel2020,Kim2021}. The importance of~\cref{eq:yrho_unc_nonaffine0} and~\cref{eq:yrho_unc_affine0} cannot be overstated, as these basically determine all influences that will be explicitly compensated for in the inverse model core. Even though the term \(\vec{\delta}(\cdot)\) in (\cref{eq:yrho_unc_nonaffine}) and (\cref{eq:yrho_unc_affine}) is most commonly disregarded at this point, it may be derived from nominal model equations if its compensation is useful.

The inverse model equations relate the input \( \vec{\mathfrak{u}} \) that is needed to realize the pseudo control input \( \vec{\nu} \) by means of model equations with arguments that stem from measured or estimated values for states, disturbances and parameters. As already discussed in~\cref{sec:architecture}, the NDI core normally does not come with any differential equations: \(\hat{f}\) and \(\hat{G}\) are used for derivation of \(\vec{y}^{(\rho)}\), but the states are estimated from signals contained in $\vec{y}_S$ (by {\em Signal Processing} algorithms \cref{fig:NDItypicalArchitecture}). The same holds for disturbances and parameters; these arguments are therefore denoted by a hat \(\hat{\cdot}\). Obviously, deviations from the actual values may cause additional inaccuracy, but these will arguably be small as compared with model-based computation (~\cref{sec:architecture}) and integration. The state vector \(\hat{\vec{x}}\) is consequently restricted to entries that can actually be measured or estimated from $\vec{y}_S$.

The inverse model core (cf.~\cref{fig:NDIinverseModel}) consists of an algebraic inverse ($\mat{\Beta}^{-1}(\vec{x},\vec{d})$) between commanded variable derivatives $\vec{y}^{(\rho)}$  and available system control inputs $\vec{u}$, as well as compensatory terms for all relevant influences on $\vec{y}^{(\rho)}$ other than $\vec{u}$, collected in $\vec{\alpha}(\vec{x},\vec{d})$. Both $\vec{\alpha}(\vec{x},\vec{d})$ and $\mat{\Beta}^{-1}(\vec{x},\vec{d})$ are directly derived from model equations.
A key design decision in an NDI-based design is to what extent and in what detail influences on $\vec{y}^{(\rho)}$ can and are to be compensated for, i.e., the \( \vec{\alpha} \) term. This decision is made in the derivation of a dedicated system model representation for the inversion, which may involve simplifying assumptions, approximations, etc. These, in turn, lead to a model representation that contains all aspects that are considered relevant and is considered sufficiently accurate for compensation.

\begin{figure}[!htb]
    \centering
    \includegraphics{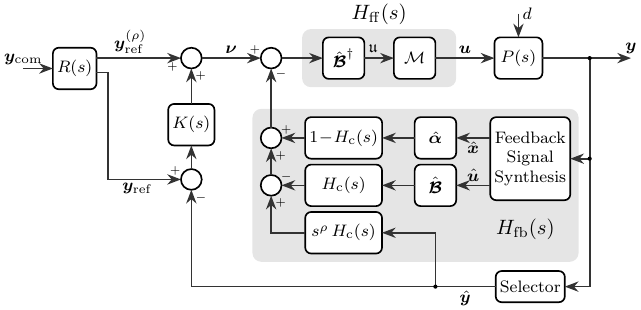}
    \caption{Diagram of the linearized closed-loop system for a generic hybrid NDI architecture.}
    \label{fig:robust_architecture}
\end{figure}

At this point, it is also important to realize that the inverse mapping from $\vec{\nu}$ to the control inputs $\vec{u}$ is of static nature: there is no feedback mechanism that ensures that $\vec{y}^{(\rho)}$ follows $\vec{\nu}$. This task is completely left to the feedback control laws in the outer loop functions (cf.~\cref{fig:NDItypicalArchitecture}). Current literature also focuses on combining the NDI architecture with (structured) robust control synthesis~\cite{Milz2024i,Encarnacao2026,Pollack2024,Pollack2026}. To do so, the control architecture is divided into a feed-forward and feedback block for the inversion and compensation of the system, wrapped by an outer loop feedback control function, as shown in~\cref{fig:robust_architecture}.
Some publications have taken a different approach at this point, allowing for elegant integration of, e.g., structural damping and load alleviation functions~\cite{Gregory2005,Stalla2026}.

A very important aspect is that the estimates need to be more or less synchronous, and overall time delay should be as small as reasonably possible. Computation of especially \(\hat{\vec{\alpha}}(\cdot)\) will otherwise be inconsistent, respectively ``too late'' to be compensated for. 

In flight control applications it is not common to explicitly estimate \(\hat{\vec{d}}\). Still, atmospheric disturbances are inherently part of airdata referenced state estimation and feed into the equation accordingly. As mentioned before, full compensation is not always desirable for control activity reasons, as will be addressed in the next section. 

\subsubsection*{Inversion error}
For \cref{eq:ndi_affine}, the inversion error can be obtained by substitution in \cref{eq:yrho_unc_affine}
\begin{align}
	\vec{e}_\mathrm{NDI} &= \vec{\nu} - \left( \hat{\vec{\alpha}}(\vec{x},\vec{d},\vec{p}) + \Delta \vec{\alpha}(\vec{x},\vec{d},\vec{p}) \right) - \left( \hat{\mat{\Beta}}(\vec{x},\vec{d},\vec{p}) + \Delta \mat{\Beta}(\vec{x},\vec{d},\vec{p}) \right) \ \vec{\mathfrak{u}} - \vec{\delta}(\cdot) \notag \\
	& = \left( \hat{\vec{\alpha}}(\hat{\vec{x}},\hat{\vec{p}},\hat{\vec{d}}) - \hat{\vec{\alpha}}(\vec{x},\vec{d},\vec{p}) \right) - \Delta\vec{\alpha}(\vec{x},\vec{d},\vec{p}) + \left( \hat{\mat{\Beta}}(\hat{\vec{x}},\hat{\vec{p}},\hat{\vec{d}}) - \hat{\mat{\Beta}}({\vec{x},\vec{d},\vec{p}}) \right) \vec{\mathfrak{u}} - \Delta {\mat{\Beta}}({\vec{x},\vec{d},\vec{p}}) \vec{\mathfrak{u}} - \vec{\delta}(\cdot) \notag \\
	& \approx - \Delta \vec{\alpha}(\vec{x},\vec{d},\vec{p}) - \Delta \mat{\Beta}(\vec{x},\vec{d},\vec{p}) \ \vec{\mathfrak{u}} - \vec{\delta}(\cdot) \label{eq:NDI_err}
\end{align}
where sufficiently good state, disturbance and parameter estimations are assumed in the last step, i.e., \( \hat{\vec{x}} = \vec{x},\  \hat{\vec{d}} = \vec{d},\  \hat{\vec{p}} = \vec{p} \). The inversion accuracy is then solely determined by the unmodeled effects and disturbances. Without the assumption, the estimation errors obviously add up to the inversion error. 

\subsubsection*{Algorithmic Complexity}
The NDI control law in~\cref{eq:ndi_affine,eq:ndi_nonaffine} gives considerable design degrees of freedom to make trade-offs between inverse model accuracy and complexity. Maximizing accuracy will unavoidably result in more complex formulations of \(\hat{\vec{\alpha}}\) and \(\hat{\vec{\Beta}}\) (or \(\hat{\vec{\beta}}\)), any simplification will leave more compensatory work to the outer feedback control function. The form of~\cref{eq:ndi_affine} and~\cref{eq:ndi_nonaffine} will be the reference for comparison with subsequently derived NDI variants.

Interestingly, the Robust Inverse Dynamics Estimation (RIDE) method presented in \cite{Muir1997} takes a different implementation approach to NDI by linearizing the aircraft model and applying NDI to the linear state space models. This results in computationally efficient gain matrices that are scheduled as a function of flight condition and configuration parameters.

\subsubsection*{Disturbance Compensation}
The system information that effectively feeds into the control algorithms is directly reflected by~\cref{eq:ndi_affine,eq:ndi_nonaffine} and can be influenced by design of signal processing of the estimated states, parameters, and disturbances. In combination with outer feedback control functions, considerable leverage exists in trading (disturbance rejection) performance, robustness, and control activity. 

\subsubsection*{Controls Failure Handling}
From an inversion point of view, the failure of a control device has two main consequences: (1) reduced capability to make \(\vec{y}^{(\rho)}\) follow the given virtual control inputs, and (2) the effect of the inoperable control device on the output derivatives \(\vec{y}^{(\rho)}\) as a potential additional disturbance (if moving freely, runs away, or gets stuck at some deflection). A third effect may be loss of trim loading, but this can be interpreted as a constant negative disturbance of the original trimmed value. Without detection of a fault, NDI has no means to compensate for either consequence. In case of detection, control allocation can be adapted to realize generalized controls \(\vec{\mathfrak{u}}\) in an alternative way. In case the faulty surface position can still be measured, its effect on \(\vec{y}^{(\rho)}\) can be explicitly compensated for by means of an added contribution to \(\hat{\vec{\alpha}}\). In the affine case, compensation of the effect of the failed control device \(\vec{u_i}\) may then for example look like:
\begin{align}
	\vec{\mathfrak{u}} &= \hat{\mat{\Beta}}^{-1}\left(\hat{\vec{x}},\hat{\vec{p}},\hat{\vec{d}}\right) \left[ \vec{\nu} - \underbrace{\left(\hat{\vec{\alpha}} ( \hat{\vec{x}},\hat{\vec{p}},\hat{\vec{d}}) + \hat{\mat{\Beta}}_{u_i}(\hat{\vec{x}},\hat{\vec{p}},\hat{\vec{d}}) \, u_i\right)}_{\tilde{\vec{\alpha}}(\hat{\vec{x}},\hat{\vec{p}},\hat{\vec{d}})} \right]
\end{align}

Reduced overall control power due to faults will obviously result in hitting control limits considerably quicker than before. This situation is not NDI-specific and therefore addressed in~\cref{sec:design}.

\subsubsection*{Zero Dynamics}
The internal dynamics based on the reduced model in~\cref{eq:abnormal_nonaffine_system0} is given by: 
\begin{align}
	\dot{\vec{x}}_I & = \vec{{f_I}}(T^{-1}(\vec{\xi_0},\vec{\eta}),\vec{p})   - \vec{{G_I}}(T^{-1}(\vec{\xi_0},\vec{x_I}),\vec{p}){\mat{\hat{\Beta}}}^{-1}\left({T^{-1}(\vec{\xi_0},\vec{x_I}),\vec{p}}\right){\vec{\hat{\alpha}}} ( {T^{-1}(\vec{\xi_0},\vec{x_I}),\vec{p}} )
\end{align}
hereby assuming that states and parameters are known exactly.
This equation is not really revealing, but application to the full system model in~\cref{eq:full_system} is:
\begin{align}
	\dot{\vec{\mathfrak{x}}}_I & = \vec{F}_I(\vec{\mathfrak{x}},\vec{u},\vec{d},\vec{p})
\end{align}
with
\begin{align}
	u & = \mathcal{M}(\vec{\mathfrak{u}},...) \\
	\vec{\mathfrak{u}} &= {\mat{\hat{\Beta}}}^{-1}\left({T^{-1}(\vec{\xi_0},\vec{\mathfrak{x}}_I),\vec{p}}\right){\vec{\hat{\alpha}}} ( {T^{-1}(\vec{\xi_0},\vec{\mathfrak{x}}_I),\vec{p}} ) \notag \\
	&\approx {\mat{\hat{\Beta}}}^{-1}\left({T^{-1}(\vec{\xi_0},\vec{{x}}_I),\vec{p}}\right){\vec{\hat{\alpha}}} ( {T^{-1}(\vec{\xi_0},\vec{{x}}_I),\vec{p}} )
\end{align}
The interesting thing is that \(\vec{\mathfrak{x}}_I\) includes the entries of \(\vec{x_I}\), as well as all states that have been truncated or residualized in the process of simplification towards the representation in~\cref{eq:abnormal_nonaffine_system0}. The NDI control law intends to only feed back \(\vec{x_I}\), which makes quite a difference on its effect on internal stability. In practice, this may not be completely true, as in the estimation of measured states, there may be influence from the neglected ones, hence the initial use of \(\vec{\mathfrak{x}}_I\) as an argument. A well-known example are aeroelastic dynamics. Given the fact that sensors are attached to the structure in some way, scooping up some contributions from structural dynamic states is unavoidable. Usually considerable signal processing and sensor positioning efforts are made to minimize this. 

\subsubsection{Sensory Nonlinear Dynamic Inversion}\label{sec:method:sndi}

Sensory, or sensor-based NDI employs sensor measurements of commanded variable derivatives to reduce dependence on the on-board plant model, as proposed in~\cite{Smith1998}. Starting from~\cref{eq:yrho_unc_nonaffine0,eq:yrho_unc_affine0}, the core principle is to estimate the drift dynamics \( \hat{\vec{\alpha}} \) using direct measurements of \( \hat{\vec{y}}^{(\rho)} \), subtracting the part that is contributed by the current control inputs via \( \hat{\vec{\beta}}_u (\hat{\vec{x}},\hat{\vec{p}},\hat{\vec{d}}, \hat{\vec{u}}) \), or \( \hat{\vec{\Beta}}_u (\hat{\vec{x}},\hat{\vec{p}},\hat{\vec{d}}) \hat{\vec{u}} \):
\begin{align}
	{\vec{\alpha}}({\vec{x}},{\vec{p}},{\vec{d}}) &\approx \hat{\vec{y}}^{(\rho)} - \hat{\vec{\beta}}_u(\hat{\vec{x}},\hat{\vec{p}},\hat{\vec{d}},\hat{\vec{u}})  \label{eq:yrho_unc_nonaffine2} \\ 
	{\vec{\alpha}}({\vec{x}},{\vec{p}},{\vec{d}}) &\approx \hat{\vec{y}}^{(\rho)} - \hat{\mat{\Beta}}_u(\hat{\vec{x}},\hat{\vec{p}},\hat{\vec{d}})\hat{\vec{u}}  \label{eq:yrho_unc_affine2}
\end{align}
Note that the formulations use measured control inputs \(\hat{\vec{u}}\). These are usually more straight-forward to determine than the actually realized generalized controls \(\hat{\vec{\mathfrak{u}}}\), as this would effectively require an inverse of the allocation function. The input terms \(\hat{\vec{\beta}}_u\), \(\hat{\vec{\Beta}}_u\) are therefore configured accordingly. Actual control deflections are typically measured for use in actuator control loops, so that their availability for flight control laws is usually not an issue. 

The sensory NDI control law for the non-affine system \( \Sigma \) can be stated as:
\begin{equation}\label{eq:sensory_ndi_nonaffine}
	\vec{\mathfrak{u}} = \hat{\vec{\beta}}^{-1} \left( (\hat{\vec{x}},\hat{\vec{p}},\hat{\vec{d}}), \ \vec{\nu} - \hat{\vec{y}}^{(\rho)} + \hat{\vec{\beta}}_u (\hat{\vec{x}},\hat{\vec{p}},\hat{\vec{d}}, \hat{\vec{u}}) \right)
\end{equation}
For the affine system \( \Sigma_G \), the control law simplifies to
\begin{equation}\label{eq:sensory_ndi_affine}
	\vec{\mathfrak{u}} = \hat{\mat{\Beta}} ( \hat{\vec{x}},\hat{\vec{p}},\hat{\vec{d}} )^{-1} \left( \vec{\nu} - \hat{\vec{y}}^{(\rho)} + \hat{\mat{\Beta}}_u ( \hat{\vec{x}},\hat{\vec{p}},\hat{\vec{d}} ) \hat{\vec{u}} \right)
\end{equation}
Note that \( \hat{\mat{\Beta}} \) relates the \emph{generalized} controls \( \vec{\mathfrak{u}} \) to the output derivatives, whereas \( \hat{\mat{\Beta}}_u \) relates the \emph{physical} controls \( \vec{u} \); the two matrices consequently differ in their input dimension (\( n_y \) versus \( n_u \)).

Obviously, measurements of \(\hat{\vec{y}}^{(\rho)}\) and especially \(\hat{\vec{u}}\) must be of sufficient quality, have sufficient sampling rate, and necessarily be \emph{synchronous}. The controls effect that is subtracted, must closely match the actual contribution as sensed in \(\hat{\vec{y}}^{(\rho)}\), or, in other words: measured \(\hat{\vec{y}}^{(\rho)}\) must, apart from the other contributions, have been caused by the measured controls~\cite{Grondman2018}. If this is not the case, the estimate is not valid and closed loop behavior may quickly become unpredictable. In sensory NDI controllers, synchronization is especially critical due to the direct connection between sensor measurements and control commands. These issues are well known and addressed in, e.g.,~\cite{Kier2020,Steffensen2022,Steffensen2023,Kumtepe2022,Autenrieb2025}.

\subsubsection*{Inversion error}
The error term for a sensory NDI law can be stated based on~\cref{eq:sensory_ndi_affine} as
\begin{align}
	\vec{e}_\mathrm{sNDI} & = \vec{\nu} - \hat{\vec{\alpha}}(\vec{x},\vec{d},\vec{p}) - \Delta\vec{\alpha} (\vec{x},\vec{d},\vec{p}) - \left( \hat{\mat{\Beta}}(\vec{x},\vec{d},\vec{p}) + \Delta\mat{\Beta}(\vec{x},\vec{d},\vec{p}) \right) \vec{\mathfrak{u}} - \vec{\delta}(\cdot)\notag \\
	& \approx - \Delta\mat{\Beta}(\vec{x},\vec{d},\vec{p})\  \hat{\mat{\Beta}}(\hat{\vec{x}},\hat{\vec{d}},\hat{\vec{p}})^{-1} \left( \vec{\nu} - \hat{\vec{y}}^{(\rho)} \right) + \left( \hat{\vec{y}}^{(\rho)} - \vec{y}^{(\rho)} \right) \label{eq:sNDI_err}
\end{align}
when assuming sufficiently good control input and state estimations in the last step, i.e., \( \hat{\vec{u}} = \vec{u} \), and \( \hat{\vec{x}} = \vec{x} \). For an exact measurement \( \hat{\vec{y}}^{(\rho)} = \vec{y}^{(\rho)} \), collecting the error terms in~\cref{eq:sNDI_err} gives 
\[ \left( \mat{I}_{n_y} + \Delta\mat{\Beta} \, \hat{\mat{\Beta}}^{-1} \right) \vec{e}_\mathrm{sNDI} = \hat{\vec{y}}^{(\rho)} - \vec{y}^{(\rho)} \]
where \( \mat{I}_{n_y} + \Delta\mat{\Beta} \, \hat{\mat{\Beta}}^{-1} = \mat{\Beta} \, \hat{\mat{\Beta}}^{-1} \). In this case, the equation consequently only has the trivial solution \(\vec{e}_\mathrm{sNDI}=0\), provided the true control effectiveness \( \mat{\Beta} \) is nonsingular. 
Otherwise the error is amplified by \( \mat{\Beta} \, \hat{\mat{\Beta}}^{-1} \), which is the origin of the sufficient condition \( \Vert \mat{I}_{n_y} - \mat{\Beta} \, \hat{\mat{\Beta}}^{-1} \Vert < 1 \) imposed in stability analyses of the incremental laws~\cite{Wang2019}. The exact measurement case does not match reality, but shows that sensory NDI mainly depends on the measurement accuracy of \( {\vec{y}}^{(\rho)} \) and the accurate estimation of the control input and states, but less so on the control effectiveness estimation residuum \( \Delta {\mat{\Beta}} \). %

It is interesting to note that this formulation inherently reduces sensitivity to modeling errors for the outer control functions, as the intended linearization and decoupling are more likely to be reliably realized. This  reduces uncertainty of the system model as seen by the outer function. 

\subsubsection*{Algorithmic Complexity}
The control law in~\cref{eq:sensory_ndi_nonaffine} or~\cref{eq:sensory_ndi_affine} is surprisingly simple as compared with the NDI variant in~\cref{eq:ndi_nonaffine}, resp.~\cref{eq:ndi_affine}, as only \(\hat{\vec{\beta}}\) or \(\hat{\mat{\Beta}}\) need to be computed as model-based components.
\vskip 1ex

Absolutely critical in implementation is the \emph{synchronization between measured or estimated \(\hat{\vec{y}}^{(\rho)}\) and \(\hat{\vec{u}}\)}. 
\vskip 1ex

This may require some adjustments during implementation testing, as precise hardware delay differences between sensors may be difficult to predict in advance. Great care during first-time testing is recommended, because mismatches may cause unpredictable behavior: very fast oscillations in control commands are notorious. An overall, but synchronous delay is less of an issue and comparable to model-based NDI \cite{Vlaar2014, Grondman2018}.

\subsubsection*{Disturbance Compensation}
Measurement of \( {\vec{y}}^{(\rho)} \) in the control law in~\cref{eq:sensory_ndi_nonaffine} resp.~\cref{eq:sensory_ndi_affine} can be modeled by substituting~\cref{eq:yrho_unc_nonaffine}, resp.~\cref{eq:yrho_unc_affine}:
\begin{align}
	\vec{\mathfrak{u}} & = {\hat{\mat{\Beta}} ( \hat{\vec{x}},\hat{\vec{d}},\hat{\vec{p}} ) }^{-1} \left( \vec{\nu} - \underbrace{\left[ \hat{\vec{\alpha}}(\vec{x},\vec{d},\vec{p}) + \Delta \vec{\alpha}(\vec{x},\vec{d},\vec{p}) + \left( \hat{\mat{\Beta}}(\vec{x},\vec{d},\vec{p}) + \Delta\mat{\Beta}(\vec{x},\vec{d},\vec{p}) \right) \vec{\mathfrak{u}} + \vec{\delta} \right]}_{{\hat{\vec{y}}}^{(\rho)}} + \hat{\mat{\Beta}}_u ( \hat{\vec{x}},\hat{\vec{d}},\hat{\vec{p}} ) \hat{\vec{u}} \right) \label{eq:sNDI_dist}
\end{align}
While the influence of model uncertainty decreases, sensory NDI directly feeds through any external disturbance due to its contribution to measurements of the output derivatives \( \hat{\vec{y}}^{(\rho)} \). While this may be very attractive in case fast and accurate trajectory tracking is required, it may pose serious problems on larger aircraft with the flight control system operating for hours at a stretch. For example, \cite{Kier2020} reports excessive loads on the aircraft empennage due to the built-in tendency to aim for full compensation of atmospheric disturbances. The problem is even magnified due to inherent flexibility of the airframe. The issue here is that the assumed residualization (or truncation) of aeroelastic dynamics in~\cref{sec:modelcomplexity} is not exercised here, as these will unavoidably come through via \( \hat{\vec{y}}^{(\rho)} \). Dynamic deformation will be picked up as well and unintendedly provided to the control law for compensation.
In the case of model-based NDI the model equations in \(\hat{\vec{\alpha}}\) do apply the above assumption, and complementary filtering can be applied to state measurements to effectively balance disturbance rejection with control activity~\cite{Looye2001b}. %

An obvious remedy for sensory NDI would be the use of filters on \( \hat{\vec{y}}^{(\rho)} \). However, being positioned in the innermost core of the control system, this immediately leads to bandwidth limitations towards outer loop functions. Adding a model-based complement solves this issue and has given rise to hybrid NDI variants, as will be discussed subsequently. 

\subsubsection*{Controls Failure Handling}
As for NDI, the main effects of a faulty control device are in (1) reduced control effectiveness, and in (2) the effect of the faulty device behavior on the system dynamics. The effect of reduced control effectiveness is not different from NDI, as the same inversion of \(\hat{\vec{\beta}}\) or \(\hat{\mat{\Beta}}\) is applied. The use of measured \(\vec{y}^{(\rho)}\) in principle helps to address the second effect. However, it is necessary to know that the control device is inoperable. This is best explained by having a closer look at the individual terms of the control law in~\cref{eq:sensory_ndi_affine}:
\begin{equation}
	\vec{\mathfrak{u}} = \underbrace{\hat{\mat{\Beta}} ( \hat{\vec{x}},\hat{\vec{d}},\hat{\vec{p}} )^{-1}}_{\parbox{2.8cm}{Compute new \\ control input}} \ \ \ (  \underbrace{\vec{\nu} - \hat{\vec{y}}^{(\rho)}}_{\parbox{2.3cm}{Increment due to difference}} \ \ \ +\ \ \  \underbrace{\hat{\mat{\Beta}}_u ( \hat{\vec{x}},\hat{\vec{d}},\hat{\vec{p}} ) \hat{\vec{u}}}_{\parbox{2.9cm}{Control deflections applied so far}})
\end{equation} 
The computed control input is a summation of its current value and increments needed to compensate for the difference between measured \(\vec{y}^{(\rho)}\) and commanded \(\vec{\nu}\). 
In a perfectly trimmed equilibrium condition, where \(\vec{\nu} - \hat{\vec{y}}^{(\rho)} = 0\) (affine case):
\begin{align}
	\vec{\mathfrak{u}} &= \hat{\mat{\Beta}} ( \hat{\vec{x}},\hat{\vec{d}},\hat{\vec{p}} )^{-1} \left( 0 + \hat{\mat{\Beta}}_u ( \hat{\vec{x}},\hat{\vec{d}},\hat{\vec{p}} ) \hat{\vec{u}} \right)  \label{eq:sndi_erroru} \\
	&= \hat{\mat{\Beta}} ( \hat{\vec{x}},\hat{\vec{d}},\hat{\vec{p}} )^{-1}  \hat{\mat{\Beta}}_u ( \hat{\vec{x}},\hat{\vec{d}},\hat{\vec{p}} ) \hat{\vec{u}}  \notag
\end{align}
Allocation of \(\vec{\mathfrak{u}}\) should then again result in the measured equilibrium value of \(\hat{\vec{u}}\). 
However, in the case of an undetected failure of the \(i^{\mathrm{th}}\) control device, combined with an uncommanded deflection, i.e. \(u_i = \tilde{u}_i(t)\), there will be an immediate effect on \(\hat{\vec{y}}^{(\rho)}\). A typical example scenario is an actuator run-away. It is assumed that the the inoperable control device has a redundant companion. Making the failed controls contribution explicit, the sensor output can be written as \(\hat{\vec{y}}^{(\rho)}=\hat{\vec{y}}^{(\rho)}_0+\hat{\Beta}_{u_i}( \hat{\vec{x}},\hat{\vec{p}},\hat{\vec{d}} )\tilde{u}_i\). Substitution into~\cref{eq:sensory_ndi_affine} results in:
\begin{align}
	\vec{\mathfrak{u}} 
	&= \hat{\mat{\Beta}} ( \hat{\vec{x}},\hat{\vec{p}},\hat{\vec{d}} )^{-1} \left(  \vec{\nu} - (\hat{\vec{y}}^{(\rho)}_0 + \mat{\Beta}_{u_i}( \hat{\vec{x}},\hat{\vec{p}},\hat{\vec{d}} )\tilde{u}_i)  +  \hat{\mat{\Beta}}_u ( \hat{\vec{x}},\hat{\vec{p}},\hat{\vec{d}} ) \hat{\vec{u}}\right) \notag \\
	&= \hat{\mat{\Beta}} ( \hat{\vec{x}},\hat{\vec{p}},\hat{\vec{d}} )^{-1} \left(  \vec{\nu} - (\hat{\vec{y}}^{(\rho)}_0 + \mat{\Beta}_{u_i}( \hat{\vec{x}},\hat{\vec{p}},\hat{\vec{d}} )\tilde{u}_i)  + \mat{\Beta}_{u_i}( \hat{\vec{x}},\hat{\vec{p}},\hat{\vec{d}} )\tilde{u}_i + \hat{\mat{\Beta}}_u ( \hat{\vec{x}},\hat{\vec{p}},\hat{\vec{d}} ) \hat{\vec{u}}_{u_i = 0}\right) \notag \\
	&= \hat{\mat{\Beta}} ( \hat{\vec{x}},\hat{\vec{p}},\hat{\vec{d}} )^{-1} \left(  \vec{\nu} - \hat{\vec{y}}^{(\rho)}_0 + \hat{\mat{\Beta}}_u ( \hat{\vec{x}},\hat{\vec{p}},\hat{\vec{d}} ) \hat{\vec{u}}_{u_i = 0}\right) \notag 
\end{align} 
Note that \(\tilde{u}_i\) may include any missing trim deflection as well. As can be seen, \(\tilde{u}_i\) is left fully uncompensated for. Furthermore, the system is out of trim, as \(\hat{\mat{\Beta}}_u ( \hat{\vec{x}},\hat{\vec{p}},\hat{\vec{d}} ) \hat{\vec{u}}_{u_i = 0} \neq \hat{\mat{\Beta}}_u ( \hat{\vec{x}},\hat{\vec{p}},\hat{\vec{d}} ) \hat{\vec{u}}\). In addition the allocation algorithm will use the device as if still operable, so that recovery from the fault will be slow or even not feasible. Just the detection of the inoperable device allows it to be removed from the control law:
\begin{align}
	\vec{\mathfrak{u}} 
	&= \hat{\mat{\Beta}} ( \hat{\vec{x}},\hat{\vec{p}},\hat{\vec{d}} )^{-1} \left(  \vec{\nu} - (\hat{\vec{y}}^{(\rho)}_0 + \mat{\Beta}_{u_i}( \hat{\vec{x}},\hat{\vec{p}},\hat{\vec{d}} )\tilde{u}_i) \cancel{ + \mat{\Beta}_{u_i}( \hat{\vec{x}},\hat{\vec{p}},\hat{\vec{d}} )\tilde{u}_i} + \hat{\mat{\Beta}}_u ( \hat{\vec{x}},\hat{\vec{p}},\hat{\vec{d}} ) \hat{\vec{u}}_{u_i = 0}\right) \notag \\
	&= \hat{\mat{\Beta}} ( \hat{\vec{x}},\hat{\vec{p}},\hat{\vec{d}} )^{-1} \left(  \vec{\nu} - \hat{\vec{y}}^{(\rho)} + \hat{\mat{\Beta}}_u ( \hat{\vec{x}},\hat{\vec{p}},\hat{\vec{d}} ) \hat{\vec{u}}_{u_i = 0}\right) \notag 
\end{align} 
Control allocation can be adapted just by excluding effectiveness of \(\mat{\Beta}_{u_i}\) in realizing \(\vec{\mathfrak{u}}\).

As compared with NDI, if only the inoperable device can be detected as such, it can be excluded from \(\hat{\vec{u}}\) and its effect as a disturbance, including loss of its trim contribution, is automatically addressed. The effect of the faulty device on \(\vec{\mathfrak{u}}\) has to be removed in the model equations and allocation function, just like in the case of model-based NDI.

\subsubsection*{Zero Dynamics}
The internal dynamics based on the reduced model in~\cref{eq:abnormal_nonaffine_system0} results from substitution of \cref{eq:sNDI_dist}: 
\begin{align}
	\dot{\vec{x}}_I 
    =& \vec{f}_I(T^{-1}(\vec{\xi}_0,\vec{\eta}),\vec{p})- \mat{G}_I(T^{-1}(\vec{\xi}_0,\vec{x}_I),\vec{p}) \notag \\  
	&{\hat{\mat{\Beta}} (T^{-1}(\vec{\xi}_0,\vec{x}_I), \hat{\vec{p}} ) }^{-1} 
	\left(\left[ {\vec{\alpha}}(T^{-1}(\vec{\xi}_0,\vec{x}_I),\vec{p}) +  {\mat{\Beta}}(T^{-1}(\vec{\xi}_0,\vec{x}_I),\vec{p}) \vec{\mathfrak{u}} \right] - \hat{\mat{\Beta}} (T^{-1}(\vec{\xi}_0,\vec{x}_I),\hat{\vec{p}} ) \hat{\vec{\mathfrak{u}}} \right) \notag \\
	\approx & \vec{f}_I(T^{-1}(\vec{\xi}_0,\vec{\eta}),\vec{p})- \mat{G}_I(T^{-1}(\vec{\xi}_0,\vec{x}_I),\vec{p})\notag \\ &{\hat{\mat{\Beta}} (T^{-1}(\vec{\xi}_0,\vec{x}_I), \hat{\vec{p}} ) }^{-1}
	\left[{\vec{\alpha}}(T^{-1}(\vec{\xi}_0,\vec{x}_I),\vec{p}) +  {\Delta\mat{\Beta}}(T^{-1}(\vec{\xi}_0,\vec{x}_I),\vec{p}) \vec{\mathfrak{u}} \right]
\end{align}
hereby again assuming that states, parameters, and realized generalized controls are known exactly. It is more revealing to look at application to the full system model (\cref{eq:full_system}), but this time with
\begin{align}
	\vec{\mathfrak{u}} &= {\hat{\mat{\Beta}} (T^{-1}(\vec{\xi}_0,\vec{x}_I), \hat{\vec{p}} ) }^{-1} 
	\left(-\left[ {\vec{\alpha}}(T^{-1}(\vec{\xi}_0,\vec{x}_I),\vec{\mathfrak{x}_I},\vec{p}) +  {\mat{\Beta}}(T^{-1}(\vec{\xi}_0,\vec{x}_I),\vec{\mathfrak{x}_I},\vec{p}) \vec{\mathfrak{u}} \right] + \hat{\mat{\Beta}} (T^{-1}(\vec{\xi}_0,\vec{x}_I),\hat{\vec{p}} ) \hat{\vec{\mathfrak{u}}} \right)
\end{align}
As in the derivation for NDI, accurate estimation of \(\vec{x}_I\) is assumed here. However, measurement of \(\vec{y}^{(\rho)}\) will unavoidably result in picking up unmodeled dynamics in \(\vec{\mathfrak{x}_I}\) and, as argued before, filtering these out may be challenging. 

\subsubsection{Incremental Nonlinear Dynamic Inversion}\label{sec:method:indi}

A further reduction of the dependence on the on-board plant model can be achieved using a Taylor series expansion or linearization. This is especially useful as an approximation of the non-affine control law.
There are several ways to derive the incremental NDI control law~\cite{Sieberling2010,Wang2019,Steffensen2022}. The different derivations lead to similar control laws but allow a different view.

\paragraph{Taylor Series Approximation}
The most widespread derivation performs a first-order Taylor series approximation of the dynamics and uses time-scale separation to neglect the slower state dynamics over the faster input dynamics. This leads to the following incremental control law~\cite{Sieberling2010,Wang2019} for the non-affine system~\cref{eq:yrho_unc_nonaffine0} %
{\small
	\begin{align}
		&\hat{\vec{y}}^{(\rho)}  = \hat{\vec{y}}^{(\rho)}_0 + \nabla_\vec{u} \left. \hat{\vec{\beta}}(\vec{x},\vec{d},\vec{p},\vec{u}) \right\vert_{0} \underbrace{\left( \vec{u} - \vec{u}_0 \right)}_{\Delta \vec{u}} + \notag \\
		& \underbrace{ \left. \Bigl( \nabla_\vec{x} \hat{\vec{\alpha}}(\vec{x},\vec{d},\vec{p}) + \nabla_\vec{x} \hat{\vec{\beta}}(\vec{x},\vec{d},\vec{p},\vec{u}) \Bigr) \right\vert_{0} \, \underbrace{\left( \vec{x} - \vec{x}_0 \right)}_{\Delta \vec{x}} + 
			\left. \Bigl( \nabla_\vec{d} \hat{\vec{\alpha}}(\vec{x},\vec{d},\vec{p}) + \nabla_\vec{d} \hat{\vec{\beta}}(\vec{x},\vec{d},\vec{p},\vec{u}) \Bigr) \right\vert_{0} \, \underbrace{\left( \vec{d} - \vec{d}_0 \right)}_{\Delta \vec{d}} + \mathcal{O}(\Delta \vec{x}^2, \Delta \vec{d}^2, \Delta \vec{u}^2) }_{ \mathcal{O}(\Delta \vec{x}, \Delta \vec{d}, \Delta \vec{u}^2) } \notag
	\end{align}
}
Ref.~\cite{Sieberling2010} at this point makes the following assumptions:
\begin{itemize}
	\item \( \Delta \vec{u} \gg \Delta \vec{x} \), the same holding for the corresponding terms in the above equation, due to time scale separation principle (\(\Delta \vec{x}\) only evolves after integration);
	\item higher order terms \( \mathcal{O}(\Delta \vec{x}, \Delta \vec{u}^2) \) are small and can be compensated by the feedback controller in outer functions in the longer term;
	\item the effect of \( \Delta \vec{d} \) is neglected and compensation is left to the outer functions as well.
\end{itemize}
Then, using the current measurements as the expansion point leads to the affine control law
\begin{equation} \label{eq:indi_nonaffine}
	\Delta \vec{\mathfrak{u}} = \left( \nabla_\vec{\mathfrak{u}} \hat{\vec{\beta}} \ (\hat{\vec{x}},\hat{\vec{d}},\hat{\vec{p}},\hat{\vec{\mathfrak{u}}}) \right)^{-1} \left( \vec{\nu} - \hat{\vec{y}}^{(\rho)} \right)
\end{equation}
For a control-affine system, the control law becomes
\begin{equation} \label{eq:indi_affine}
	\Delta \vec{\mathfrak{u}} = {\hat{\mat{\Beta}}(\hat{\vec{x}},\hat{\vec{d}},\hat{\vec{p}})}^{-1} \left( \vec{\nu} - \hat{\vec{y}}^{(\rho)} \right)
\end{equation}

The control law computes increments \( \Delta \vec{\mathfrak{u}} \) on the current control inputs \( \hat{\vec{\mathfrak{u}}} \). These may either be obtained from measurement or estimation. The latter typically requires on-board models of the actuators. Both approaches have been successfully applied and tested, but measurements appear to work better~\cite{Grondman2018}.

In order to obtain the absolute control inputs, \( \Delta \vec{\mathfrak{u}} \) must be appropriately allocated, i.e., \( \vec{u} =  \hat{\vec{u}} + \mathcal{M} ( \Delta \vec{\mathfrak{u}} ) \). This requires an incremental approach to control allocation, which is for example addressed in~\cite{Matamoros2018}. The optimization problem is inherently more complicated, as cumulative control input limits and coordination are to be addressed via constraints. For this reason, the procedure is often changed to \( \vec{u} =  \mathcal{M} \left(  \mathcal{M}^{-1} ( \hat{\vec{u}} ) + \Delta \vec{\mathfrak{u}} \right) \), which effectively resembles the sensory NDI control law.

An important issue to keep an eye on is the interaction with actuator control loops. In case \(\vec{\mathfrak{u}} = \vec{u}\) (no allocation), the control command is:
\begin{equation}
	\vec{u} = \vec{u}_0 + \Delta \vec{u}
\end{equation}
This implies positive feedback of the measured current control setting, which may be simultaneously used negatively by the actuator control loops. As a consequence, this control loop may be opened by the incremental NDI law. This may not pose an immediate problem for the control laws at hand, but the actuator control loop is also opened for any other function operating on the same control device, as will be addressed in~\cref{sec:design}.

\paragraph{Continuous-Time Total Derivatives}
Derivation through the continuous-time total derivatives is presented in~\cite{Raab2019}. There, instead of a Taylor series approximation, the authors use the total derivative of the normal form~\cref{eq:yrho_unc_nonaffine0} to derive a continuous-time incremental NDI control law
\begin{equation}
	\dot{\hat{\vec{y}}}^{(\rho)} = \Bigl( \frac{\partial \hat{\vec{\alpha}}(\vec{x},\vec{d},\vec{p})}{\partial \vec{x}} + \frac{\partial \hat{\vec{\beta}}(\vec{x},\vec{d},\vec{p},\vec{u})}{\partial \vec{x}} \Bigr) \dot{\vec{x}} + \Bigl( \frac{\partial \hat{\vec{\alpha}}(\vec{x},\vec{d},\vec{p})}{\partial \vec{d}} + \frac{\partial \hat{\vec{\beta}}(\vec{x},\vec{d},\vec{p},\vec{u})}{\partial \vec{d}} \Bigr) \dot{\vec{d}} + \frac{\partial \hat{\vec{\beta}}(\vec{x},\vec{d},\vec{p},\vec{u})}{\partial \vec{u}} \dot{\vec{u}}
\end{equation}
Assuming linear actuator dynamics \( \dot{\vec{u}} = \mat{K}_\mathrm{act} \mat{F}_\mathrm{act}(s) \Delta \vec{u} \) and neglecting the \textit{disturbance} terms yields conceptually
\begin{equation}
	\Delta \vec{u} = \mat{F}^{-1}_\mathrm{act}(s) \left( \frac{\partial \vec{\beta}(\vec{x},\vec{d},\vec{p},\vec{u})}{\partial \vec{u}}  \mat{K}_\mathrm{act} \right)^{-1} \mat{K}_\nu \mat{F}_\nu(s) \left( \vec{\nu} - \hat{\vec{y}}^{(\rho)}  \right)
\end{equation}
where a linear pseudo control dynamic for \( \vec{\nu} \) similar to the actuator's is assumed.

\paragraph{Sensory View}
A \emph{sensory view} is gathered by modifying~\cref{eq:sensory_ndi_affine} to
\begin{align}
    \vec{\mathfrak{u}} & = {\hat{\mat{\Beta}} ( \hat{\vec{x}},\hat{\vec{d}},\hat{\vec{p}} )}^{-1} \left( \vec{\nu} - \hat{\vec{y}}^{(\rho)} + \hat{\mat{\Beta}}_u ( \hat{\vec{x}},\hat{\vec{d}},\hat{\vec{p}} ) \, \hat{\vec{u}} \right) \notag \\
	& = {\hat{\mat{\Beta}} (  \hat{\vec{x}},\hat{\vec{d}},\hat{\vec{p}} )}^{-1} \left( \vec{\nu} - \hat{\vec{y}}^{(\rho)} \right) + \hat{\vec{\mathfrak{u}}} \label{eq:indilaw}
\end{align}
This shows that the incremental NDI law is often achieved with an intrinsic control allocation. 
Typical implementations of incremental NDI controllers comprise a filtered feedback of the commanded control inputs, instead of measured deflections as resulting from control allocation and actual actuator behavior.

The practical applicability of the time scale separation principle for incremental NDI is demonstrated in numerous simulations and flight tests, e.g.,~\cite{Vlaar2014,Smeur2016,Grondman2018}.
Wang \textit{et al.}~\cite{Wang2019} point out that the separation applied with incremental NDI is not the one of singular perturbation theory. They reformulate the incremental law without it, for systems of arbitrary relative degree and with the internal dynamics retained, and on that basis provide a framework of stability and robustness analyses using Lyapunov methods.
The crucial part lies in measuring the state derivatives. Measuring, filtering, and estimating the angular accelerations presupposes a trade-off between noise reduction and phase delay, possibly leading to desynchronization with the control measurements or estimates and other signals. Incremental NDI requires higher sample rates to cope with high bandwidth signals. Sample rates of practical applications lie in the magnitude of \SIrange{40}{1000}{\hertz}~\cite{Grondman2018,Keijzer2019,Pollack2019,Gaessler2025,Pfeifle2022} for aircraft attitude control.

\subsubsection*{Inversion Error}
The inversion error for incremental NDI is identical to the one for the sensory variant.

\subsubsection*{Algorithmic Complexity}
The algorithmic complexity for incremental NDI is even lower than with the sensory variant, as only the inverse of the control effectiveness matrix is left.

\subsubsection*{Disturbance Compensation}
The disturbance compensation for incremental NDI is identical to the one for the sensory variant.
As for sensory NDI:
\begin{align}
	\vec{\mathfrak{u}} & = {\hat{\mat{\Beta}} ( \hat{\vec{x}},\hat{\vec{d}},\hat{\vec{p}} ) }^{-1} \left( \vec{\nu} - \underbrace{\left[ \hat{\vec{\alpha}}(\vec{x},\vec{d},\vec{p}) + \Delta \vec{\alpha}(\vec{x},\vec{d},\vec{p}) + \left( \hat{\mat{\Beta}}(\vec{x},\vec{d},\vec{p}) + \Delta\mat{\Beta}(\vec{x},\vec{d},\vec{p}) \right) \vec{\mathfrak{u}} + \vec{\delta} \right]}_{{\hat{\vec{y}}}^{(\rho)}} \right) +  \hat{\vec{\mathfrak{u}}} \label{eq:iNDI_dist}
\end{align}
The same remarks apply as made previously.

\subsubsection*{Controls Failure Handling}
The incremental formulation of NDI has a clear advantage in handling control failures.
Looking at~\cref{eq:indilaw} in the trimmed situation, the computation of the actual control input vector \(\vec{u}\) from the current measured deflection does not go via computation of \(\vec{\mathfrak{u}}\) and reallocation, but is simply set to \(\hat{\vec{u}}\). The faulty \(\hat{u}_{i}\) measurement is applied to the faulty device without effect to the system. The disturbance effect on \(\hat{\vec{y}}^{(\rho)}\), including missing trim offset, is retained and therefore automatically compensated for. The reduced overall control effectiveness only affects the increment computation and is therefore handled more robustly~\cite{Vlaar2014,Grondman2018}. A detailed comparison is also made in \cite{Hafner2026}, although this reference does not address the point made above.

\subsubsection*{Zero Dynamics}
The internal dynamics based on the reduced model in~\cref{eq:abnormal_nonaffine_system0} results from substitution of \cref{eq:iNDI_dist}: 
\begin{align}
	\dot{\vec{x}}_I 
	=& \vec{f}_I(T^{-1}(\vec{\xi}_0,\vec{x}_I),\vec{p})- \mat{G}_I(T^{-1}(\vec{\xi}_0,\vec{x}_I),\vec{p}) \notag \\  
	&\left({\hat{\mat{\Beta}} (T^{-1}(\vec{\xi}_0,\vec{x}_I), \hat{\vec{p}} ) }^{-1}
	\left[ {\vec{\alpha}}(T^{-1}(\vec{\xi}_0,\vec{x}_I),\vec{p}) +  {\mat{\Beta}}(T^{-1}(\vec{\xi}_0,\vec{x}_I),\vec{p}) \vec{\mathfrak{u}} \right] -\hat{\vec{\mathfrak{u}}}\right)
\end{align}
hereby again assuming that states, parameters, and realized generalized controls are known exactly. It is again more revealing to look at application to the full system model (\cref{eq:full_system}), but this time with
\begin{align}
	\vec{\mathfrak{u}} &= {\hat{\mat{\Beta}} (T^{-1}(\vec{\xi}_0,\vec{x}_I), \hat{\vec{p}} ) }^{-1} 
	\left(-\left[ {\vec{\alpha}}(T^{-1}(\vec{\xi}_0,\vec{x}_I),\vec{\mathfrak{x}_I},\vec{p}) +  {\mat{\Beta}}(T^{-1}(\vec{\xi}_0,\vec{x}_I),\vec{\mathfrak{x}_I},\vec{p}) \vec{\mathfrak{u}} \right] \right)  +\hat{\vec{\mathfrak{u}}}
\end{align}
Again, measurement of \(\vec{y}^{(\rho)}\) will unavoidably result in picking up unmodeled dynamics in \(\vec{\mathfrak{x}_I}\). Although not addressing this issue, a detailed analytical example of internal, structural dynamics in relation to the application of incremental and model-based NDI is provided in~\cite{Diz2026}.

\subsubsection{Hybrid Nonlinear Dynamic Inversion}\label{sec:method:hndi}

Incremental and sensory NDI have clear advantages relative to model-based NDI regarding disturbance rejection, robustness and fault tolerance. The main drawbacks are the crucial synchronization between output derivative and control deflection measurements, as well as reduced means to balance disturbance compensation with control activity and aspects like structural coupling and controls-induced flight loads. It is for this reason that hybrid forms have been proposed in various publications~\cite{Kumtepe2022,Jiali2016,Milz2024e,Kim2021,Akkinapalli2018}. Hybrid NDI addresses the above issues by combining the lower frequency content of sensory NDI, and the higher frequency content of NDI commands in a complementary way. The formulations in literature differ slightly, in this work complementary filtering is applied to the \( \hat{\vec{\alpha}}(\hat{\vec{x}},\hat{\vec{d}},\hat{\vec{p}}) \) term.

The generic hybrid NDI control law for \( \Sigma \) is formulated as 
\begin{equation}\label{eq:hndi}
	\vec{\mathfrak{u}} = {\hat{\vec{\beta}}}^{-1} \left( \hat{\vec{x}},\hat{\vec{d}},\hat{\vec{p}}, \ \vec{\nu} - \hat{\vec{\nu}}_\mathrm{compl} \right)
\end{equation}
for an arbitrary complementary filter function \( \mathcal{F}_\mathrm{compl} \):
\begin{equation}
    \hat{\vec{\nu}}_\mathrm{compl} = \mathcal{F}_\mathrm{compl} \Biggl\{ \begin{array}{ll}  \hat{\vec{\alpha}} (\hat{\vec{x}},\hat{\vec{d}},\hat{\vec{p}}), & \text{high-pass} \\ \hat{\vec{y}}^{(\rho)} - \hat{\vec{\beta}}_u (\hat{\vec{x}},\hat{\vec{d}},\hat{\vec{p}},\hat{\vec{u}}), & \text{low-pass} \end{array} \Biggr. 
\end{equation}
Assuming a linear low-pass complementary filter function \( H_\mathrm{c} (s) \), \cref{eq:hndi} can be written as
\begin{equation} \label{eq:hndi_laplace}
	\vec{\mathfrak{u}} = {\hat{\vec{\beta}}}^{-1} \Bigl( \hat{\vec{x}},\hat{\vec{d}},\hat{\vec{p}}, \ \vec{\nu} - \mathcal{L}^{-1} \left\{ H_\mathrm{c} (s) \right\} * \left( \hat{\vec{y}}^{(\rho)} - \hat{\vec{\beta}}_u ( \hat{\vec{x}},\hat{\vec{d}},\hat{\vec{p}}, \hat{\vec{u}} ) \right) - \mathcal{L}^{-1} \left\{  1 - H_\mathrm{c} (s) \right\} * \hat{\vec{\alpha}}(\hat{\vec{x}},\hat{\vec{d}},\hat{\vec{p}}) \Bigr)
\end{equation}
using the inverse Laplace transformation \( \mathcal{L}^{-1} \).
For the affine system \( \Sigma_G \), the control law becomes
\begin{equation}\label{eq:hindi_affine}
	\vec{\mathfrak{u}} = {\hat{\mat{\Beta}} \left( \hat{\vec{x}},\hat{\vec{d}},\hat{\vec{p}} \right)}^{-1} \Bigl( \vec{\nu} - \mathcal{L}^{-1} \left\{ H_\mathrm{c} (s) \right\} * \left( \hat{\vec{y}}^{(\rho)} - \hat{\mat{\Beta}}_u (\hat{\vec{x}},\hat{\vec{d}},\hat{\vec{p}}) \hat{\vec{u}} \right) - \mathcal{L}^{-1} \left\{  1 - H_\mathrm{c} (s) \right\} * \hat{\vec{\alpha}} (\hat{\vec{x}},\hat{\vec{d}},\hat{\vec{p}}) \Bigr)
\end{equation}

\begin{figure*}[!htb]
	\centering
	\includegraphics[width=\textwidth]{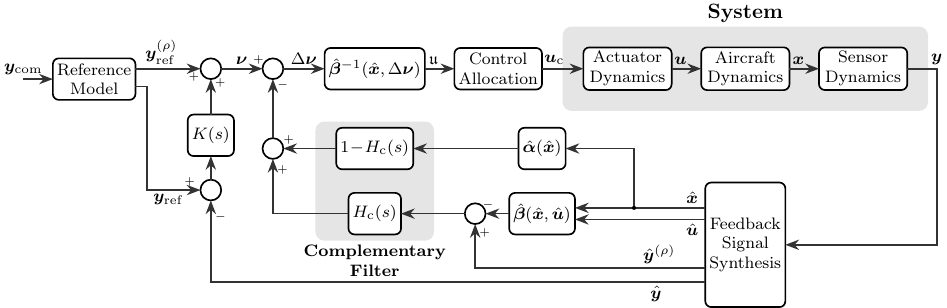}
	\caption{Schematics of hybrid NDI control structure with the complementary filter function \( H_\mathrm{c} (s) \) and a two degrees of freedom control design.}\label{fig:hndi}
\end{figure*}

\Cref{fig:hndi} shows the implementation schematically, where the shown compensator \( \vec{y}_\mathrm{ref} \) needs to be \( \rho \)-times continuously differentiable.
The complementary filter is commonly implemented in an integrated way with shared states between \( H_\mathrm{c} (s) \) and \( \left( 1 - H_\mathrm{c} (s) \right) \) as shown, e.g., in~\cite{Milz2024e,Kumtepe2022,Jiali2016}.
By using this complementary filter implementation on a system of relative degree 1, we can additionally circumvent the differentiation of the estimated state vector \( \hat{\vec{x}} \) or the need for measured accelerations by ``pulling \( \hat{\dot{\vec{x}}} \) through the integrator'', resulting in a simpler and potentially more robust implementation of the control law~\cite{Milz2026jgcd}.

By reformulating the complementary filter equation, another view on the hybrid NDI law is obtained
\begin{gather}
    \mathcal{L}^{-1} \bigl\{ \left[ H_\mathrm{c} (s) \ \ 1-H_\mathrm{c} (s) \right] \bigr\} * \begin{bmatrix} \hat{\vec{y}}^{(\rho)} - \hat{\vec{\beta}}_u ( \hat{\vec{x}},\hat{\vec{d}},\hat{\vec{p}}, \hat{\vec{u}} ) \\  \hat{\vec{\alpha}}(\hat{\vec{x}},\hat{\vec{d}},\hat{\vec{p}}) \end{bmatrix} \notag \\
	= \hat{\vec{\alpha}}(\hat{\vec{x}},\hat{\vec{d}},\hat{\vec{p}}) + \mathcal{L}^{-1} \bigl\{ H_\mathrm{c} (s) \bigr\} * \Bigl( \hat{\vec{y}}^{(\rho)} - \hat{\vec{\beta}}_u ( \hat{\vec{x}},\hat{\vec{d}},\hat{\vec{p}}, \hat{\vec{u}} ) - \hat{\vec{\alpha}}(\hat{\vec{x}},\hat{\vec{d}},\hat{\vec{p}})  \Bigr)
\end{gather}
This equation indicates that the hybrid NDI law uses \( \hat{\vec{\alpha}}(\hat{\vec{x}},\hat{\vec{d}},\hat{\vec{p}}) \) similar to NDI but corrects it with the low-pass filtered residuum of the sensor measurements of the commanded variable derivatives \( \hat{\vec{y}}^{(\rho)} \) and the estimated ones using the reconstructed state vector, measured inputs, and on-board plant model \( \hat{\vec{\alpha}}(\hat{\vec{x}},\hat{\vec{d}},\hat{\vec{p}}) + \hat{\vec{\beta}}_u ( \hat{\vec{x}},\hat{\vec{d}},\hat{\vec{p}}, \hat{\vec{u}} ) \).
This requires a reasonable choice of the complementary filter and its cutoff frequency.

Consequently, direct sensor measurements are low-pass filtered, cutting off higher frequency signal content, while model-dependent terms dominate in this domain. Simultaneously, the model-based terms providing noise-free higher frequency commands are corrected in the lower frequency domain by the measured aircraft state. The drawback of this approach is the necessity of both an estimated model and sensor measurements. However, the blend strongly increases robustness to uncertainties of either.

\subsubsection*{Inversion error}
The error of the hybrid NDI control law combines those of~\cref{eq:NDI_err} and~\cref{eq:sNDI_err}:
\begin{align}
	\vec{e}_\mathrm{hNDI} & %
	= \mathcal{L}^{-1} \left\{ 1 - H_\mathrm{c} (s) \right\} * \vec{e}_\mathrm{NDI} + \mathcal{L}^{-1} \left\{ H_\mathrm{c} (s) \right\} * \vec{e}_\mathrm{sNDI}                                                                                                                                                                  %
\end{align}
Applying the hybrid NDI law~\cref{eq:hindi_affine} to~\cref{eq:yrho_unc_affine} and assuming ideal sensor measurements \( \vec{y}^{(\rho)} = \hat{\vec{y}}^{(\rho)} \) yields the error:
\begin{equation}
	\vec{e}_\mathrm{hNDI} = \vec{\nu} - \vec{y}^{(\rho)} = -\mathcal{L}^{-1} \bigl\{ 1 - H_\mathrm{c} (s) \bigr\} * \Bigl( \Delta \vec{\alpha}(\vec{x}) + \Delta \mat{\Beta}(\vec{x}) \vec{\mathfrak{u}} + \vec{\delta}(\cdot)\Bigr) %
\end{equation}
The error of the sensory lower frequency complement is small, right where tracking accuracy of \(\vec{y}^{(\rho)}\)is most relevant. At higher frequencies the error will be larger due to the uncertainty in \(\vec{\alpha} (\vec{x},\vec{d},\vec{p})\). This is also where tracking accuracy is nearly always traded against control activity and stability margins anyway. The cut-off frequency of the complementary filter is thus an important design degree of freedom.

\subsubsection*{Algorithmic Complexity}
Complexity of hybrid NDI is more or less the sum of model-based and sensory NDI, see~\cref{eq:hindi_affine}. Given the fact that complexity of the latter is relatively low, the order of magnitude arguably compares with model-based NDI. 

\subsubsection*{Disturbance Compensation}
The advantage of the hybrid NDI approach lies in the filtering, especially in the low-pass filtering of disturbances and unmodeled dynamics.
Combining equations~\cref{eq:ndi_affine} and~\cref{eq:sNDI_dist} this effect becomes apparent~\cite{Pollack2024}:
\begin{align}
	\vec{\mathfrak{u}} & =  H_\mathrm{c} (s) {\hat{\mat{\Beta}} ( \hat{\vec{x}},\hat{\vec{d}},\hat{\vec{p}} ) }^{-1} \left( \vec{\nu} - \underbrace{\left[ \hat{\vec{\alpha}}(\vec{x},\vec{d},\vec{p}) + \Delta \vec{\alpha}(\vec{x},\vec{d},\vec{p}) + \left( \hat{\mat{\Beta}}(\vec{x},\vec{d},\vec{p}) + \Delta\mat{\Beta}(\vec{x},\vec{d},\vec{p}) \right) \vec{\mathfrak{u}} + \vec{\delta}(\cdot) \right]}_{{\hat{\vec{y}}}^{(\rho)}} + \hat{\mat{\Beta}}_u ( \hat{\vec{x}},\hat{\vec{d}},\hat{\vec{p}} ) \hat{\vec{u}} \right) \notag \\
	& + (1-H_\mathrm{c} (s)) \hat{\mat{\Beta}}^{-1}\left(\hat{\vec{x}},\hat{\vec{d}},\hat{\vec{p}}\right) \left[ \vec{\nu} - \hat{\vec{\alpha}} ( \hat{\vec{x}},\hat{\vec{d}},\hat{\vec{p}} ) \right] \notag \\
	& = \underbrace{\hat{\mat{\Beta}}^{-1} \left(\vec{\nu} - \hat{\vec{\alpha}} ( \hat{\vec{x}},\hat{\vec{p}},\hat{\vec{d}} ) \right)}_{\mathrm{NDI}} + H_\mathrm{c} (s) 
	\underbrace{ {\hat{\mat{\Beta}} ( \hat{\vec{x}},\hat{\vec{d}},\hat{\vec{p}} ) }^{-1} \left( -\left[ \Delta \vec{\alpha}(\vec{x},\vec{d},\vec{p}) + \Delta\mat{\Beta}(\vec{x},\vec{d},\vec{p}) \vec{\mathfrak{u}} + \vec{\delta}(\cdot) \right] \right)}_{\mathrm{additionally\ compensated\ terms}}
	\notag %
\end{align} %
This equation should be interpreted somewhat differently from the error estimate. The terms are those in the physical model that are effectively compensated for by the algorithm in~\cref{eq:hindi_affine}. Compared with~\cref{eq:ndi_affine}, the hybrid formulation is identical, but additionally and intentionally compensates for the uncertain and disturbance terms in a frequency range defined by \(H_\mathrm{c} (s)\). The latter thus directly affects control activity.

\subsubsection*{Reconfiguration during Air Data Sensor Failure}

A handy feature of hybrid NDI is its capability to seamlessly degrade or reconfigure to an inertial-measurements-only control law in case of an air data failure~\cite{Milz2024e} while maintaining similar handling characteristics. If sensors fail, which might especially occur with air data sensors, it is possible to blend out the NDI part and solely rely on the sensory NDI part~\cite{Milz2024e}.
By ``cutting off'' the NDI signal from the complementary filter, the control law seamlessly degrades to a sensory NDI-based control law, which in many flight control designs relies on inertial measurements including the angular velocity measurements and is less dependent on air data sensors. The only remaining dependency is the dynamic pressure \( \bar{q} \). Due to low sensitivity to uncertainty in the control effective matrix (\cref{eq:sNDI_err}), even this signal is allowed to have certain inaccuracies, or may possibly be held at a constant value. Alternatively, the hybrid NDI core may be replaced with a sensory one all together, maintaining performance, but with the disadvantages that originally triggered the hybrid formulation.
Retaining the complementary filter allows an easy seamless switch and graceful degradation. A flight test demonstration documented in \cite{Milz2024e} showed that handling characteristics of the aircraft remain essentially unchanged, allowing the pilot to focus on handling the failure condition itself. This simple-to-implement feature qualifies hybrid NDI as a promising approach for sensor-related fault-tolerant control.

\subsubsection*{Zero Dynamics}
The internal dynamics under hybrid NDI follow directly from the complementary structure of the control law in~\cref{eq:hindi_affine}: substituting the control law into the full system model yields the low-pass filtered terms of the sensory NDI law complemented by the high-pass filtered terms of the model-based NDI law. Consequently, the observations made for NDI and sensory NDI above apply here per frequency range. Within the bandwidth of \( H_\mathrm{c} (s) \), the measured \( \hat{\vec{y}}^{(\rho)} \) feeds back unmodeled dynamics in \( \vec{\mathfrak{x}}_I \), just as in the sensory and incremental variants. Beyond the cut-off frequency, the model-based path dominates, for which the residualization assumptions from~\cref{sec:modelcomplexity} are effectively exercised. The cut-off frequency of the complementary filter is thus also a design degree of freedom with respect to unmodeled internal dynamics: choosing it well below the lowest structural eigenfrequencies avoids feeding dynamic airframe deformation, unavoidably picked up in \( \hat{\vec{y}}^{(\rho)} \), back to the controls (cf. the flight loads example in~\cref{sec:design:structural}). 

\subsection{Procedure for derivation and implementation of NDI inverse model equations}\label{sec:procedureNDI}
\begin{figure}[!htb]
	\centering
	\includegraphics{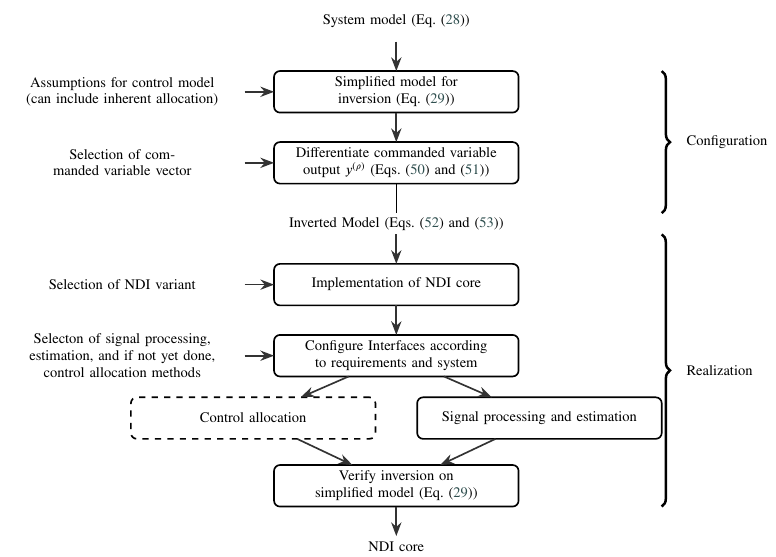}
	\caption{Procedure for NDI core derivation}
	\label{fig:NDI_Process_Methodology}
\end{figure}

The derivations of the four NDI variants follow an identical path that is well applicable as a generic design procedure for the inverse model core of any NDI-based architecture. This procedure is visualized in~\cref{fig:NDI_Process_Methodology}.

Starting point is the system model, which usually originates from a separate process. The first step is to derive a reduced model as described in~\cref{sec:modelsimplification}, which eventually will be the On-Board AirCraft (OBAC) model \cite{Enns2006}. In case of model-based NDI the important design decision here is in that this reduced model decides which effects will eventually be explicitly compensated for, an how accurately. Any neglected effect or simplification reduces control law complexity, but is to be addressed by robustness or disturbance rejection capability in the outer control functions, likely at the cost of overall control law performance. Simplifications may also be used to reduce control activity and flight loads, as well as improved ride comfort. At this point, NDI based on explicit model equations gives the design team valuable design degrees of freedom to find a compromise solution. Finally, also generalized controls are decided upon and implemented accordingly. 

The commanded variables result from the adopted overall control law architecture and need to be formulated as outputs. These in turn are differentiated until an invertible relation with the generalized control inputs is found, see~\cref{sec:differentiatemodel}. The reduced model is then mathematically inverted as detailed in~\cref{sec:invmodelrealization}. At this point the specific NDI variant is to be decided upon. The most suitable variant is selected based on the control law design problem at hand, in relation to the variants' specific strengths and weaknesses. In the derivations, the aspects of inversion error, algorithm complexity, disturbance compensation, and handling of controls faults have been addressed. These have been summarized in \cref{tab:ndi_comparison}.
\begin{table*}[!htb]
	\centering
	\caption{Comparison of NDI Methods}
	\label{tab:ndi_comparison}
	\begin{tabular}{p{2.8cm}p{3.6cm}p{4.2cm}p{3.2cm}}
		& NDI & Incremental/Sensory NDI & Hybrid NDI  \\ \specialrule{1.5pt}{1pt}{1pt}
		Model Dependency & High & Low (control effectiveness) & High \\ \hline
		Inversion Error & Driven by model uncertainty (\( \Delta \vec{\alpha} \), \( \Delta \mat{\Beta} \)) and disturbances \( \vec{\delta} \) & Driven by measurement and synchronization errors of \( \vec{y}^{(\rho)} \) and \( \vec{u} \); insensitive to \( \Delta \vec{\alpha} \) & Frequency-weighted blend of both \\ \hline
		Sensitivity to Sensor Noise/Delay & Lower (state feedback) & High (output derivatives) & Adjustable \\ \hline
		Disturbance Rejection Characteristics & Adjustable (state estimation) & Aggressive & Adjustable via the complementary filter \\ \hline
		Control Activity & Adjustable (state estimation and selection of compensated model terms) & High & Adjustable via the complementary filter \\ \hline
		Implementation Complexity & Scales with model & Low, but signal processing of output derivative measurement may be required & Highest; scales with model \\ \hline
		Failure Handling & Not inherent, but can be facilitated if detected & Inherent, but sensory NDI requires detection & Inherent in sensory complement \\ \bottomrule
	\end{tabular}
\end{table*}
The next step is to configure and design all signals that feed into the NDI core. Design of estimation algorithms and (complementary) filters to ensure good quality signals is key to a successful design and therefore a critical step in the process. The same holds for control allocation. It has to be ensured that the selected generalized controls can be accurately realized by this function, which needs a careful design approach. 

It is good practice to first verify the resulting NDI core on the reduced model from the very first step, preferably also assuming perfect state, commanded variable, and measured controls feedback. The result should be an exact match of \(\vec{\nu}\) as in~\cref{eq:inverror}. This is a good way to ensure correct implementation of the inverse model, before proceeding with detailed design of the outer and other functions in the system.

\section{A Design Perspective}\label{sec:design}

The raison d'être of Nonlinear Dynamic Inversion is to facilitate the design of control algorithms for dynamically complex systems with coupled command response behavior. In the previous section NDI has been reviewed from a methodological point of view, involving the derivation and implementation of an inverse model core that facilitates tracking of commanded variables \(\vec{y}\) by compensating for nonlinear system behavior and kinematics, varying operating conditions, coupled command responses, and, if desired, direct compensation of disturbances upfront. However, requirement specifications, especially for flight control laws include many more aspects than can be handled by a design methodology single-handedly. This section therefore starts from an overall design point of view instead and qualitatively discusses how the challenges as listed at the beginning of \cref{sec:introduction} can be addressed. The intention is to explore how NDI can be effectively used in architecture development and as a supporting methodology in the overall design of flight control law functions. Useful references in this respect are \cite{Ito2002,Looye2008,Honeywell1996,Robustcontrol1997}. Design will hereby be addressed from three points of view:
\begin{enumerate}
	\setlength{\itemsep}{0pt}
	\item {\em Flow} (or Process): how NDI influences the overall sequence of steps and interrelations involved in realizing the design, as well as how the methodology may reduce (or shift) effort in fulfilling design requirements;
	\item {\em Form} (or Implementation): how NDI integrates into a control law architecture that provides and arranges required functions in an effective way; 
	\item {\em Function}: how NDI facilitates and integrates into the detailed design of the actual control law functions.
\end{enumerate}

\subsection{A Systems Engineering Perspective ({Flow})}
\begin{figure*}[!htb]
    \centering
	\resizebox{12cm}{!}{%
		\includegraphics{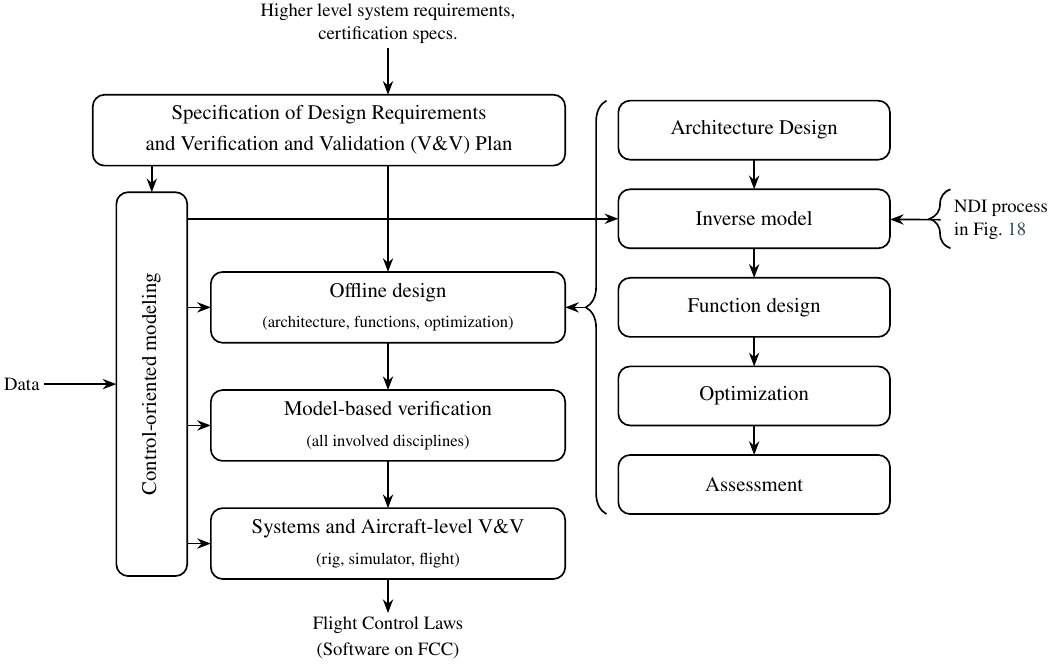}}
	\caption{Simplified Design Procedure (design loops have been omitted for clarity)}
	\label{fig:proc}
\end{figure*}

It is useful to first outline the flight control law design process in most general terms (see~\cref{fig:proc}, left). More detailed versions can be found in~\cite{Irving1995,Fielding2000,Looye2008}, which are in turn based on \emph{SAE ARP4754A  - Guidelines for Development of Civil Aircraft and Systems}~\cite{SAE_ARP4754A_2010}. Reference \cite{Angelov2022} at this point takes an MBSE (Model-Based Systems Engineering) approach, working out a full design example from this perspective.

The primary stages of the process are as follows (cf.~\cref{fig:proc}):
\begin{itemize}
	\item{\em Specification of Design Requirements and Verification and Validation (V\&V) Plan.} The main thread is built by the design requirements and their eventual verification and validation. 
	The specification of requirements in turn traces back to higher-level system requirements, Concepts of Operation (ConOps) and certification specifications. Each design requirement is to be verified computationally and evaluated through testing. To this end, verification metrics are defined, along with analysis methods and acceptable assumptions for their computation as part of the V\&V. Design and verification of flight control laws, in turn, pose requirements on the modeling process.%
	
	\item{\em Control-oriented Modeling. (cf.~\cref{sec:method:mod})} Effective analysis requires models that encompass all relevant physical and algorithmic aspects while supporting the computation of necessary metrics \cite{Looye2008,Kier2009}. Requirements derived from certification specifications typically include proposed analyses and metrics as Acceptable Means of Compliance. Aircraft model development is generally conducted in collaboration with or within the control design team, utilizing structures and data from disciplines such as flight mechanics, aerodynamics, propulsion, systems, and aeroelasticity. Model development also includes parameterization, trimming, linearization, and uncertainty modeling. A key requirement is loop-capability. Due to the large number of flight and operating conditions to perform the verification analyses at, an appropriate balance between accuracy and run-time efficiency is to be found. The models are updated throughout the process to reflect the current aircraft design status or to incorporate results from flight testing. It is important to note that there may be a multitude of models, depending on the type of analysis. The models feed all design stages, including the derivation of the NDI inverse model core that will be discussed subsequently.
	
	\item{\em Offline Design.} The core control design is typically performed offline using the previously developed models. The specific details of this process will be addressed subsequently.
	
	\item{\em Model-based Verification.} Following the design phase, model-based verification is conducted to ensure that all requirements are satisfied. Part of the verification coincides with verification analyses in other (flight physical) disciplines, like flight loads, aeroservoelasticity, and stability \& control. %
	
	\item{\em Systems and Aircraft level V\&V.} Alongside model-based verification, the flight control laws subsequently integrate into Flight Control Computer (FCC) software, usually via automatic code generation tools. They in turn enter testing at software,  component (FCC), Flight Control System (FCS) rig, and eventually aircraft-level (simulator, on ground, in flight) as part of overall aircraft development. As far as the verification and validation of the flight control laws is concerned, these stages extend the aforementioned model-based verification as part of the aforementioned V\&V plan, and design loops may feedback to all previous development phases. %
\end{itemize}

The {\em offline design} phase is detailed to the right in~\cref{fig:proc}. Typical stages include architecture design, detailed function design, overall design optimization, and off-line assessment that may feedback to preceding stages in the form of design iterations. As already discussed from a conceptual perspective in~\cref{sec:conceptual}, it is in the architecture and functional design where the features of NDI may be of great value from the aforementioned {\em flow}, {\em form}, and {\em function} points of view. 
In the off-line process, NDI becomes apparent in the derivation of the {\em inverse model} as a connecting step between {\em architecture} and {\em function} development. This is where the methodological aspect and the underlying process discussed in~\cref{sec:method} (\cref{fig:NDI_Process_Methodology}) integrate into~\cref{fig:proc}. The influence on {\em design optimization} and {\em assessment} is rather influenced by the control design methods that NDI integrates with in {\em function design}. 
\vskip 2ex

NDI typically assigns key functions to individual components of the architecture, facilitated by the decoupling between commanded variables in its inner core. This functional breakdown enables straightforward traceability and linkage between requirements and design degrees of freedom, supporting seamless integration into systems engineering workflows.
Key decisions at this point include the functional breakdown, selection of the most suitable commanded variables for the NDI core, and the choice between an integrated or multi-loop design, which addresses fast dynamics in the inner core and slower dynamics in the outer loops~\cite{Snell1992,Bugajski1992,Adams1994,Lane1988}. 
Once the architecture and NDI core are defined, the next stage is to develop individual functions. These can be directly integrated into the architecture and configured for analyses of design metrics.
Design optimization is usually best performed on the integrated system to address coordination and interactions. This may require a structured procedure, especially for multi-loop designs~\cite{Looye2006}. Detailed design of functions and overall optimization will be addressed subsequently. Design Assessment is performed after optimization to identify design weaknesses across the operating envelope that need to be addressed in the selection of design parameters or, possibly, the control law architecture.

\subsection{Architecture Design (Form)}\label{sec:Design:arch}

The main functionalities to be provided by the flight control laws are essentially derived from Concepts of Operation (ConOps) related to the flight control and/or auto flight system. Different missions, or operational scenarios (nominal, off-nominal) result in a requirement for multiple sets of concurrently operating functionalities~\cite{Crawley2015}. At this point, there are two implications from an NDI design point of view. First of all, the functions are usually tasked with providing tracking of suitable commanded variables and thus constitute the entries of the virtual controls input \(\vec{\nu}\) to the NDI core (cf. \cref{sec:invmodelrealization},~\cref{fig:NDItypicalArchitecture}). Different allowed combinations may thus give rise to different variants of \(\vec{\nu}\) and different inverse model realizations. Detailed examples can be found in~\cite{Lane1988,Azam1994}. ConOps will also reveal combinations that should not occur, like combining autopilot and manual functions \cite{Branch1998}.

A second point of attention is behavior of the so-called zero dynamics. Although these are not unique to NDI, the methodology hides these better than any other method due to its intentional compensation through the \(\alpha(x,d,p)\) term (\cref{fig:NDIinverseModel}). Resulting unwanted or even unstable dynamics are usually addressed by means of supporting functions, or by adapting the commanded variables. A good example is augmented manual control of the vertical load factor for a fixed-wing aircraft, with the intention to maintain flight path angle once the cockpit control inceptor is released. This unavoidably results in speed instability. One solution is to require simultaneous speed control via an auto throttle and to provide active protection against stall (i.e. additional functions) \cite{Favre1994}, or to compromise pitch control with stabilizing feedback of airspeed (i.e. adapting the commanded variable (``C\textsuperscript{*}-U'', see \cite{Niedermeier2012}). 

\subsubsection{Time Scale Separation}
Complexity of the control law architecture with multiple sets of commanded variables can be considerably reduced by making use of the general physical principles based on which the aircraft is controlled. Flight control laws in fixed-wing aircraft typically make use if a division between so-called inner and outer loop functions that exploit the separation in time scales between rotational and translational dynamics. The same principle is very frequently used in NDI-based designs. Meyer and Cicolani~\cite{Meyer1981} already formulate assumptions that allow the aircraft model to be brought into block triangular form. Its inversion then naturally leads to an inner and outer loop structure. Elgersma formalized the principle in his thesis and introduced the concept of \textit{Partial Dynamic Inversion}~\cite{Elgersma1988}. Mulgund and Stengel apply this to escape maneuvers from wind shear events~\cite{Mulgund1995}, designating it \textit{Sequential Dynamic Inversion}.

\begin{figure}[!htb]
	\centering %
	\includegraphics{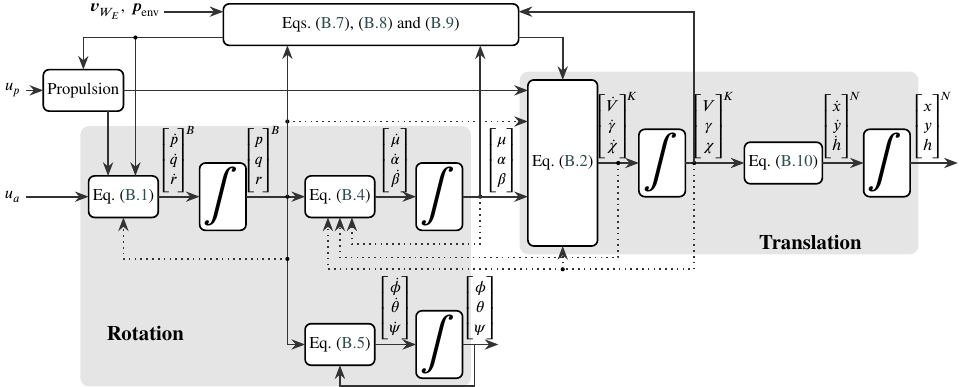}
	\caption{Aircraft flight dynamics model in graphical form} %
	\label{fig:EQM_NDI}
\end{figure}

\cref{fig:EQM_NDI} graphically structures the dynamics of a fixed wing aircraft in a similar way as presented in \cite{Meyer1981}. Starting from the right, control of the aircraft flight path, velocity, and accelerations is done by adjusting aerodynamic, propulsion, and possibly other forces (e.g. landing gears). Propulsion forces are delivered by the engines (top left). At a given airspeed \(V_a\), aerodynamic ones are controlled by changing the airframe attitude relative to the airflow (\(\mu_a, \alpha_a, \beta_a\)), and by means of aerodynamic control devices like moving surfaces or spoilers, collected in \(\vec{u_a}\). For these devices a distinction has to be made between those that intendedly generate forces, like in the case of Direct Lift Control (DLC) \cite{Varriale2025}, and those that do so unintendedly, in order to generate moments around the aircraft body axes. The aircraft attitude in turn, is adjusted by means of aerodynamic, propulsive and other moments, shown on the lower left in \cref{fig:EQM_NDI}. 

Time scale separation assumes that the aircraft attitude can be adjusted sufficiently faster than velocity and direction of the flight path. This allows separate control functions for the attitude, typically by means of aerodynamic control devices, and functions for direction and speed by means of propulsion, and the realized attitude~\cite{Kato1986,Mulgund1993,Elgersma1988,Menon1991,Lombaerts2020,Kawaguchi2011}. It is hereby assumed that the attitude commands can be realized almost instantly. Even attitude control can be divided into functions that control angular rates \(\vec{\omega_B}\), and attitude angles via those. Attitude angles may be the aforementioned (\(\mu_a, \alpha_a, \beta_a\)), or alternatively Euler angles (\(\Theta = [\phi, \theta, \psi]^T\)), or any sensible combination of those. It must be added though, that the bandwidth of the attitude angles control loops must be reduced in order to be sufficiently slower than the angular ones. This performance limitation may become problematic in the case of fast maneuvers~\cite{Lombaerts2011c}.

The advantage of exploiting time scale separation is in two aspects:
\begin{itemize}
    \item Inner functions can be used with various outer ones, simplifying switching and reducing code overlaps between outer functions. Differentiation (\cref{sec:method}) of different commanded variable candidates will usually go through the same moment equations to establish the algebraic link with aerodynamic devices. As a result, duplicate model code will be present between inverse models for interchangeable commanded variables. Common functions for attitude rate, or even attitude angles will avoid this and reduce control law complexity~\cite{Looye2006,Weiser2024}.
    \item The relative degrees of the outer loop functions is reduced. Differentiation for inversion of flight path related commanded variables is halted once attitude angles appear in the equation. Further differentiation adds terms due to the chain rule, increasing complexity of the inverse model. In the case of incremental or sensory NDI formulations, the lower relative degree reduces signal processing effort in estimating and filtering commanded variable derivatives.
\end{itemize}
Obviously, the artificial reduction of relative degree introduces inaccuracy that unavoidably limit achievable bandwidth and performance. Although conceptualized for high performance aircraft \cite{Kato1986,Elgersma1988}, the principle is therefore better suited for transport aircraft.

Apart from inherent reduction in achievable performance, it has also to be considered that the choice of commanded attitude variables (aerodynamic or inertial) may have to be compromised to serve different types of outer loop functions. This is for example the case in the automatic landing system presented in~\cite{Looye2006}. The same reference proposes a procedure for tuning design parameters in the attitude loops to work well both with approach, and with landing functions.

\subsubsection{NDI Function Architecture}\label{sec:design:generalarch}

\begin{figure*}[!htb]
	\centering %
	\includegraphics[]{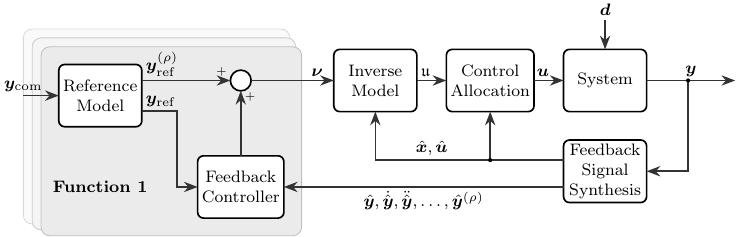}
	\caption{Generalized NDI control law architecture}
	\label{fig:arch}
\end{figure*}

So far, the detailed design of the outer functions has received only limited attention. The reason is that their realization may be done using any control design method of preference. However, there are three standard solutions that are tailored to, and frequently applied in combination with NDI: 
\begin{itemize}
    \item \textit{Pole placement.} This control law looks as follows:
    \begin{align}
        \nu_i = k_{i_0} \left(y_{c_i} - \hat{y}_{i}\right) - k_{i_1} \dot{\hat{y}}_{i} -\cdots- k_{i_{\rho_i-1}} \hat{y}_{i}^{(\rho_i-1)}
    \end{align}
    In the ideal case, where \(y_i^{(\rho_i)} = \nu_i,\ \hat{y}_i = y_i\), this results in the following transfer function:
    \begin{align}
        \frac{y_i(s)}{y_{c_i}(s)} &= \frac{k_{i_0}}{s^{\rho_i} + k_{i_{\rho_i-1}}s^{\rho_i-1} + \cdots + k_{i_0}} 
    \end{align}
    The poles can be placed at preferred locations, which is often used to obtain desired dynamic command response behavior, i.e. for command shaping. This structure was mostly used in early applications \cite{Lane1988,Bugajski1992}, but tend to lack robustness to non-exact inversion due to simplifications or model uncertainties. A fix for steady state tracking errors is the addition of integral feedback: 
    \item \textit{Pole placement with integral control.} This control law adds an integral term:
    \begin{align}
        \nu_i = k_{i_I} \int \left(y_{c_i} - \hat{y}_{i}\right) dt +  k_{i_f} y_{c_i} - k_{i_0} \hat{y}_{i} - k_{i_1} \dot{\hat{y}}_{i} -\cdots- k_{i_{\rho_i-1}} \hat{y}_{i}^{(\rho_i-1)}
    \end{align}
    Resulting in the following transfer function:
    \begin{align}
        \frac{y_i(s)}{y_{c_i}(s)} &= \frac{k_{i_f}s + k_{i_I}}
        {s^{\rho_i+1} + {k}_{i_{\rho_i-1}}s^{\rho_i} + \cdots + k_{i_{0}}s + {k}_{i_I}} 
    \end{align}
    The feed-forward gain \(k_{i_f}\) is usually picked such, that the additional real pole \(p_{i_I}\) that comes with integral feedback, is canceled exactly \cite{Osterhuber2004,Ito2002}:
    \begin{align}
        \frac{y_i(s)}{y_{c_i}(s)} &= \frac{k_{i_f} \left(s + \frac{k_{i_I}}{k_{i_f}} \right)}
        {(s+p_{i_I})(s^{\rho_i} + \hat{k}_{i_{\rho_i-1}}s^{\rho_i-1} + \cdots + \hat{k}_{i_0})} \label{eq:1DOF_i}
    \end{align}
    For \(\rho_i=1\) analytical formulations can be derived \cite{Ito2002}.
    \item \textit{Two-Degrees Of Freedom (2DOF) design.} In this case a reference model is defined that provides smooth references for the commanded variable and its derivatives at least up to \(y_i^{(\rho_i)}\), see \cref{fig:arch}. The feedback controller adds to the commanded variable derivative as follows:
    \begin{align}
        \nu_i = y_{i_\mathrm{ref}}^{(\rho_i)} +
        k_{i_I}\int(y_{i_\mathrm{ref}}-\hat{y}_{i})dt +
        k_{i_0}(y_{i_\mathrm{ref}}-\hat{y}_{i}) + k_{i_1} (\dot{y}_{i_\mathrm{ref}}-\dot{\hat{y}}_{i}) +\cdots+ k_{i_{\rho_i-1}}(y_{i_\mathrm{ref}}^{(\rho_i-1)}- \hat{y}_{i}^{(\rho_i-1)}) \label{eq:2DOF}
    \end{align}
    In the ideal case, \(y_{i}\) will match \(y_{i_\mathrm{ref}}\) and this the dynamic behavior of the reference model exactly. 
\end{itemize}
The 2DOF design offers separate degrees of freedom command shaping (via the reference model), and disturbance rejection (feedback controller) \cite{Kreisselmeier1999}. In addition, it allows for effective handling of control saturation by means of Pseudo Control Hedging (PCH)~\cite{Johnson2000,Holzapfel2004,Lombaerts2010,Lombaerts2012}, effectively choking the output of the Reference Model to stay within physical control limits. Industry applications tend to prefer the second structure, see for example \cite{Osterhuber2004}. 

The system outputs and derivatives \(\hat{y}_{i}, \dot{\hat{y}}_{i},\cdots,\hat{y}_{i}^{(\rho_i-1)}\) may be obtained directly, or computed and/or estimated from available sensor outputs. Analytical expressions for output derivatives come with the differentiation process as described in \cref{sec:method} and may be of use at this point \cite{Lin1994}. In the case of aircraft, it is important to apply appropriated filtering in order to sufficiently reduce structural dynamics and noise content in the synthesized signals. 

The gains in the feedback control laws may be synthesized to best compromise command following performance, disturbance rejection, and robustness. In the architecture in \cref{fig:arch} this is done for each function separately. An important degree of freedom that is lacking in this setting is compensation of couplings between commanded variable responses that are inaccurately or not compensated for by the inverse model core. This may be addressed by cross-feedback loops, or by combined design of the functions by means of multivariable controller synthesis methods~\cite{Adams1993}. The disadvantage of course is that functions become inherently interdependent. 

From an architectural perspective, it is important to co-ordinate placement of integral feedback loops between commanded variables that are simultaneously active. A practical approach is to interpret integrators as trimming devices that are capable to hold control surface positions in steady state, without the need for sustained tracking errors (i.e. \(u_i = K_i (y_{i_{\mathrm{ref}}}-\hat{y}_i)\)). Integrators should therefore be placed where static offsets may arise. As an example, asymmetry of the ATTAS aircraft caused the NDI design without integral feedback described in \cite{Looye2001} to fly with a fixed offset from the localizer reference . Even though noticeable in the outer most loop, the problem was fixed by integral control of roll and yaw rates to generate rudder and aileron commands that correctly trim for the asymmetry.

An interesting situation arises when \(\rho_i = 0\) and the relation between commanded variable and control input is a matter of (variable) scaling. This is for example the case when controlling vertical or lateral acceleration via aircraft attitude. In this case, direct command scaling can be combined with integral feedback of the acceleration error, see for example \cite{Lombaerts2013,Keijzer2019}. The integrator will then compensate for any scaling mismatches and trim offsets.

\subsubsection{Selection of commanded Variables}\label{sec:method:design:cv}

As mention at the beginning of this section, commanded variables (CVs) of the main control law functions often result from the requirement specification as derived from ConOps. Some design freedom may be left, allowing selection for example to meet handling quality related ones~\cite{Enns1994}. Also system characteristics may drive the selection of commanded variables, for example, in the case of non-minimum phase systems~\cite{Hauser1992,Hauser1992vstol}.

In case of a cascaded multi-loop structure (i.e. partial dynamic inversion), it has already been noted that the selection of internal CVs may have important architectural implications. It is difficult to derive formal guidelines at this point, relying on physical considerations. Instead of continued differentiation of outer function CVs, also static considerations may be of use. A good example can be found in \cite{Lambregts2013a}. 

The force equation in direction of the velocity vector is given in~\cref{eq:forceeq} (\cref{sec:eqm}) and may be written as \cite{Meyer1981}:
\begin{align}
    m \dot{V} &\approx T - \bar{q}S C_D - mg \sin\gamma
\end{align}
where $T$ is the thrust force. During operation, sustained acceleration and flight path angle will require thrust as follows:
\begin{align}
    T  &\approx m \dot{V} + mg \sin\gamma + \bar{q}S C_D, \ \ \text{or:} \\
    \frac{T-D}{W} &\approx \frac{\dot{V}}{g} + \sin\gamma
\end{align}
where the left-hand term is known as the specific excess power, \(D\) is total drag, and \(W = mg\) the aircraft weight. Although the derivation in~\cite{Lambregts2013a} starts from energy principles, the author eventually chooses \(\frac{\dot{V}}{g} + \sin\gamma\) as commanded variable. As the relative degree obviously equals zero, the feedback control loop initially is integral feedback only (with \(\sin\gamma \approx \gamma\)):
\begin{align}
    T_{\mathrm{com}}  &= W (\frac{\dot{V}_c}{g} + \sin\gamma_c) + D \ \ \text{with:} \\
    \frac{\dot{V}_c}{g} + \sin\gamma_c &= k_I \int \left(\frac{\dot{V}_c-\dot{V}}{g} + \gamma_c-\gamma\right) \mathrm{d}t
\end{align}
This is effectively the thrust path of the Total Energy Control System (TECS), using integral control to obtain desired dynamic command response behavior. In its practical implementation, proportional feedback is applied to address the engine dynamics neglected above. TECS does not estimate drag, but initializes the integral at the current thrust setting. It further assumes that \(D\) varies only slowly, allowing integral feedback to keep pace. The system uses an inverse thrust map to actually compute required throttle commands to obtain  \(T_{\mathrm{com}}\) \cite{Looye2008}.

The elegance is in that integral feedback directly helps to trim engine thrust setting. When applying the standard differentiation approach as detailed in \cref{sec:method}, control on \(\gamma\) would pass through pitch attitude, and elevator. Trimming as provided by integral control would not immediately command thrust.

In the selection of commanded variables and subsequent differentiation for inversion (\cref{sec:method}), air mass referenced ones require special attention. The reason is that the equations of motion are usually represented in body axes and inertially referenced. At the same time, atmospheric disturbances contribute considerably to relative rotation of aerodynamic or stability axes. Tracking of such CVs is only desired up to a given frequency. 

A common example is control of the attitude relative to the airspeed vector, as this enables the so-called velocity vector roll. The kinematic equation is given by~\cref{eq:muab}. Even though the angles \(\mu,\ \alpha,\ \beta\) are based on the aerodynamic definitions (the inertial ones are quite different, see for example \cite{Brockhaus2011}), it must be kept in mind that in the equation these are relative to the inertial velocity vector. For NDI, the equation this must be expressed in its aerodynamic form:
{\small
\begin{equation} \label{eq:aeroattitude2}
\begin{bmatrix}
	\dot{\mu}_a \cos\beta_a \\ \dot{\alpha}_a \cos\beta_a \\ \dot{\beta}_a
\end{bmatrix}
=
\begin{bmatrix}
	{\cos\alpha_a} & 0 & {\sin\alpha_a} \\
	-\cos\alpha_a\sin\beta_a & \cos\beta_a & -\sin\alpha_a\sin\beta_a \\
	\sin\alpha_a & 0 & -\cos\alpha_a
\end{bmatrix}
\vec{\omega}_{a_B}
+
\begin{bmatrix}
	\cos\mu_a\sin\beta_a \\
	-\cos\mu_a \\
	-\sin\mu_a
\end{bmatrix}\,\dot{\gamma}_a
\;+\;
\begin{bmatrix}
	\sin\gamma_a \cos\beta_a + \sin\mu_a\cos\gamma_a\sin\beta_a \\
	-\sin\mu_a\cos\gamma_a \\
	\cos\mu_a\cos\gamma_a
\end{bmatrix}
\dot{\chi}_a 
\end{equation}
}
Using partial dynamic inversion here for clarity, this equation is normally inverted towards \(\vec{\omega}_{a_B} \approx \vec{\omega}_{B}\), as the wind contribution to the latter will be difficult to determine \cite{Bugajski1990,Snell1992}. Furthermore, it is typically assumed that \(\dot{\gamma}_a \approx \dot{\gamma}\) and \(\dot{\chi}_a \approx \dot{\chi}\), thus leaving the effects of wind gradients to the outer function feedback control laws. In addition, the angles \(\mu_a\) (necessarily estimated), \(\alpha_a\) and \(\beta_a\) are filtered complementarily, as fast disturbances cannot possibly be compensated for (to be discussed in \cref{sec:design:dist}). The resulting inverse model control law then becomes:
{\small
\begin{equation} \label{eq:aeroattitudeinv}
\vec{\omega}_{B_c}
=
\begin{bmatrix}
	{\cos\hat{\alpha}_a}/\cos\hat{\beta}_a & 0 & {\sin\hat{\alpha}_a}/\cos\hat{\beta}_a \\
	-\cos\hat{\alpha}_a\tan\hat{\beta}_a & 1 & -\sin\hat{\alpha}_a\tan\hat{\beta}_a \\
	\sin\hat{\alpha}_a & 0 & -\cos\hat{\alpha}_a
\end{bmatrix}^{-1}
\left\{
\begin{bmatrix}
	\nu_{\mu_a} \\ \nu_{\alpha_a}  \\ \nu_{\beta_a} 
\end{bmatrix}
-
\begin{bmatrix}
	\cos\hat{\mu}_a\tan\hat{\beta}_a \\
	-\cos\hat{\mu}_a/\cos\hat{\beta}_a \\
	-\sin\hat{\mu}_a
\end{bmatrix}\,\dot{\gamma}
\;-\;
\begin{bmatrix}
	\sin\hat{\gamma}_a  + \sin\hat{\mu}_a\cos\hat{\gamma}_a\tan\hat{\beta}_a \\
	-\sin\hat{\mu}_a\cos\hat{\gamma}_a/\cos\hat{\beta}_a \\
	\cos\hat{\mu}_a\cos\hat{\gamma}_a
\end{bmatrix}
\dot{\chi} 
\right\}
\end{equation}
}
where \(\hat{..}\) implies estimated values and \(\nu_{\mu_a},\ \nu_{\alpha_a},\ \nu_{\beta_a}\) are the virtual commands for aerodynamic roll, attitude, and side slip rates. In some applications, the \(\dot{\gamma}\) and \(\dot{\chi}\) terms are neglected \cite{Acquatella2020,DaCosta2003}.

In summary, commanded variable selection for the outer functions are likely to be dictated by the functional requirement specification for the control system. For internal CVs there is more freedom of design, where interoperability between different sets of outer functions, physical considerations, non-minimum phase behavior, as well as steady-state considerations may be taken into account to obtain an effective and efficient architecture. Air-mass referenced variables need special care, as the effects of atmospheric disturbances need to be incorporated by means of appropriately filtered inertial and airdata measurements.

\subsection{Addressing Design Requirements in \textit{Function} design}\label{sec:design:reqs} 

The third aspect listed at the beginning of this section --- addressing design requirements with NDI --- can be well addressed by looking at the subsequent design steps in \cref{fig:proc} from architecture to design optimization. Extensive design guidelines on addressing flight control requirements with multivariable control laws, including NDI-based ones, are provided in~\cite{Honeywell1996}.
The intention hereby is not to discuss requirements in detail, but rather how requirements of different nature can be effectively addressed with available design degrees of freedom. 
Without claiming completeness, a substantial set of requirements will be addressed in the following.

\subsubsection{Handling Qualities and Tracking Requirements}

Meeting handling quality requirements starts with the selection of best-suitable commanded variables~\cite{Enns1994}. The choice of the latter may further be influenced by the type of control inceptors.
Pilot modeling may hereby support the evaluation of handling quality metrics in the design loop~\cite{Colgren1997}. Most metrics relate to the dynamic command response behavior of the CV. Parameter selection for shaping of the latter is relatively straight forward using the approaches sketched in~\cref{sec:design:generalarch}, as flight condition dependency, nonlinearities and couplings are accounted for by the inverse model core. In more complex cases, alternative synthesis methods may be applied, like multi-objective optimization. Tools for this purpose, like CONDUIT~\cite{Tischler1997} and MOPS~\cite{Joos2002}, come with ready-to-use criterion libraries that implement various handling quality metrics.

\subsubsection{Envelope Protection}
NDI comes with various features that help to ensure flight within safe envelope bounds. First of all, commanded variables have to be limited to safe value ranges. The use of a reference model (\cref{sec:design:generalarch}) allows this to be done dynamically in a straight-forward way. The assumption hereby is that tracking performance is sufficient in order to prevent violations due to external disturbances. In case the envelope is defined by variables that are not directly commanded, additional measures need to be taken. Some applications like described in \cite{Lombaerts2017,Lane1988} map envelope bounds onto pitch and roll command limitations. Alternatively, the architecture may be augmented to switch to commanded variables that match the envelope definition. This is particularly helpful in case severe disturbances cause a violation of the bounds. Stengel shows this for an angle of attack limiter in \cite{Lane1988}. This approach is commonly taken in transport aircraft fly-by-wire systems ~\cite{Favre1994}.

An alternative approach that is currently gaining attraction uses \emph{control barrier functions} (CBFs), which provide mathematical guarantees of forward invariance of the safe flight envelope~\cite{Ames2019,Autenrieb2025quadratic}. CBFs act at the controller output using estimated or measured states to determine compliance with a function that defines the safe set. This formulation is structurally close to the main NDI inversion law, as CBFs likewise build on Lie derivatives of a model function along the plant dynamics: the function defining the safe set is differentiated until the control input appears, so that the forward-invariance condition becomes affine in the control and can be enforced by a quadratic program. Thus, the methodology offers an alternative to traditional reference filtering methods.

Nominal envelope bounds and performance limitations come with the aircraft operating manual (AOM). In the case of system failures or airframe damage, these bounds shrink to an unknown level, unless anticipated in the AOM. Over the last decade, quite some research has been performed on off-line and in-flight determination of bounds, see for example \cite{Tang2009,Lombaerts2015,Yin2019}. 

\subsubsection{Addressing and Rejection of Disturbances}\label{sec:design:dist}

One characteristic aspect of flight control law design is in handling of disturbances. The modeling of disturbances by means of an unknown signal addition to the control inputs, which is common in textbooks on control, is insufficient by far. For aircraft, aerodynamic forces and moments are determined by the airframe motion relative to the surrounding airmass, mainly the attitude and velocity vector. The airmass may in turn be highly dynamic by itself and move, for example, at constant speed in the horizontal plane (steady wind), in the vertical plane (thermals), as well as change speed (wind shear), and show fast fluctuations (gusts and turbulence). Steady (tail) wind or thermals are often not even classified as disturbances. Flight control systems must continuously balance two competing objectives: maintaining safe velocity and attitude relative to the surrounding air (i.e. riding along) for performance and controllability reasons, and fulfilling inertial constraints. Examples of the latter may vary from arriving at the intended geographical destination, to precise tracking of a landing trajectory, or meeting with a moving target. From an NDI perspective, the first objective poses the challenge of appropriately accounting for atmospheric effects in model computations, the second objective involves actively compensating for those.   

In ~\cref{sec:method:mod} and \cref{sec:invmodelrealization} disturbances (collected in \(\vec{d}(x,t)\)) have been explicitly retained in the derivations. The vector \(\vec{d}(x,t)\) actually appears as such in aircraft flight dynamics models (in quite a nonlinear way, see Appendix~\ref{sec:eqm}), as the model states \(\vec{x}\) are usually inertially referenced (see Appendix~\cref{sec:eqm}, as well as \cite{Linden1998} for an exception). For the computation of aerodynamic and propulsion induced forces and moments, the states and disturbances are combined to compute the aforementioned airmass referenced variables of relevance. A great thing in flight controls is that the most relevant of these variables are sensed by the airdata system. As already discussed in \cref{sec:conceptual}, a characteristic aspect of NDI is that the model states in the inverse equations are obtained from measurement, rather than from integration of internal state equations. At this point, the combined availability of airdata and inertially referenced systems is of great help and usually completely avoids the need for estimating \(\vec{d}(x,t)\). An example of exploiting this for the \textit{riding along} to maintain safe flight and controllability has been discussed in the derivation of airmass referenced commanded variables (\cref{eq:aeroattitude2}). 

At the same time, it is usually not desirable to compensate for each atmospheric fluctuation, as fast ones simply cannot be reduced due to aircraft inertia and actuator and engine response times. Such fluctuations rather tend to cause unnecessary control activity, airframe loads, and ride discomfort. 
A balance is typically achieved by integrating both airmass-referenced sensors (airdata) and inertial-referenced sensors (Inertial Reference System, IRS) using complementary filtering techniques. In this approach, the airdata signal undergoes low-pass filtering, while the IRS provides complementary high-frequency information~\cite{Lambregts2013a,Looye2006}. As an example, the airspeed feedback signal for use in the model may be composed using first order filtering as follows:
\begin{align}
    \hat{V} &= \frac{1}{\tau s + 1} V_{\mathrm{tas}} + \frac{\tau s}{\tau s + 1} V = \frac{1}{\tau s + 1} (V - V_{\mathrm{wind}}) + \frac{\tau s}{\tau s + 1} V = V - \frac{1}{\tau s + 1} V_{\mathrm{wind}}
\end{align}
where \(V_{\mathrm{tas}} = V - V_{\mathrm{wind}}\). The complementary filtered speed thus retains its airmass reference at lower frequencies.
The time constant \(\tau\) in turn should be carefully picked in order to make sure that performance degrading wind shears remain compensated for ~\cite{Lambregts1983,Looye2006}.

In the case of the sensory or incremental formulation, drift dynamics are computed based on (angular) acceleration measurement. The latter unavoidably contains the effect of any atmospheric disturbance, independent of its frequency. This in principle leads to excellent disturbance rejection performance ~\cite{Smeur2016,Rota2026}. This is useful for smaller and agile unmanned aircraft. For passenger aircraft always a trade-off between tracking accuracy and aspects like control activity has to be made. At this point an important design degree of freedom is lost~\cite{Kier2020}.

In addition to disturbances addressed in the NDI inverse model core, the feedback controller has disturbance rejection as its main task. For the two-degrees-of-freedom design shown in~\cref{sec:design:generalarch}, command shaping and disturbance rejection can be addressed relatively independently. For the design of this function, any control method may be used. With the inverse model largely handling nonlinearities and decoupling, classical linear methods are usually employed. If there is a strong necessity to maximize performance and handle large uncertainties, robust control methods can be applied~\cite{Balas1992,Adams1993,Looye1998,Reigelsperger1998,Wang2005,Goman1998}.

It is interesting to note that NDI-based design allows for balancing rejection of disturbances via inverse-model computation and resolution via feedback. The first is effectively very fast, comparable to feed-forward control, and achievable through a balanced use of air data and inertial measurements. Even explicitly estimated disturbances can be compensated for in the inverse model (\cref{sec:method}) . 
The feedback controller is an obvious necessity for disturbance rejection and providing robustness with respect to uncertainties, errors, or simplifying assumptions in the inverse model. Here, a potential trade-off arises: increased model fidelity may increase control law complexity, whereas each simplifying assumption may offload the compensation for deviations from actual system dynamics to the feedback controller, unavoidably trading tracking accuracy for robustness~\cite{Looye1998,Adams1993}.

In classical aircraft configurations, rejection of atmospheric disturbances can be considered a somewhat unfair match. These tend to affect aerodynamic forces and moments equally in a direct way, as can be seen immediately from the aerodynamic equations. Control surfaces at the aircraft empennage or at the front (e.g., canards) are primarily intended to generate moments, allowing the control and trim the aircraft's angles of attack and side slip with relatively low control power and small control surfaces. Forces tend to be small, even causing unwanted non-minimum phase effects in the case of controlling the vertical acceleration (see the discussion on the physical relative degree of controls in~\cref{sec:modelcomplexity}).
Compensation of disturbance-induced aerodynamic forces, therefore, comes with an inherent delay, requiring the buildup of aerodynamic angles via moment dynamics first. This inherent time scale separation is often exploited to reduce the complexity of the inverse model (cf.~\cref{sec:method}). Although studied in the 1970s and implemented on the Lockheed L-1011 TriStar, direct means of controlling lift (Direct Lift Control, DLC) are currently receiving renewed attention with evolving aircraft and wing control configurations~\cite{Varriale2025}.
From an NDI point of view, this may be an effective means of improving disturbance rejection and command-tracking accuracy by adding additional commanded variables to the inverse model, which, in turn, if available, commands DLC surfaces in a coordinated and decoupled way~\cite{Lombaerts2013,Denham2016,DiasMartins2026}. 
DLC tends to be more comfortable for passengers, as short-term atmospheric disturbances can be reduced by avoiding fast changes of attitude.

\subsubsection{Control Activity}\label{sec:design:reqs:ctrlact}

Control activity encompasses the magnitude, frequency, and bandwidth utilization of control commands generated by the (NDI) controller. It is a major concern for transport aircraft, as it directly affects wear, the energy consumption of the actuation system and the engines, and overall system robustness, especially during operations at off-nominal operating points and frequencies and during disturbance attenuation. The trade-off against disturbance rejection performance and command response bandwidth has been discussed in the previous sections. Excessive activity may even impose higher-than-necessary performance requirements on control system hardware in aircraft design. The frequency content of NDI control commands is fundamentally shaped by the reference model dynamics and the closed-loop bandwidth requirements. %
Effective degrees of freedom are versatile and can be found in all control law components: in the feedback controller, reference model, inverse model equations, control allocation, and sensors and signal processing. 
However, while the control allocation seems to be the suitable place to address the control activity, it already needs to be considered in the preceding controller blocks. While distribution of control power is done in the (effectiveness-based) control allocation, they often neglect signal frequencies or treat them indirectly~\cite{Johansen2013,Harkegard2004}.
Important measures include the closed-loop cut-off frequency of the overall control system, the roll-off at higher frequencies, RMS values from control responses in simulations with external disturbances, and peak values in position and rate in simulated command-response analyses \cite{Looye2006}.

\subsubsection{Structural Interaction}\label{sec:design:structural}

Structural interaction may be divided into influencing damping and stability of structural airframe modes and the effect of the control system on flight loads induced by maneuvering, gusts, or failure cases. By now, a substantial number of publications have appeared on the application of NDI to flexible aircraft~\cite{Gregory1998,Wang2019b,Shearer2008,Wang2022,Wu2018,Beyer2024,Caverly2016}. In practice, there is no such thing as a rigid aircraft, albeit often assumed in the scope of the model derivation (\cref{sec:method:mod}). It is for this reason that flight control algorithms are necessarily taken into account in flutter and flight loads analyses as part of acceptable means of compliance for aircraft certification~\cite{Kier2024}.
Each industrial application of NDI is an application example of flexible aircraft~\cite{Pratt2000,Tauke2002,Kim2021}.

Even though accurate models are necessary to perform required aeroelastic analyses, their explicit use in inverse models (e.g.,~\cite{Wang2019b}), especially in combination with estimated aeroelastic states, is questionable from a sensor point of view. Aeroelastic states (modal coordinates, lag states~\cite{Looye2008,Kier2009,Kier2024}) tend, by definition, to be artificial from a physical point of view. The common approach, therefore, is to use rigid aircraft models for the actual design and to integrate structural filters to avoid the pickup of critical airframe modes~\cite{Miller2011}.
As these modes change under the influence of feedback (NDI) control, tuning these filters is an iterative process~\cite{Irving1995}. Applying angle of attack as a commanded variable with relative order of 2, the angular rate (which requires load factor measurement) is replaced with inertial pitch rate feedback. The effect of the approximation error on command response is compensated for by the feedback controller. 

During the flight tests with DLR's VFW-614 ATTAS (\cref{fig:ATTAS}) reported in~\cite{Lombaerts2012}, a so-far unpublished severe structural coupling was encountered that resulted in violent shaking of the airframe due to an unstable lateral fuselage mode. %
This was caused by a so-called hidden feedback, which applied a lateral load factor to the feedback signal synthesis. A simple addition low-pass filter would have avoided the occurrence. With no structural model of the aircraft type available, a critical review in this respect is compulsory, even for seemingly rigid airframes and even for inertial sensors located close to the center of gravity.

The first application of incremental NDI to a large, flexible passenger aircraft~\cite{Kier2020} revealed a significant influence of angular acceleration feedback on empennage loads. Even though incremental NDI has been proposed for load alleviation in a passenger aircraft~\cite{Wang2019b}, flight loads need to be monitored at all critical airframe locations. Especially, structural loading inboard of applied control surfaces may increase considerably.

\begin{figure*}[!htb]
	\centering
	\includegraphics{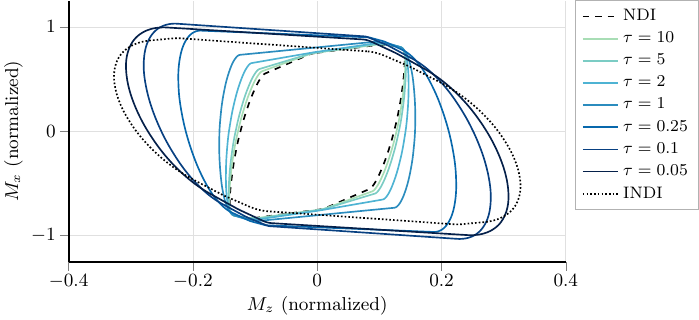}
	\caption{Load envelope of the moments \( M_x \) and \( M_z \) at the vertical tailplane root, taken as the maximum over gust loads and the MLU353 maneuver, using NDI, incremental NDI, and hybrid NDI with the complementary filter time constant \( \tau \). A long time constant keeps the sensory path narrow-band and thus stays close to model-based NDI, whereas a short one approaches incremental NDI. Both moments are normalized to their respective maximum values.}\label{fig:load}
\end{figure*}

\cref{fig:load} shows the effect of bending and torsion loads at the vertical fin of a passenger aircraft under lateral gust encounters. The setup is similar to~\cite{Kier2020}, but modified with a tunable hybrid NDI controller. Using incremental NDI, the effect of lateral gust encounters is immediately picked up by the yaw acceleration (derived) measurement. In the case of classical NDI, the encounter would be sensed by the sideslip vane (or estimator), allowing complementary filtering as an effective degree of freedom to shape the extent of disturbance rejection. This has been the main reason to propose hybrid NDI in applications like these (cf.~\cref{sec:method:hndi}). \cref{fig:load} shows the effect of the complementary filter time constant on the gust load envelope from fully incremental to fully classical NDI. Feedback signal synthesis and inverse model computation are thus effective degrees of freedom in addressing flight loads interactions.

\subsubsection{Stability and Margins}

Stability for dynamic inversion-based control laws can be approached in different ways, where the nonlinearity of the overall control law is often the driving factor leading to Lyapunov-based stability analyses~\cite{Wang2019,Schumacher1998,Hovakimyan2005,Lavretsky2005}.
However, demands on stability and stability margins for NDI control laws are no different from those for classical flight control law designs, and there seems to be no reason not to apply traditional metrics to compute them~\cite{Looye2001,Miller2011}. Admittedly, the nonlinear nature of the underlying system needs to be borne in mind, as for classical control law designs on nonlinear systems (which is virtually always the case). This can, however, be achieved by using linearization over a grid of operating points, linear parameter-varying (LPV) representation, or linear time-varying (LTV) representation.

However, the use of inverse models always raises concerns regarding dynamics made invisible from commanded variable responses, i.e., the zero dynamics (cf.~\cref{sec:mathematical:normal}).
At this point, two observations can be made:
\begin{enumerate}
	\setlength{\itemsep}{0pt}
	\item NDI control laws are often used in the inner core of manual and automatic flight control systems, i.e., for the attitude (rate) control. As has been mentioned in \cref{sec:Design:arch} and will be shown in \cref{sec:application:dyninv}, the zero dynamics that occur there are taken care of by outer loop functions. An overall stability margin analysis of the full model, with relevant combinations of manual and automatic control law functions, addresses this aspect by definition.
	\item Stability problems with zero dynamics can be largely avoided by a good understanding of aircraft flight dynamics. The classic example is using acceleration or load factor as commanded variables: Direct application of the inversion procedure yields a relative degree of zero,	since control surfaces produce (small) aerodynamic forces directly, in addition to the moments they are intended to generate (cf.~\cref{sec:modelcomplexity}). As the load factor response to control surface inputs at the empennage is non-minimum phase, the resulting zero dynamics will unavoidably be unstable. A similar example is provided in~\cite{Hauser1992}.
	From a flight mechanics point of view, it is clear that these control surfaces are intended to influence moments. Simply neglecting the force effects in the model to be inverted will result in taking the intended indirect path via aircraft attitude control. Even in the case of canards, where the load factor response is minimum-phase, this principle must be kept in mind to avoid excessive control activity.
\end{enumerate}

\subsubsection{Initialization and Trim Loading}

NDI control laws inherently compute trim solutions in case commanded variable inputs are zero. 
However, even due to small model deviations (e.g., modeling errors, or changes in aircraft loading, or differences between aircraft tail numbers) inverse model computations will hardly ever result in the actual trimmed state.
The badly performing example provided by Looye in~\cite{Looye1998}, is, in hindsight, predominantly caused by this trim error.

Trim loading is of concern in operational as well as experimental flights, as a failure of the control system may render the aircraft massively out of trim once steady physical control commands vanish. It is important to offload static control deflections to means that the aircraft is equipped with. This may involve adjusting the horizontal tailplane or the trim tabs on the control surfaces. Thus, an alternative approach is to address trim loading in control allocation. For example, the Airbus A320 offloads elevator deflection by slow, i.e., low gain, integration and limitation to create stabilizer commands~\cite[p. 858]{Brockhaus2011}. By feeding the stabilizer position back into the NDI control law, the elevator deflection will automatically reduce. In the case of incremental, sensory, or hybrid NDI, this is not even necessary, as it is simply inferred from pitch acceleration estimation.

During the flight tests reported in~\cite{Lombaerts2013}, offloading was done by engaging the basic and independent aircraft autotrim system. This improved safety during test, but also caused adverse interactions that compromised tracking accuracy.

A known approach for flight testing is to calculate the trim points of the closed-loop system for the test cases beforehand, manually bring the aircraft close to this trimmed test point, and then initialize the controller~\cite{Morelli1999}. Though the initial solution will never exactly trim the aircraft, for unmanned flight tests, this approach can be sufficient. However, for larger aircraft, this initial deviation and control step can be unwanted. A further problem arises when testing very experimental controllers. Because the closed-loop system must correct during initialization, it is impossible to tell in the first second whether sudden aircraft movements are due to trim-state correction or to controller instability.

In case an NDI control law is switched on in mid-flight (this is a common occurrence in flight experiments), it is common practice to determine commands at initialization time \( \vec{u}(t_0) \) and to replace those with actual control deflections at this time point \( \hat{\vec{u}}(t_0) \), see~\cite{Looye2001}, i.e., the corrected control commands \( \vec{u}^\prime \) become
\begin{equation}
	\vec{u}^\prime(t) = \vec{u}(t) + (\hat{\vec{u}}(t_0) - \vec{u}(t_0))
\end{equation}

A more refined approach is proposed by Kaminer~\cite{Kaminer1995} and Osterhuber~\cite{Osterhuber2004} by pushing the integrators through to the output of the control algorithms. This involves differentiating all sensor inputs and, if necessary, the model equations to calculate the control command derivative \( \dot{\vec{u}} \). This derivative is integrated at the output of the control law, similar to the canonical control form (cf.~\cref{sec:mathematical:normal}). This ensures that all internal signals in the control law are zero in steady flight, so that no incorrect static control outputs result. At the same time, the integrators at the controller output can be initialized with the current control deflections \( \hat{\vec{u}}(t_0) \), making the transition smooth, apart from the desired dynamic response to attain the commanded variable references. Thus, the corrected control commands \( \vec{u}^\prime \) are
\begin{equation}
	\vec{u}^\prime(t) = \int\limits_{t_0}^t \dot{\vec{u}}(\tau) \mathrm{d}\tau + \hat{\vec{u}}(t_0)
\end{equation}
Osterhuber~\cite{Osterhuber2004} extends the principle to the so-called Differential PI implementation, smartly integrating the transformation from aerodynamic velocity referenced into body axes with the integration algorithm. A similar approach is done in some continuous NDI derivatives, e.g.,~\cite{Raab2019}.

\subsubsection{Interaction with External Functions}

When using measured control deflections in incremental NDI, it is important to note that the actuator may use the same sensor in a negative feedback path of the servo control law. Under certain circumstances and as already noted in \cref{sec:method:indi}, the servo control loop may be effectively canceled, likely resulting in integrator-like behavior. This is exploited, for example, by~\cite{Raab2019}, but it must be borne in mind that this cancellation will also affect other control functions that use the same surfaces. An interesting example is static aileron droop in the Airbus high lift system~\cite{Lulla2011}. Applying incremental NDI in a scenario like this would cause the high-lift system to drive the aileron deflections to their limits, as these become effectively rate commands. This has been one of the main reasons for Milz and Looye to propose sensory NDI, allowing control allocation using full rather than incremental commands (see~\cref{sec:method:sndi}).

\begin{figure}[!htb]
	\centering
	\includegraphics{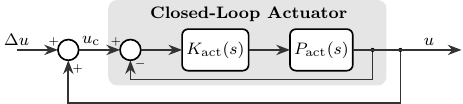}
	\caption{Actuator control loop in combination with incremental NDI.}\label{fig:actuatorcontrol}
\end{figure}

A combination of incremental NDI and the actuator position control loop feedback is shown conceptually in~\cref{fig:actuatorcontrol}. Assume a linear closed-loop actuator with a position control loop, then the dynamics can be approximated as
\begin{equation}
	\frac{u(s)}{u_c(s)} = \frac{P_\mathrm{act}(s) K_\mathrm{act}(s)}{1 + P_\mathrm{act}(s) K_\mathrm{act}(s)}
\end{equation}

By adding the feedback of the current actuation position to the increment, the position is canceled:
\begin{equation}
	\frac{u(s)}{\Delta u(s)} = P_\mathrm{act}(s) K_\mathrm{act}(s)
\end{equation}

This fact already indicates that there must be some effect occurring through this phenomenon. However, the overall impact has not yet been adequately investigated. Furthermore, the effect may be reduced by delays between actuator and flight controllers and faster high-frequency actuator control loops which allow for the application of time-scale separation. 
Note that actual actuators do not behave as a linear system and are influenced by the loads that occur. However, the schematics show the principal effect of absolute control allocation compared to incremental control allocation. A major advantage of this absolute allocation is that, if commands from secondary control functions (working in the null-space) are added to the command signal, these are handled as intended. Incremental NDI may require compensatory measures to combine effective rate, and position commands.

\subsubsection{Robustness to Parametric Uncertainties and Unmodeled Dynamics}\label{sec:design:robustness}

The most basic reasons for using feedback control are the rejection of disturbances (\cref{sec:design:dist}), and the achievement of satisfactory system performance in the presence of uncertainty. The latter presents a classical trade-off in control system design~\cite{Doyle1990}.

Robustness is commonly mentioned as a topic of concern in the application of NDI. This is not fully justified though. As already discussed in \cref{sec:conceptual}, NDI aims to linearize, decouple, and normalize responses to commanded variable derivatives. 
The inverse model is a mapping and no attempt is even made to achieve the above robustly, as feedback control with the intention to this end is only added once the outer control law functions are integrated. The main effect of model uncertainty is thus in the fact that the above objectives will be met less accurately. The latter is usually anticipated anyway, as simplifying assumptions are commonly made to reduce complexity of the inverse model. In addition, and as mentioned in \cref{sec:method}, NDI uses measured states and does not self integrate own state equations.

As in the case of disturbance rejection, NDI allows robustness to be addressed in all of its components. A first design degree of freedom, obviously, is the inverse model itself. Any improvement in model quality will leave less effort to the outer feedback control laws, and therefore positively influence system performance. Incremental and sensory NDI largely avoid the use of model data, making the method more robust in this respect. 
Also, in the presence of parametric uncertainties in the design model, it is not mandatory to invert the equations for the nominal case. In \cite{Looye2001} uncertain model parameters are optimized to values that offer a better compromise over all assumed tolerance ranges.

Accuracy of the inverse model can be improved drastically by adjusting model parameters as soon as flight test validated model data becomes available. With NDI this is very straight-forward. A major step further obviously is in the application of adaptive control as proposed by \cite{Johnson2000b,Holzapfel2004,Lombaerts2011a,Lombaerts2011b}, adapting inverse model equations in flight.

Robustness to remaining uncertainties or unmodeled dynamics is usually addressed by the feedback controller.
If necessary, the controller can be designed using methods that specialize in balancing performance and robustness to model uncertainty. Robust control methods offer means to optimize robust performance directly~\cite{Reiner1996,Adams1993,Valasek2001,Pollack2023,Looye1998,Snell1998b,Oort2006,Wang2015,Fer1997,Wang2005,Hachem2026,Pineau2026,Kawaguchi2011,Steffensen2023b,Tipan2020,Tamaskani2026,Goman1998}. Furthermore, (global) multi-objective optimization can be used for compromise tuning between different off-nominal model cases. The latter may be found by means of anti-optimization~\cite{Looye2001}.

An important issue to address is sensor signal delay. As stated in~\cref{sec:inversearchitecture:di}, the use of measured rather than self-integrated states in the model equations introduces sensitivity to sensor accuracy and signal delays. It has been found that incremental NDI is particularly sensitive to delays between acceleration and control measurements~\cite{Vlaar2014}. During a first flight test campaign, Looye and Vlaar found that simple synchronization between these signals is key to successful implementation~\cite{Vlaar2014}. Common signal delays, e.g., due to filtering, turned out to affect performance and stability margins in ways comparable to those of classical NDI control laws. This was later formally analyzed by Smeur~\cite{Smeur2016}.
To a lesser extent, the synchronization also needs to apply to any NDI control law. In this respect, it is important to realize that airdata sensors may exhibit more delay than inertial ones. This somewhat affected flight test results reported in~\cite{Lombaerts2012}. Synchronization of multiple signals in an incremental NDI design remains an ongoing topic in recent publications, including~\cite{Kumtepe2022,Steffensen2023,Smeur2016}. The remaining model dependency of incremental NDI, i.e., the control effectiveness, and its influence on robustness are analyzed in, e.g.,~\cite{Cao2025}.

Apart from accuracy of the inverse model, \emph{Adaptive NDI}~\cite{Lombaerts2011a,Lombaerts2011b} methods use online system identification and fault detection algorithms to adapt the reference model parameters to the current state of the flight vehicle. This is mainly achieved by reducing the limits.

Finally, it is noted that robustness of NDI-based design may be done using any commonly applied analysis method. Linear analysis, including structured singular value (\( \mu \)) analyses~\cite{Brinker1996}, but also nonlinear Monte-Carlo simulations~\cite{Gaessler2025,Wang2019}, and anti-optimizations~\cite{Looye2001} are established methods to quantify the robustness.

\subsubsection{Control Limit Handling}

The assumption of ideal, i.e., fast, or absent actuator dynamics becomes invalid as soon as physical rate and position limits are reached. If no control power via other means can be allocated, tracking performance will degrade, and integrator windup and potential stability loss may result. This is of particular concern in the case of failure scenarios~\cite{Lombaerts2010}.

The best approach, at least under nominal conditions, is to minimize control activity relative to the tracking accuracy and disturbance rejection levels to be achieved (cf.~\cref{sec:design:reqs:ctrlact}), and to operate well within control limits. The use of reference models in combination with, for example, pseudo control hedging (PCH)~\cite{Johnson2000,Holzapfel2004,Lombaerts2010,Lombaerts2012} is a powerful means to safely exploit these control limits as far as possible. PCH effectively hedges the commanded reference signal to levels achievable by the constrained actuator dynamics. 

PCH aims to avoid, but does not handle saturations, especially if caused by severe external disturbances.
Measures for handling consequences like integrator wind-up~\cite{Astrom1989} are needed, not much different from control laws designed using other methods. The differential PI~\cite{Osterhuber2004} makes handling of integrator wind-up considerably easier, as integrators are positioned at the control law outputs.

In some cases, architectural measures may have to be taken, especially in the case of physically unstable systems, or variables that have to be kept at bay to stay within safe flight envelope limits. This may set new priorities and require a different combination or reduced set of commanded variables. An according inverse model core may have to be kept available for such a scenario.

Apart for trade-offs between control law performance and control activity, modern flight control systems, particularly those with redundant control surfaces, allow for sophisticated control allocation algorithms. Those are an important design degree of freedom, aiming at maximizing control power or offloading steady deflections to trim devices. Techniques like daisy chaining have also proven to be very effective in case complementary means of control are at hand, like thrust vectoring~\cite{Enns1994,Steinhauser2004}. Furthermore, optimization-based control allocation schemes and attainable moment set methods do also cover these topics~\cite{Durham2017,Johansen2013}.

Recent advances in incremental NDI treat the case of non-negligible actuator dynamics by explicitly incorporating actuator dynamics within the control law derivation~\cite{Steffensen2023}. This is particularly relevant in the case of complementing controls that have different bandwidths.

\subsubsection{Handling of Failure Cases}

Failure cases that need to be addressed depend on higher-level design requirements derived from the overall flight control system architecture and the preceding Failure Mode and Effects Analysis (FMEA). Sensor and actuator systems are usually designed with redundancy to meet certification requirements, but it may be required that the control algorithms be capable of handling complete failure, or at least reduced control power. 

Control system failures are best handled by control allocation. Their effects on various variants of NDI have been discussed in \cref{sec:method}.

On the sensor front, an inertial reference system seems to be the bare minimum, unless attitude and velocity components can be estimated by other means (airdata, GPS). As incremental and sensory NDI tend to be inherently less dependent on estimation of states, having these structures as a backup can be a very effective means to handle airdata failures~\cite{Milz2024e}, without loss of internal control law functionality. Obviously, tracking of air mass referenced variables becomes problematic but this is not NDI-specific. 

Aircraft-specific properties are largely captured by the inverse model core. This allows for updating this core to failure cases that have been anticipated in advance in a straight-forward way. As mentioned previously, advanced extensions of NDI-based controllers implement on-line estimations algorithms that allow for determining and adapting the model in flight \cite{Lombaerts2010b,Holzapfel2004,Johnson2000b}. One of the findings of these studies has been that the changed envelope bounds may be considerably more critical, requiring these to determined as well \cite{Lombaerts2017}.

\subsection{Design Optimization}

Due to its modular structure, clear functional breakdown, and effective linearization and normalization by the inverse model core, tuning of NDI-based control laws to meet the above requirements tends to be relatively straightforward. However, in cases of considerable uncertainties or simplifications in the inverse model implementation, the risk of structural couplings and prevailing conflicts in requirements, or even between functions, may necessitate a combined tuning of the overall system. To this end, various approaches are available.
In~\cite{Reiner1996,Pollack2024}, this is achieved by applying robust control methods in the design of the feedback control laws. 
In~\cite{Tischler2002,Looye2006}, multi-objective optimization is proposed to find best-compromise solutions. This methodology uses the maximum over a potentially large number of evaluated and scaled criteria as its optimization function \cite{Kreisselmeier1979}. These criteria may be computed from basically any type of analysis that is deterministic and can be computed in affordable time \cite{Joos2002}. In \cite{Looye2006}, even risk values obtained from on-line Monte-Carlo analysis were used in optimization. This reference also proposes sequential optimization procedures, eventually leading to overall tuning and compromising of the complete system.  

Controller synthesis methods are typically characterized by performance, robustness, stability, or all three. They are optimized for some form of weighting (functions) or by direct adjustment of controller parameters. For example, in the case of backstepping~\cite{Krstic1995}, the Lyapunov function is used, and the proposed control law may be inversion-based. In robust control, a controller is synthesized based on weighted induced system norms. %

This, in turn, is highly applicable to the verification and validation process, as agreed-on acceptable means of compliance are likely to remain applicable as before~\cite{Looye2006}.

\subsection{(Re-)Design Effort}

As Enns already stated in~\cite{Enns1994}, a great advantage of NDI is that the redesign effort is relatively straightforward, as after adapting the inverse model, the outer control functions again effectively see the same command-response dynamics (see also~\cite{Wacker2001}). 

Taken together, the properties discussed in this section amount to a form of scalability of the design process itself. Because the inversion presents a decoupled interface to the outer loop, separate functions can be partitioned into blocks that are developed largely independently against that common interface and merged afterwards.
This decomposition runs along \emph{functions}; it is not the divide-and-conquer paradigm that NDI supersedes (cf.~\cref{sec:conceptual}), which instead partitions the \emph{flight envelope} into operating points and manages the design by scheduling gains. 
Enns \textit{et al.} identified this shift as one of the principal attractions of the methodology~\cite{Enns1994}. This way, functions can be added or exchanged at a well-defined places, e.g., hedging or limiting function ahead of the inversion (cf.~\cref{sec:design:reqs:ctrlact}), or a low-pass or notch filter in a feedback path (cf.~\cref{sec:design:structural}). In the first iteration, its effects can be assessed locally rather than by re-tuning the controller as a whole. However, such interfaces only remain valid as long as the controlled variables are reasonably chosen (cf.~\cref{sec:method:design:cv}).

This idea has been further developed in~\cite{Kuchar2018}, enabling control law designs within overall aircraft optimization loops. Adaptation to the current aircraft design iteration is, in principle, simply done by adapting the inverse model core. Early provision of prototype control laws that automatically adapt to each new aircraft design configuration enables closed-loop analyses early in the design process by all disciplines involved or affected (e.g. flight loads), and can be used to derive performance requirements for control system hardware.
This makes dynamic inversion-based control laws a sensible choice in aircraft design~\cite{Ostroff1999,Kuchar2018,Looye2007,Patel1998,Johansson2022,Kiehn2022}. The prototype NDI design naturally evolves into detailed design once the main aircraft parameters have been frozen. Once flight testing of the prototype commences and improved model data becomes available, the inverse model core may be adapted in case performance is insufficient.

\section{An Application Perspective}\label{sec:application} %

This chapter shows two application examples of NDI-based control law designs. The main example describes the development of an integrated automatic and manual flight control system for a passenger aircraft from concept to implementation and flight test. The design implements all previously introduced variants of the methodology. More details on the control laws, the applied V\&V process, and successful flight test campaign can be found in~\cite{Weiser2024}. As a second example, a design for an eVTOL aircraft with non-affine control effectors is briefly discussed.

\subsection{Design, Implementation, and Testing on a Passenger Aircraft}

\begin{figure}[!htb]
	\centering
	\includegraphics[width=0.5\textwidth]{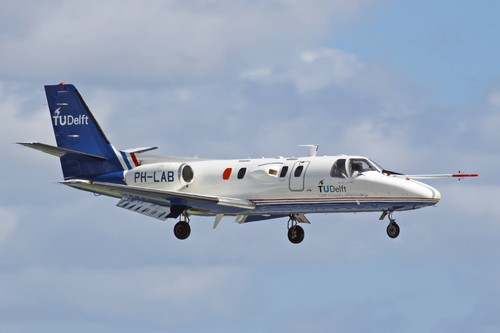}
	\caption{Cessna Citation II (Model 550) Research Aircraft \textit{PH-LAB}\ \copyright{} Photo courtesy of Malcolm Nason}\label{fig:phlab}
\end{figure}

The \textit{PH-LAB} research platform (\cref{fig:phlab}) is a Cessna Citation II (CS-25 class) that has been equipped with an experimental Fly-By-Wire (FBW) system~\cite{Zaal2009}, allowing for rapid-prototyping implementation and testing of new control algorithms. The FBW system has access to aileron, elevator, and rudder servos. Trim tabs and throttle have to be set manually, whereby the control laws may display references if necessary. A high-fidelity model for the aircraft, including control systems hardware, is available. The \textit{PH-LAB} therefore provides an excellent basis for testing of new control methods and functions in flight. In the recent past, various designs based on NDI and incremental NDI, as well as incremental back-stepping, and LPV methods have been successfully flown on the aircraft, some methods for the first time~\cite{Grondman2018,Keijzer2019,Pollack2019}. 

\subsection{Flight Dynamics Model}\label{sec:application:model}
The model of the aircraft is based on an original implementation documented in \cite{Linden1998}, later adapted to the specific aircraft in \cite{Grondman2018}. It has been implemented and configured both for simulation, as well as control law design purposes. It for example comes with excellent trimming and linearization tools. The 6-DoF rigid-body flight mechanics consist of six translational and six rotational states. These are the position in Earth-fixed co-ordinates \(\vec{r}_E\) (optionally: WGS-84 referenced latitude, longitude, altitude), the translational velocity \(\vec{v_B}\) in the body frame, the Euler attitude vector \( \vec{\Theta} = \left[ \phi \ \theta \ \psi \right]^T \), and the rotational velocity \vec{\omega_B}, also in the body frame. The state space system of the aircraft dynamics is given by (see also Appendix~\ref{sec:eqm})
\begin{subequations}\label{eq:phlab_statespace}
	\begin{alignat}{2}
		\dot{\vec{r}}_E &= \mat{R}_{\footnotesize \mathrm{EB}} \ \vec{v}_B && \\ 
		\dot{\vec{\Theta}} &= \mat{R}_{\footnotesize \mathrm{\Theta B}} \ \vec{\omega}_B && \label{eq:Theta} \\ 
		\dot{\vec{v}}_B \ &= \frac{1}{m} \mat{R}_{\footnotesize \mathrm{BA}}\vec{f}_{\mathrm{a}_A} +  \mat{R}_{\footnotesize \mathrm{EB}}^T \vec{g}_E - \vec{\omega}_B \times \vec{v}_B  &&+ \frac{1}{m} \left( \vec{f}_{\mathrm{p}_B} + \mat{R}_{\footnotesize \mathrm{BA}}\vec{f}_{\mathrm{a}_A}^\prime \right) \\  
		\underbrace{ \dot{\vec{\omega}}_B }_{ \dot{x} } &= \underbrace{ \mat{I}_B^{-1} \left( \vec{m}_{\mathrm{a}_B} - \vec{\omega}_B \times \mat{I}_B \vec{\omega}_B \right) }_{ f(x) } &&+ \underbrace{ \mat{I}_B^{-1} \left( \vec{m}_{\mathrm{p}_B} + \vec{m}_{\mathrm{a}_A}^\prime \right) }_{ G(x) u }
	\end{alignat}
\end{subequations}
where \( m \) denotes the aircraft mass, \( \mat{I}_B \) the inertia tensor, and \( \mat{R} \) the corresponding rotation matrices.
The forces and moments include the aerodynamic control moments \( \vec{m}_{\mathrm{a}_A}^\prime \approx \bar{q} S \bar{c} \mat{C}_{m_\delta} \delta_\mathrm{cs} \) with the dynamic pressure \( \bar{q} = \frac{1}{2} \rho V_\mathrm{tas}^2 \), the reference area \( S \), the reference length \(\bar{c}\), and the aerodynamic moment derivatives w.r.t.\ the control surface deflections \( \mat{C}_{m_\delta} \).%
The direct aerodynamic input forces \( \vec{f}_\mathrm{a}^\prime \) can often be neglected, as the control surfaces are primarily intended to generate moments and their direct force contribution is small (cf.\ the discussion on the physical relative degree in~\cref{sec:modelcomplexity}).
The propulsive forces and moments, \( \vec{f}_\mathrm{p} \) and \( \vec{m}_\mathrm{p} \), are often neglected as an input as no control over the engine is available in many primary flight control applications. 
The gravitational acceleration \( \vec{g}_E \) is assumed to be constant, and the aerodynamic forces and moments, \( \vec{f}_\mathrm{a} \) and \( \vec{m}_\mathrm{a} \), solely depend on the aircraft state (disregarding atmospheric disturbances momentarily).

The control surface deflections \( \delta_\mathrm{cs,com} \in \mathbb{R}^3 \) represent aileron, elevator and, rudder deflection. 
The actuator models, detailed in~\cite{Pollack2019}, are neglected in the inversion process, because these are presumably considerably faster than the aircraft rotational dynamics. Obviously, the actuators are considered in subsequent design analyses. 

For the application of NDI, a dedicated model has been configured along the lines of \cref{sec:modelsimplification} and as a first step in \cref{fig:NDI_Process_Methodology}. More details can be found in ~\cite{Grondman2018}.

\subsection{Flight Control Law Architecture}\label{sec:application:structure}

\begin{figure*}[!htb]
	\centering
	\includegraphics{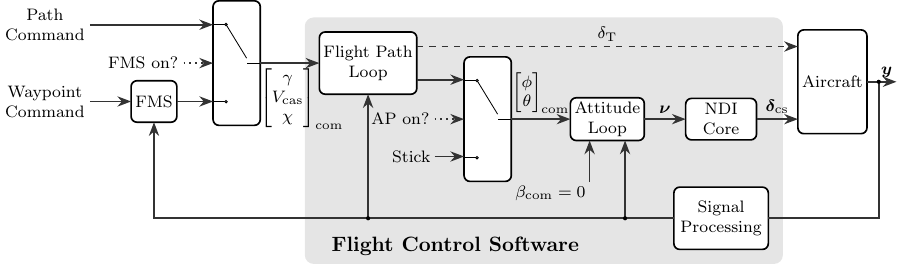}
\caption{Flight control law architecture with NDI-based attitude loop.}\label{fig:cascade}
\end{figure*}

As a first step in offline design of the control laws (\cref{fig:proc}), the architecture is defined. The present system follows a cascaded structure, separating the control loop into an inner and an outer loop. The inner loop captures the faster rotational dynamics (attitude control), while the outer loop captures the slower flight path dynamics (flight path control). The two loops can be separated by the time scale separation principle and thus adjusted sequentially. \Cref{fig:cascade} sketches the cascaded control structure (see~\cite{Weiser2024}).%
The flight path control loop computes engine \(N_1\) setting references that are presented on a display. As the aircraft does not have a throttle servo, these are to be set accordingly by the safety pilot (flight director-like).

The attitude control laws are based on NDI, whereas the flight path and speed tracking functions use the Total Energy and Total Heading Control systems respectively, see \cref{sec:method:design:cv} and \cite{Lambregts2013a,Lambregts2013b}. Combined with the option of manual augmented control, \({\phi}\,\ {\theta}, \beta\) have been selected as commanded variables that serve both intended modes of operation well. 

Differentiation (as the second step in \cref{fig:NDI_Process_Methodology}) of the first two results in \cref{eq:Theta}, which, in turn, is differentiated one more time to arrive at the angular acceleration vector \(\vec{\dot{\omega}_B}\). The derivative of \(\mat{R}_{\footnotesize \mathrm{\Theta B}}\) is hereby explicitly and necessarily taken into account and becomes a compensatory part in the inverse model. Inversion of the side slip angle will be addressed in \cref{sec:sideslip_control}. The three variables have a relative degree of 2, where the second differentiation and subsequent inversion goes via angular accelerations \(\vec{\dot{\omega}_B}\). 

The aircraft has no redundant means of control around it axes, so that control allocation does not play a role in this application.

\subsection{Dynamic Inversion and Zero Dynamics}\label{sec:application:dyninv} 

When implementing a dynamic inversion-based flight controller following~\cref{sec:application:structure}, it is important to remember that the inversion is part of an integrated system consisting of outer-loop functions. Although solely the rotational rates \( \vec{\omega} \) are inverted, the remaining system equations remain. Having an explicit look at the internal and zero dynamics is essential.%

The formal dynamic inversion or feedback linearization control design (see \cref{fig:proc} and \textit{Realization} steps in  \cref{fig:NDI_Process_Methodology}) starts with a look on the canonical control form~\cref{eq:canonicalcontrolform} of the system and the selection of the commanded variables \( \vec{y} \). 
As the system has 6 degrees of freedom (3 translational and 3 rotational), control authority in all axes is required in order to have a fully invertible system. However, most fixed-wing aircraft have control surfaces that can control the 3 rotational degrees of freedom and a thrust channel for the forward translation. Additionally, the velocity control is in many cases decoupled from the primary flight control system, leading to a substitution of the thrust as an input with a decoupled velocity.

For a cascaded flight control law, the rotational dynamics for \( \vec{\omega}_B \) are chosen to be inverted according to~\cref{sec:application:structure}, resulting in \( \vec{y} = \mat{C}_c \ \xi = \vec{\omega}_B \) with a relative degree of 1.
As the states \( r_x \) and \( r_y \) do generally not contribute to the flight dynamics (and are marginally stable), those are neglected in the following.
The canonical control form can be stated using the pseudo control inputs \( \nu_\omega \) for the external states \( \xi_\omega \) and with the internal states \( \eta_{r_z} \), \( \eta_\Theta \), and \( \eta_{v} \): \begin{subequations}
	\begin{align}
		\dot{\xi}_\omega &= \mathcal{L}_{\vec{f}_\omega} \vec{h}(x) + \mathcal{L}_{\vec{g}_\omega} \vec{h}(x,u) = \nu_\omega \label{eq:impl:externalstates} \\
		&= \nabla_\omega \vec{h} (\vec{z}) \mat{I}_B^{-1} \left( \vec{m}_a (\vec{z}) - \xi_\omega \times \mat{I}_B \xi_\omega \right) + \nabla_\omega \vec{h}(\vec{z}) \mat{I}_B^{-1} \left( \vec{m}_p(\vec{z},\vec{u}) + \vec{m}_a^\prime(\vec{z},\vec{u}) \right) \notag \\
		\dot{\eta}_{r_z} &= -\sin\eta_\theta \eta_{v_x} + \cos\eta_\theta \sin\eta_\phi \eta_{v_y} + \cos\eta_\theta \cos\eta_\phi \eta_{v_z} \\
		\dot{\eta}_\Theta &= \mat{R}_{\Theta B}(\eta_\Theta) \ \xi_\omega \\
		\dot{\eta}_{v} &= \frac{1}{m} \mat{R}_\mathrm{BA}\vec{f}_{\mathrm{a}_A} (\vec{z}) + \mat{R}_\mathrm{EB}^T \, \vec{g}_E - \xi_{\omega} \times \eta_v + \frac{1}{m} \left( \vec{f}_{p_B} + \mat{R}_\mathrm{BA}\vec{f}_{a_B}^\prime \right)
	\end{align}
\end{subequations} 
where \( \vec{f}_{a_B}^\prime \) and \( \vec{f}_{p_B} \) are considered as side-effects and neglected as inputs.
Assuming that the only the aerodynamic input moments are significant and can be described linearly, i.e., \( \vec{m}_a^\prime(\vec{z},\vec{u}) = \vec{M}_a(\vec{z}) \ \vec{u}_{1..3} \), the external states can be described as
\begin{align}
	\dot{\xi}_\omega &= \mat{I}_B^{-1} \left( \vec{m}_a (\vec{z}) - \xi_\omega \times \mat{I}_B \xi_\omega \right) + \mat{I}_B^{-1} \vec{M}_a(\vec{z}) \ \vec{u}_{1..3} \tag{\ref{eq:impl:externalstates}\textquotesingle}
\end{align}

To be invertible, it must hold that \( \mathrm{rank} \left\{ \mat{I}^{-1} \vec{M}_a(\vec{z}) \right\} = 3 \) throughout the envelope. 
Furthermore, the internal dynamics \( \dot{\eta} = f_0(\xi=0,\eta) \) need to be asymptotically stable, which is not fulfilled on this system.
However, stability of the zero dynamics can be neglected here. The illustrative answer is that the attitude, flight path, and altitude are controlled and stabilized by outer loop functions. The control theoretic answer is that the actual system we are investigating is no longer the open-loop aircraft but the closed-loop one. Thus, although controlling only \( \vec{\omega}_B \) on an open-loop aircraft leads to unstable zero dynamics, we are here looking at the closed-loop system that already has control cascades on the outer loop. Thus, we actually need to investigate the augmented zero dynamics, which will be stable (as the outer loop control functions stabilize them).
This is achieved through the cascaded control structure. \( \eta_\phi \) and \( \eta_\theta \) as well as \( \beta \), which directly determines \( \eta_{v_y} \), are controlled and stabilized through the attitude control loop via \( \xi_\omega \). The flight path control loop controls and stabilizes \( \eta_{v_x} \) via autothrottle functions, as well as \( \eta_\psi \) and \( \eta_{v_z} \) through \( \eta_{r_z} \).

\subsection{Hybrid Nonlinear Dynamic Inversion}
As explained in the previous section, the cascaded structure motivates the inversion of the rotational dynamics, i.e. simplified,
\begin{equation}\label{eq:eom_omega}
	\dot{\vec{\omega}}_B \approx \mat{I}_B^{-1} \left( \vec{m}_{\mathrm{a}_B}(\vec{x}) - \vec{\omega}_B \times \mat{I}_B \vec{\omega}_B \right) + \bar{q} S \bar{c} \mat{I}^{-1} \mat{C}_{m_\delta} \vec{\delta}_\mathrm{cs}
\end{equation}
The remaining states can be controlled by means of a cascaded control structure (see~\Cref{fig:cascade}). Based on~\cref{eq:eom_omega}, the hybrid NDI control law can be derived as
\begin{align}
	\delta_\mathrm{cs,com} & = \frac{1}{\hat{\bar{q}} S \bar{c}} {\hat{\mat{C}}_{m_\delta}}^{-1} \mat{I}_B \cdot \left( \vec{\nu}_\omega - \dot{\vec{\omega}}_\mathrm{compl} \right) \\ \notag
    \text{with } & \dot{\vec{\omega}}_\mathrm{compl} = \mathcal{L}^{-1} \bigl\{ H_\mathrm{c} (s) \bigr\} * \Bigl( \hat{\dot{\vec{\omega}}}_B - \hat{\bar{q}} S \bar{c} \, \hat{\mat{I}}_B^{-1} \hat{\mat{C}}_{m_\delta} \hat{\vec{\delta}}_\mathrm{cs} \Bigr) + \mathcal{L}^{-1} \bigl\{ 1 - H_\mathrm{c} (s) \bigr\} * \hat{\mat{I}}_B^{-1} \left( \hat{\vec{m}}_{\mathrm{a}_B}(\hat{\vec{x}}) - \hat{\vec{\omega}}_B \times \hat{\mat{I}}_B \, \hat{\vec{\omega}}_B \right)
\end{align}

The complementary filter uses the first-order low-pass filter
\begin{math}
	H_\mathrm{c} (s) = \frac{1}{\tau s + 1}
\end{math}
with the time constant \( \tau = \SI{0.5}{\second} \) following the implementation from~\cite{Milz2024e}.

\subsection{Attitude Control Function Design}
\begin{figure}[!htb]
	\centering
	\includegraphics{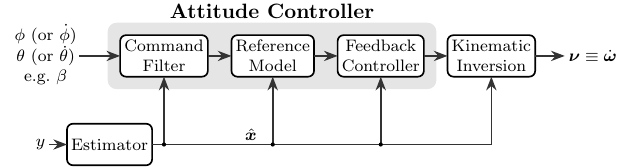}
	\caption{Attitude control architecture}\label{fig:architecture}
\end{figure}

The commanded angular acceleration, represented by the pseudo control input \( \vec{\nu} \equiv \dot{\vec{\omega}}_B \), is calculated in an attitude control loop as shown, for instance, in~\cite{Milz2022}. 
This loop consists of three functions that that are based on 2-DOF structures as described in~\cref{eq:2DOF} for roll and pitch attitude tracking, and based on~\cref{eq:1DOF_i} for side slip angle control. For the first,
reference models followed by a feedback controllers are used to compute \( \vec{\nu} \) from attitude commands. The feedback controller compensates for errors and uncertainties through an integrative element. Furthermore, the reference model ensures a smooth trajectory by smoothing sharp commands and limiting them to the aircraft's performance. As already indicated in~\cref{fig:hndi}, the reference model should output a signal that is \( \rho \)-times differentiable. This is often achieved by selecting a reference system of \( \rho \)\textsuperscript{th} order or higher. The design of the attitude control loop is shown in~\Cref{fig:architecture}. Since in this design (\(\rho = 2\)), the command filters for pitch and roll are of second order.

The flight path and speed control functions that constitute the flight path loop in \cref{fig:cascade} command pitch and roll angles directly. For manual control, the respective rates of these angles are commanded via a sidestick. The manual functions hereby work based on the so-called rate-command-attitude-hold principle. The sideslip angle \( \beta \) is regulated to zero, as no control inceptor for the yaw axis is available to the crew.

\subsubsection{Pseudo Control Hedging}
Pseudo control hedging (PCH), initially introduced in~\cite{Johnson2000a}, is a method to handle input dynamics and limitations, especially in adaptive or dynamic inversion-based control functions. It hedges the reference model in case of exceeding commands due to actuator dynamic limits, such as control surface saturation. The PCH signal \( \vec{\nu}_\mathrm{hdg} \) to be hedged or compensated is defined as the difference between the commanded and the actual pseudo control input:
\begin{equation}
	\vec{\nu}_\mathrm{hdg} = \vec{\nu} - \vec{\nu}_\mathrm{act} = \vec{\nu} - \left( \hat{\vec{\alpha}}\left(\vec{x}\right) + \hat{\mat{\Beta}}\left(\vec{x}\right) \hat{\vec{u}} \right)
\end{equation}
with the actual control surface deflections \( \hat{\vec{u}} \).
The PCH signal is used on the reference model to hedge the reference signal to be realizable in a system with relevant actuator dynamics. \Cref{fig:referencemodel} shows an exemplary implementation.
However, it is important to note that PCH should not be active during nominal operations, where actuator limits should not be reached by design but just for edge cases, where saturations might be reached.

\subsubsection{Feedback Controller}
For the two degrees-of-freedom control design applied in many NDI control structures (see \cref{sec:design:generalarch}), a reference model and feed-forward signal for \(\nu\) are used for command shaping, whereas a feedback controller is used for disturbance rejection.

A common design for attitude control is to calculate the virtual control commands using decoupled PID controllers. However, as many designs use the roll and pitch angle and the sideslip angle for tracking, the sideslip or lateral load factor controller is often decoupled (see \cref{sec:sideslip_control}). The PID controllers for the roll and pitch angle are given as:
\begin{align}
	\nu_\phi   & = \left( K_\mathrm{\phi,P} + \frac{K_\mathrm{\phi,I}}{s} \right) \left( \phi_\mathrm{ref} - \hat{\phi} \right) + K_\mathrm{\phi,D} \left( \dot{\phi}_\mathrm{ref} - \hat{\dot{\phi}} \right) + \ddot{\phi}_\mathrm{ref}                        \\
	\nu_\theta & = \left( K_\mathrm{\theta,P} + \frac{K_\mathrm{\theta,I}}{s} \right) \left( \theta_\mathrm{ref} - \hat{\theta} \right) + K_\mathrm{\theta,D} \left( \dot{\theta}_\mathrm{ref} - \hat{\dot{\theta}} \right) + \ddot{\theta}_\mathrm{ref}
\end{align}

\subsubsection{Reference Model}
\begin{figure}[!htb]
	\centering
	\includegraphics{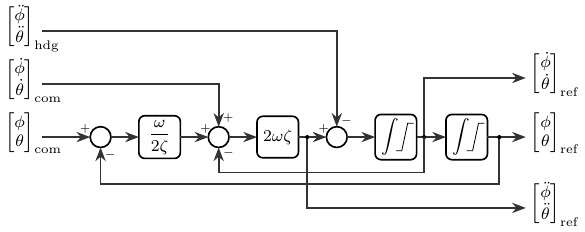}
	\caption{Second-order reference model with pseudo control hedging.}\label{fig:referencemodel}
\end{figure}

The reference signal for the feedback controller is filtered by a reference model. This ensures that, first, the commands stay inside the valid flight envelope, and second, it allows for the specification of flying qualities. In order to obey the flight envelope limits, corresponding limits are added to the integrators and signals in combination with anti-wind-up techniques.
An established approach is using a second-order low-pass filter
\begin{equation}
	H^\prime(s) = \frac{\omega_n^2}{s^2 + 2 \zeta \omega_n s + \omega_n^2}
\end{equation}
with the natural frequency \( \omega_n \) and damping ratio \( \zeta \).
The second order is especially useful since, first, the highest derivative in most cases corresponds to the acceleration, which is meaningful to limit and specified by the flight envelope, and secondly, the PCH is the highest derivative input. Thus, at least a second-order filter should be used.
Using PCH and inputting a command derivative in addition to the desired signal itself is advantageous and leads to the established design of the reference model as shown in~\cref{fig:referencemodel}.

In the case of manual control, deflecting the stick results in proportional angular rate commands, which are integrated into commanded angles that are to be held once the stick is released. In order to avoid command response delay, feedforward compensation is integrated into the command filters. In addition, for the roll angle logic has been designed that switch between roll angle (\(\lvert\phi\rvert > 27.5 \deg\)) and roll angular rate command (\(\lvert\phi\rvert\leq 27.5 \deg\)), see~\cite{Lombaerts2011c,Weiser2024}.

\subsubsection{Sideslip Control}~\label{sec:sideslip_control} %
The side slip control function is similar to the one shown in~\cite{Grondman2018}, and calculates a reference body yaw rate \( r_\mathrm{com} \) to track a commanded sideslip angle command \( \beta_\mathrm{com} \). To this end, the third row in \cref{eq:muab} is inverted~\cite{Lombaerts2011c}:
\begin{align}
    \dot{\beta_i} &= p_B\ \sin{\alpha} - r_B\ \cos{\alpha} -\sin\mu\ \dot{\gamma} + \cos\mu\cos\gamma\ \dot{\chi} \notag \\
                  &approx p_B\ \sin{\alpha} - r_B + \frac{(n_y + \sin\phi \cos\theta) g}{V} 
\end{align}
where \( n_y \) is the lateral load factor, \( \alpha \) the angle of attack, \( g \) the gravitational acceleration, and \( p_B \) the roll rate. This results in:
\begin{equation}
	r_\mathrm{com} = p_B\ \sin{\alpha} + \frac{(n_y + \sin\phi \cos\theta) g}{V} - \nu_\beta
\end{equation}
Adding the PI-structured linear feedback control law, and substituting complementary filtered values (that are air-mass referenced in the lower frequency ranges), the control law becomes:
\begin{equation}
	r_\mathrm{com} = \frac{(n_y + \sin\phi \cos\theta) g}{V_\mathrm{tas}} + p_B\ \sin\hat{\alpha} + K_\beta \hat{\beta}_\mathrm{compl} - \frac{1}{s} K_{\beta,I} \left( \beta_\mathrm{com} - \hat{\beta}_\mathrm{compl} \right)
\end{equation}
The integrator is placed in the side slip feedback loop in order to compensate for aircraft asymmetry, as well as static offsets due to simplifications in the derivation above. This is an example of what has been discussed in \cref{sec:procedureNDI}, where model simplification reduces complexity of the control law, but requires consideration regarding robustness in the outer function. Also note that the control law was originally based on the estimated side slip angle, which was centered around \(n_y = 0\) \cite{Looye2001,Lombaerts2011c}, resulting in accurately coordinated turns. On the Citation aircraft also complementary filtered measured side slip has been used.
Finally, a linear controller calculates the commanded derivative as
\begin{equation}
	\nu_r = K_{r}(s) ( r_\mathrm{com} - \hat{r} )
\end{equation}

For tuning of all feedback gains, the Multi-Objective Parameter Synthesis (MOPS) has been intensively used, see \cite{Grondman2018} for more details. 

\subsection{Flight Path and Speed Control Functions}

The flight path and speed control functions are based on the Total Energy Control System (TECS)~\cite{Lambregts1983,Lambregts2013a} (\cref{sec:method:design:cv}) for vertical path and airspeed tracking, and Total Heading Control System (THCS)~\cite{Lambregts2013b} for heading tracking respectively. For the latter, a slightly modified formulation as described in~\cite{Looye2013} has been used. 
Simultaneously, the flight path control stabilizes the zero dynamics of the inversion as discussed in~\cref{sec:application:dyninv}.

For further details on the specific design the reader is referred to \cite{Weiser2024b}.

\subsection{Flight Testing}
Flight tests have been conducted to demonstrate the capabilities of the control law and show its functionalities in a relevant environment.
Before the flight tests, the control laws went through intensive verification and clearance analyses \cite{Weiser2024} and were evaluated using the DLR Robotic Motion Simulator~\cite{Seefried2019}. This evaluation was conducted to confirm the controller's operability by the pilot, verify various handling qualities metrics, ensure that command magnitudes and signs were appropriate, and assess the system's real-time capabilities. Details of the testing procedure are available in~~\cite{Weiser2024,Grondman2018,Keijzer2019,Pollack2019,Weiser2020}, where a similar approach to flight testing was applied. 
The control law was implemented on the flight hardware and tested in a hardware-in-the-loop setup, initially with the Flight Control Computer, subsequently with the aircraft on ground power.

\subsubsection{Hardware Implementation} %
The implementation utilizes the real-time framework DUECA~\cite{Paassen2000}. Flight code is exported as C code from a Simulink model via Embedded Coder and verified via PolySpace\textregistered{}. The generated code was then wrapped by a DUECA module, which exhibits interfaces similar to those of the generated code (i.e., initialize, step, and terminate functions). The modules comprise data acquisition, the fly-by-wire system including actuator controllers, extensive logging, a flight management system, and the flight controller. The base clock is at \SI{2000}{\hertz}, where most functionalities, including the flight controller, use a frequency of \SI{1000}{\hertz}. During the tests, the machine was just partly loaded, which gives margin for expanding the flight control laws in future campaigns. Furthermore, tests also showed that even a sampling frequency of~\SI{100}{\hertz} is sufficient for the developed control law. The machine is part of the total fly-by-wire system shown in~\cite{Zaal2009}. %
A more detailed explanation of the setup is given, e.g., in~\cite{Zaal2009,Scholten2020}.

\subsubsection{Flight Test Results}
\begin{figure*}[!htb]
	\centering
	\resizebox{.7\textwidth{}}{!}{%
		\subfigure{\centering
			\includegraphics[page=1]{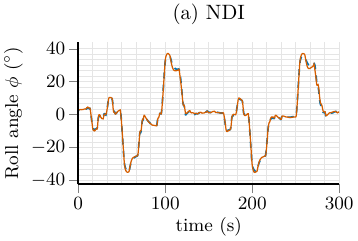}
		}
		\subfigure{\centering
			\includegraphics[page=2]{assets/tikz/flight_test_results_tikz.pdf}
		}
	} \resizebox{.7\textwidth}{!}{%
		\subfigure{\centering
			\includegraphics[page=3]{assets/tikz/flight_test_results_tikz.pdf}
		}
		\subfigure{\centering
			\includegraphics[page=4]{assets/tikz/flight_test_results_tikz.pdf}
		}
	}
	\caption{Roll angle tracking of the different NDI controllers during flight tests. The reference (\tikzline{dashed,refcolor}) and the measured roll angle (\tikzline{respcolor}) are essentially similar for all four variants.} \label{fig:flight_test_phi}
\end{figure*}

\begin{figure*}[!htb]
	\centering
	\resizebox{.7\textwidth}{!}{%
		\subfigure{\centering
			\includegraphics[page=5]{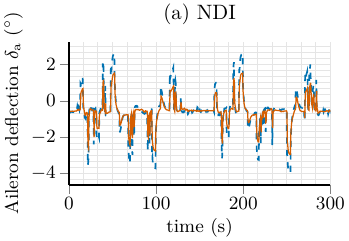}
		}
		\subfigure{\centering
			\includegraphics[page=6]{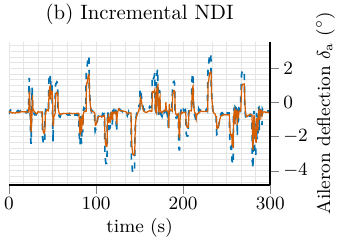}
		}
	} \resizebox{.7\textwidth}{!}{%
		\subfigure{\centering
			\includegraphics[page=7]{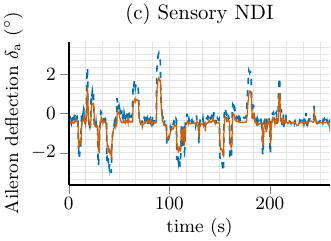}
		}
		\subfigure{\centering
			\includegraphics[page=8]{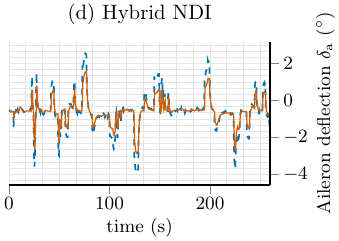}
		}
	}
	\caption{Aileron commands (\tikzline{dashed,refcolor}) and resulting deflections (\tikzline{respcolor}) of the different NDI controllers during flight tests.}\label{fig:flight_test_da}
\end{figure*}

The flight test maneuvers consist mainly of step responses in the different axes, as well as handling quality rating by means of pilot-in-the-loop maneuvers. The results of the hybrid NDI flight tests are presented and compared to other controllers from the same flight test campaign by means of flight performance of tracking a roll angle \( \phi \) in \cref{fig:flight_test_phi} as well as aileron control surface utilization in \cref{fig:flight_test_da}.

The flight test results suggest that the four dynamic inversion-based controllers perform essentially similarly. Although slight differences in performance can be seen in the flight test results, those can be neglected, since each flight test poses different environmental conditions, such as turbulence, making it hard to compare the controllers on a detailed level. Those results are mainly consistent with the assumptions drawn from the theory: In a nominal case, all controllers perform comparably equal in terms of flight performance and tracking.
Differences in the flight test results may originate from slightly different signal latencies in both controllers, as well as different test conditions.
However, if compared directly, the flight tests also strengthen the theoretical conclusions: Incremental NDI achieves the best tracking performance at the cost of the highest actuator loads. NDI does the opposite. Hybrid NDI, however, shows a good compromise in performance and actuator utilization.

\subsection{Non-affine System Control of an eVTOL aircraft} %
\begin{figure}[!htb]
	\begin{center}
		\includegraphics[width=.5\textwidth{}]{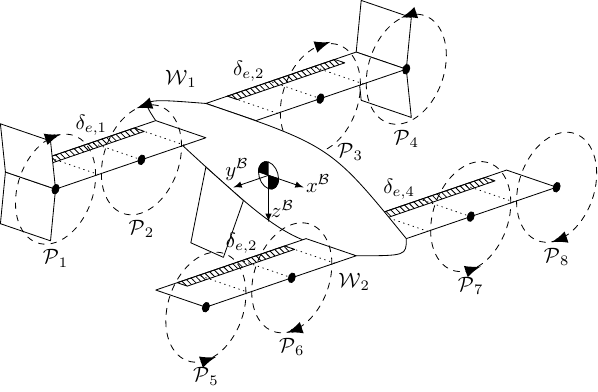}
	\end{center}
	\caption{Tandem tilt-wing configuration from~\cite{Milz2026jgcd}}\label{fig:tiltwing}
\end{figure}

New aircraft configurations, including electric vertical take-off and landing vehicles (eVTOLs), bring up new challenges for control design~\cite{Milz2026jgcd,Cheng2025,Raab2018}. Exemplary, the tandem tilt-wing configuration shown in~\cref{fig:tiltwing} and described in~\cite{Milz2026jgcd} is an overactuated (\( n_u > n_x \), actually \( n_u > n_{y_\mathrm{CV}} \)) and non-affine dynamical system with multiple nonlinear aerodynamic interactions between control inputs and states. Furthermore, this vehicle exhibits 5 degrees of freedom (3 rotational and 2 translational) to be controlled directly, leading to new opportunities. The control concept proposed in~\cite{Milz2026jgcd} is based on hybrid NDI but, opposed to~\cref{fig:cascade}, controls the longitudinal velocity and the attitude simultaneously instead of in a cascaded manner.
Furthermore, this example also shows an approach to seamlessly combine hybrid NDI with the control allocation task in the form of an optimization problem. The inverse system dynamics are approximated via local linearization.

As the control inputs (mainly the thrust and tilt-angle) influence nearly all aerodynamic effects occurring, the reduced state-space system (similar to \cref{eq:phlab_statespace}) for the commanded variables \( \vec{y}^\prime = \left[ \vec{v}_\mathrm{x}, \ \vec{v}_\mathrm{z}, \ \vec{\omega} \right] \) is:
\begin{equation}
	\vec{\alpha}^\prime (\vec{x}) = \begin{bmatrix}  v_\mathrm{y} \omega_\mathrm{z} - v_\mathrm{z} \omega_\mathrm{y} + \left( \mat{R}_\mathrm{EB}^T \vec{g} \right)_\mathrm{x} \\ v_\mathrm{x} \omega_\mathrm{y} - v_\mathrm{y} \omega_\mathrm{x} + \left( \mat{R}_\mathrm{EB}^T \vec{g} \right)_\mathrm{z} \\ \mat{I}^{-1} \left( - \vec{\omega} \times \mat{I} \vec{\omega} \right) \end{bmatrix}, \quad %
	\vec{\beta}^\prime (\vec{x}, \vec{u}) = \begin{bmatrix} \frac{1}{m} \vec{f}_\mathrm{x}(\vec{x},\vec{u}) \\ \frac{1}{m} \vec{f}_\mathrm{z}(\vec{x},\vec{u}) \\ \mat{I}^{-1} \vec{m}(\vec{x},\vec{u}) \end{bmatrix}
\end{equation}
where the forces \( \vec{f} \) and moments \( \vec{m} \) are nonlinear in the state and input and are not guaranteed to be analytically and globally invertible. 
The control variable selection yields zero dynamics not only for the attitude, but also the lateral velocity \( \vec{v}_\mathrm{y} \). Additional functionalities like Translational Rate Control~\cite{Milz2022} are thus required to guarantee stable zero dynamics. 

The resulting implicit hybrid NDI law from~\cref{eq:hndi_laplace} can be combined with the control allocation problem, which minimizes an arbitrary function \( L(\hat{\vec{x}}, \vec{u}) \), to:
\begin{subequations}\label{eq:tiltwing_optim}
	\begin{align}\label{eq:tiltwing_obj}
		\min_{\vec{u} \in \mathcal{U}} \  & L(\hat{\vec{x}}, \vec{u}) \\
		\text{s.t.} \ & \vec{\hat{\beta}}(\hat{\vec{x}}, \vec{u}) = \vec{\nu} - \mathcal{L}^{-1} \left\{ \begin{bmatrix} \hfill H_\mathrm{c} (s) \\ 1-H_\mathrm{c} (s) \end{bmatrix}^T \right\} * \begin{bmatrix} \hat{\vec{y}}^{(\rho)} - \hat{\vec{\beta}} ( \hat{\vec{x}}, \hat{\vec{u}} ) \\ \hat{\vec{\alpha}} ( \hat{\vec{x}} ) \end{bmatrix} \label{eq:tiltwing_constr}
	\end{align}
\end{subequations}
This problem can be solved, for instance, by nonlinear equation solvers, look-up tables, or by using local approximations. The latter employs the Taylor series expansion of \cref{eq:tiltwing_obj} around an arbitrary expansion point \( \vec{u}_0 \), leading to a locally affine control law as a constraint to a quadratic program. 
A detailed discussion of the approach is given in~\cite{Milz2026jgcd,Steinert2025a}. 

A simplified solution can be carried out using the sensory NDI law \( \vec{\hat{\beta}}(\hat{\vec{x}}, \vec{u}) = \nu - \hat{\vec{y}}^{(\rho)} + \hat{\vec{\beta}} ( \hat{\vec{x}}, \hat{\vec{u}} ) \) in~\cref{eq:tiltwing_constr} for simplicity and \( L(\hat{\vec{x}}, \vec{u}) = \frac{1}{2} \left( \vec{u} - \vec{u}^\ast \right)^T \mat{W} \left( \vec{u} - \vec{u}^\ast \right) \) for~\cref{eq:tiltwing_obj}. Employing the Taylor series expansion around \( \vec{u}_0 \) leads to following approximated quadratic program with a linear constraint
\begin{subequations}
	\begin{align}
		\min_{\vec{u} \in \mathcal{U}} & L(\hat{\vec{x}}, \vec{u}_0) + \overbrace{\left( \vec{u}_0 - \vec{u}^\ast \right)^T \mat{W}}^{ \nabla_u L(\hat{\vec{x}}, \vec{u}_0) } \left( \vec{u} - \vec{u}_0 \right) + \frac{1}{2} \left( \vec{u} - \vec{u}_0 \right)^T \overbrace{ \mat{W} }^{\mathclap{\nabla_{uu} L(\hat{\vec{x}}, \vec{u}_0) }} \left( \vec{u} - \vec{u}_0 \right) \\
		\text{s.t.} \ & \underbrace{ \nabla_\vec{u} \vec{\hat{\beta}}(\hat{\vec{x}}, \vec{u}_0) }_{:= \mat{B}} \vec{u} = \underbrace{ \vec{\nu} - \hat{\vec{y}}^{(\rho)} + \hat{\vec{\beta}} ( \hat{\vec{x}}, \hat{\vec{u}} ) - \hat{\vec{\beta}} ( \hat{\vec{x}}, \vec{u}_0 ) + \nabla_\vec{u} \vec{\hat{\beta}}(\hat{\vec{x}}, \vec{u}_0) \, \vec{u}_0 }_{:= \vec{\mathfrak{u}}}
	\end{align}
\end{subequations}
which can be solved analytically as
\begin{equation}
	\vec{u} = \mat{B}^+ \vec{\mathfrak{u}} + \left( \mat{I}_{n_u} - \mat{B}^+ \mat{B} \right) \left( \vec{u}_0 - \mat{W}^{-1} \left( \vec{u}_0 - \vec{u}^\ast \right)^T \mat{W} \right)
\end{equation}
with the weighted pseudo-inverse \( \mat{B}^+ = \mat{W}^{-1} \mat{B}^T \left( \mat{B} \mat{W}^{-1} \mat{B}^T \right)^{-1} \). Choosing \( \vec{u}_0 = \hat{\vec{u}} \) makes the solution locally equivalent to incremental NDI~\cite{Steinert2025a,Hafner2025}.

\section{Conclusions}
Since gaining considerable attention in the late 1980s and 1990s, and the publication of a landmark paper by Enns in 1994~\cite{Enns1994}, Nonlinear Dynamic Inversion has seen rapid development and numerous applications in industrial and research flight control law design programs. In this paper, subsequent developments towards extended capabilities, as well as incremental, sensory, and hybrid forms, have been reviewed from different perspectives. %

From a \textit{conceptual perspective} NDI and its various variants invert the relation between controlled variable derivatives and available generalized or allocated controls, explicitly canceling any other effect of influence that has been included in the model for inversion by design choice. In this way, the system is approximately decoupled, and effectively linearized. The fundamental architecture and philosophy of dynamic inversion have proven highly suitable, especially given their modularity, which elegantly accommodates key advancements across various functional areas, including control allocation, model adaptation, failure handling, and signal processing. The architecture in turn provides the basis for the detailed design of the intended control law functions and sub-functions, either with the help of standard solutions that come with NDI, or using controller synthesis methods that may address specific challenges in the design. 

It is best to compare NDI with other inversion-based control methods from an \textit{architectural} perspective. NDI is unique in that it uses measured states in the inverse model equations. This avoids part of the robustness questions that may arise. Since it does not come with state equations of its own, it is not comparable with inverting a transfer function. Also, NDI puts itself between the control law functions and the system. From a literature perspective, the role of inversion can be compared with  Douglas Adams' Babel Fish~\cite{Adams1979}, which translates input from a foreign language (the decoupled, linear coordinates) into its native tongue (the nonlinear coordinates). This allows any method used for outer-loop control functions to \textit{talk linearly with a nonlinear system}. Many applications discussed have proven that this is an efficient and effective alternative to control law gain scheduling. 

From a \textit{mathematical} perspective, NDI as a methodology is based on sound principles that in turn build on geometric control theory. At this point various forms, different designations, and various derivations do have a common theoretical basis. 
In its early stages the NDI research was focusing on the transformation of a nonlinear system into an equivalent linear system in the scope of exact feedback linearization. One step that made NDI practically applicable was the input-output linearization and approximate feedback linearization. Those reduce the strictness and complexity of the mathematical construct and allow to use the formal results in a more pragmatic way, as has been discussed from a \textit{methodological} perspective.

From this perspective, the various NDI variants have been proven to be equivalent. It is in implementation details where differences emerge that may have advantages and disadvantages relative to the application at hand. Examples are handling of failures, complexity of algorithms, dependency on model accuracy, disturbance rejection, and control activity. In general we found that a substantial number of papers do not mention or explicitly distinguish between air-mass and inertial referenced flight states. Often even, wind is assumed to be absent. Dealing with both references is one of the fundamental aspects of flight control. For this reason, the role of disturbances in NDI has been made explicit by carrying them through the derivations of classical, sensory, incremental, and hybrid NDI variants.

Incremental or sensory formulations of NDI considerably reduce model dependency and, by inherent direct sensing of disturbance effects through (angular) acceleration measurement, improved tracking accuracy. A powerful design degree of freedom has always been in the selection of the aircraft model that becomes part of the control laws (On-Board AirCraft model -- OBAC \cite{Enns2006}). This allows selective compensation of dynamic behavior and disturbances in order to reduce control law complexity, reduce control activity, address flight loads, and improve ride comfort. In incremental and sensory formulations, this design degree of freedom vanishes. This has been a key reason for the emergence of hybrid concepts.

From a \textit{design} perspective the control law structure and underlying principles of NDI almost naturally fit into design processes that are common in aerospace. Adaptation to a new or modified flight vehicle is largely covered by adapting the inverse model core accordingly. NDI should maybe not be regarded as a stand-alone design method, but rather as an architecture with the aforementioned Babel Fish capacity. One reason is that it does not come with metrics on fulfillment of design requirements. These are brought in by classical (for which several well proven standard solutions exits), robust or other advanced control methods in the design of the surrounding functions. These methods complete the design and may additionally compensate for any effects where the inverse model is inaccurate or incorporates simplifying assumptions. In other words: \vskip 1ex

\textit{In selecting a control method for a flight control system the question is not ``method x, y, or NDI'', but rather ``method x, y with or without NDI''.}\vskip 1ex %

From an \textit{application} perspective, the various NDI variants are ubiquitous today, working on fixed and tilt-wing aircraft, rotorcraft, as well as various types of drones. Four variants of NDI were used to design a control system for an experimental passenger aircraft. Flight performance of the variants was very similar, validating the conclusion drawn from the methodological view.

Finally, from a \textit{historical} perspective, the authors have observed that the essential features and structure of Dynamic Inversion have clearly endured the test of time, being fundamental to all of the developments discussed above. This durability underscores the elegance and significance of the original concept.

The foresight in the conclusions of the landmark paper by Enns \textit{et al.} \cite{Enns1994}, stating: \textit{While these discussions showed that the method's current status is already adequate for serious designs, the method will benefit substantially from additional research developments, particularly in the areas of nonlinear zero dynamics and nonlinear robustness}, has been right on the mark. 

Looking ahead, new flight vehicles will pose new challenges. Increasingly complex, over-actuated, and non-affine configurations, as exemplified by the eVTOL application in~\cref{sec:application}, stretch the classical affine formulations and tie the inversion ever closer to (online) optimization. Learning-based control is developing rapidly, but raises questions regarding stability guarantees and certification that are still open. The continued evolution and proven versatility of Dynamic Inversion, its architecture, and functional breakdown are likely to ensure its central role in addressing emerging challenges, as well as in accommodating new capabilities in flight control system design. NDI will keep on evolving.

\appendix
\numberwithin{equation}{section}
\numberwithin{figure}{section}
\numberwithin{table}{section}

\section{A Tool Perspective}
Despite the large number of publications on NDI, only a few present software tools or libraries specifically tailored for inversion-based control methodologies. Well-established and reliable modeling, simulation, analysis, and synthesis tools are available for control system design in general~\cite{Joos2002,Varga2014,MATLAB2025,Simulink2025}, enabling straightforward implementation of NDI-based control algorithms. For fixed model structures, parameterized versions can be made available that allow for adaptation to new projects by exchanging model parameters, e.g.,~\cite{Kuchar2018,Kiehn2022}.

As discussed in~\cref{sec:method}, the derivation of an inverse model core consists of five steps:
\begin{enumerate}
	\item Collect commanded variables of functions that are simultaneously active in the output vector $y$;
	\item Differentiate each output until an algebraic relation with one or more (generalized) control inputs $u$ results and ensure a well-defined relative degree vector~\cite{Zhan1991,DeJager1995} and physically meaningful independence and availability of controls. This gives rise to the output vector $y^{(\rho)}$;
	\item Invert the relation between $u$ and $y^{(\rho)}$ (the inversion should be non-singular);
	\item Formulate the inverse equation in the desired form (model-based, incremental, hybrid, sensory, etc.);
	\item Implement the inverse model core in run-time code, and integrate as part of the overall control system in the preferred simulation environment for further detailed design~\cite{Weiser2024}.
\end{enumerate}

These steps involve mathematical operations that require symbolic manipulation capabilities to allow for automation in the form of a software tool. This was recognized quite early, with the advent of symbolic mathematical tools like Maple~\cite{Maple2023} and Mathematica~\cite{Mathematica2024}. Early applications can be found in~\cite{DeJager1999,Roebenack2005,Vibet1995,DeJager1995,vandeVen2000}, and the methodology has become a standard feature in related and new modeling and simulation environments~\cite{MapleSim2023,Wolfram2025,Dyad2026}. 

\begin{figure*}[!htb]
	\centering
	\includegraphics{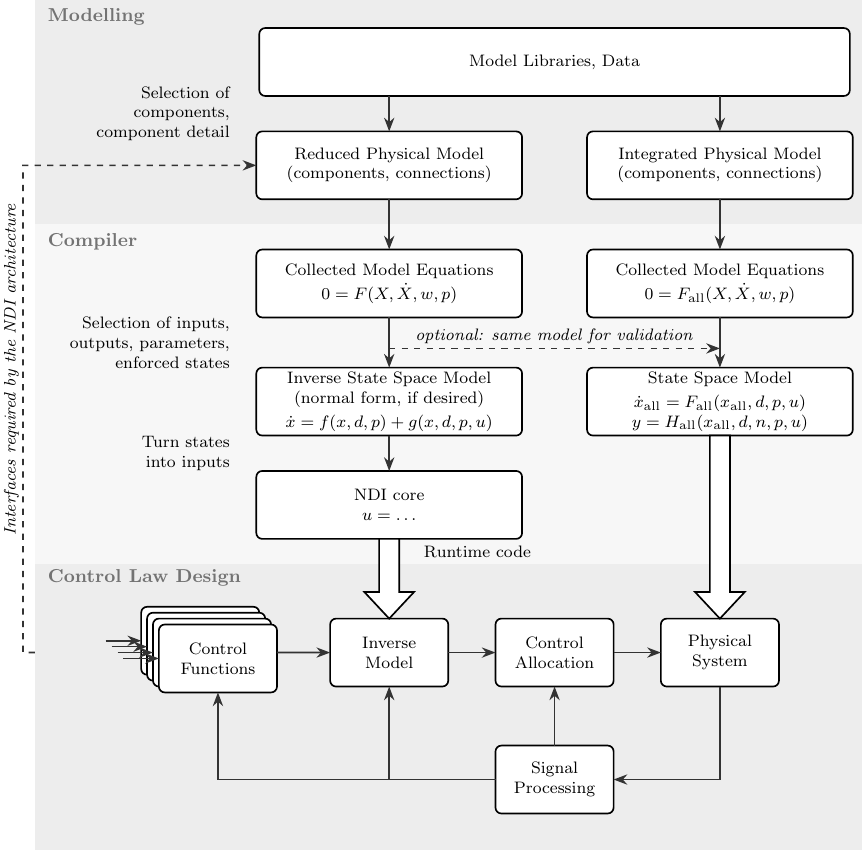}
	\caption{Automated modeling and NDI derivation process. The left branch carries the reduced model that is inverted, the right branch the integrated model used for simulation, with the control law at the bottom following~\cref{fig:arch}.}
	\label{fig:NDImodelica}
\end{figure*}

Modeling and simulation tools based on object-oriented modeling languages like Modelica~\cite{Fritzson2015} or symbolic languages~\cite{Maple2023,MapleSim2023,Mathematica2024,Wolfram2025,Dyad2026} allow for automating the entire process, including the generation of runtime code~\cite{Looye2001,Looye2007}. 	Examples of automatic derivation for different inversion-based control law architectures is given by Looye \textit{et al.} in~\cite{Looye2005}.
The full process for developing a dynamic inversion control law is shown in~\cref{fig:NDImodelica}. The foundation of these languages is the implementation of models, components, and interfaces in the form of physical equations that do not require to be arranged or sorted causally. This is different from block-oriented model implementations, where components are implemented in the form of causal state-space equations that interface via input-output connections.

Selection of inputs and outputs depends on intended model use and is subject to the constraint that the number of known and unknown variables and equations is balanced. In the case of a regular simulation model, this is usually clear from physical system knowledge. Based on this information, a model compiler automatically generates differential equations in the form of a nonlinear state-space model (\cref{eq:full_system}) or its simplifications. This compilation step is based on graph algorithms, tearing, and symbolic differentiation~\cite{Elmqvist93,Elmqvist94}. It is only at this point that the state vector $\vec{x}$ and causality in the model are defined. Equations that become obsolete in the process are automatically removed. 

The selection of \textit{commanded variables} as inputs and \textit{controls} as outputs results in automatic compilation of the desired inverse model and yields a function \( \vec{u} (\vec{x}, \vec{\nu}, \vec{p}) \). In this way, the aforementioned inversion process is automated and may be applied to any inversion-based control law architecture presented in~\cref{sec:architecture}~\cite{Looye2005,Looye2007}. After compilation, the inverse model should first be verified against the same (forward) simulation model it was derived from, before it is integrated into the overall control system (cf.~\cref{sec:invmodelrealization}). For NDI, an additional step replaces state equations with additional inputs that allow for insertion of measured or estimated states (cf.~\cref{sec:inversearchitecture:di}). A detailed reference covering the complete design process up to implementation of automatically generated NDI code on an aircraft flight control computer can be found in~\cite{Looye2001b}.

As a side note, the Modelica language provides means for the user to select specific variables as states, provided this makes sense from mathematical and physical points of view. This, in turn, allows the automatic transformation of a simulation model into the normal form of~\cref{eq:canonicalcontrolform}.

\section{A Flight Physical Perspective}\label{sec:eqm}
The aircraft flight dynamics equations of motion are included here for reference in the paper \cite{Brockhaus2011,Stevens2016,Etkin1995,Stengel2022,Schmidt2023,Cook2012}. For NDI it is often beneficial to use flight path related states for the force equations.

\subsection*{Moment equations}
The equations of motion related to rotation around the body axes are given by:
\begin{subequations}\label{eq:momenteq}
\begin{align}
	I_B\dot{\vec{\omega}}_B+\vec{\omega}_B \times I_B {\vec{\omega}}_B  
	&= \vec{m_{a_B}}(V_a, \alpha_a, \beta_a, \vec{\omega}_B, ...) + \vec{m_{a_{B}}}^\prime(V_a, \alpha_a, \beta_a, \vec{\omega}_B, ...,\vec{u}_a) \notag \\ 
	&+ \vec{m_{p_B}}(M_\mathrm{a},T_s,...,\vec{x}_p) + \vec{m_{p_{B}}^\prime}(M_\mathrm{a},T_s,...,\vec{x}_p,\vec{u}_p) \notag \\
	&+ \vec{m_{\mathrm{other}}}(...)
\end{align}
where \(\vec{m_{a_B}}\) and \(\vec{m_{a_{c_B}}}\) are computed from
\begin{align}
    \vec{m_{a_B}} &= \bar{q}S\bar{c}\left[C_l(M_\mathrm{a},\beta_a,...),\ C_m(M_\mathrm{a},\alpha_a,...),\ C_n(M_\mathrm{a},\beta_a,...)\right]^T_B \\
    \vec{m_{a_{B}}}^\prime &= \bar{q}S\bar{c}\left[C_{l_\delta}(M_\mathrm{a},\beta_a,...,u_a),\ C_{m_\delta}(M_\mathrm{a},\alpha_a,...,u_a),\ C_{n_\delta}(M_\mathrm{a},\beta_a,...,u_a)\right]^T_B 
\end{align}
\end{subequations}
with the reference length \( \bar{c} \).
The contributions of control inputs have been separated in computation of aerodynamic (subscript $a$) and propulsion-related (subscript $p$) moments.

\subsection*{Force equations}
The equations of motion related to translation along flight path axes are given by:
\begin{subequations}\label{eq:forceeq}
\begin{align}
	m \left[\dot{V}, \dot{\chi} V \cos{\gamma}, -\dot{\gamma} V\right]^T &=  \mat{R}_{\mathrm{\small FB}}(\mu,\alpha,\beta)\mat{R}_{\mathrm{\small BA}}(\alpha_a,\beta_a) \left(\vec{f}_{a_A}(V_a, \alpha_a, \beta_a, \vec{\omega}_B, ...) + \vec{f}_{a_{A}}^\prime(V_a, \alpha_a, \beta_a, \vec{\omega}_B, ...,\vec{u}_a) \right) \notag \\ &+ \mat{R}_{\mathrm{\small FB}}(\mu,\alpha,\beta) \left( \vec{f}_{p_B}(M_\mathrm{a},T_S,...,\vec{x}_p) + \vec{f}_{p_{B}}^\prime (M_\mathrm{a},T_S,...,\vec{x}_p,\vec{u}_p) \right) \notag \\ &+ \mat{R}_{\mathrm{\small FE}}(\gamma,\chi) \left[0, 0, mg\right]^T 
\end{align}
where 
\begin{align}
    \vec{f_{a_A}} &= \bar{q}S\left[-C_D(M_\mathrm{a},\alpha_a,...),\ C_Y(M_\mathrm{a},\beta_a,...),\ -C_L(M_\mathrm{a},\alpha_a,...) \right]^T_A \\
    \vec{f}_{a_{A}}^\prime &= \bar{q}S\left[-C_{D_\delta}(M_\mathrm{a},\alpha_a,...,u_a),\ C_{Y_\delta}(M_\mathrm{a},\beta_a,...,u_a),\ -C_{L_\delta}(M_\mathrm{a},\alpha_a,...,u_a) \right]^T_A 
\end{align}
\end{subequations}
Alternatively, the equations may be expressed in body axes instead:
\begin{align}
	m (\dot{\vec{v}}_B +\vec{\omega}_B \times {\vec{v}}_B)  
	&=  \mat{R}_{\mathrm{\small BA}}(\alpha_a,\beta_a) \left(\vec{f}_{a_A}(V_a, \alpha_a, \beta_a, \vec{\omega}_B, ...) + \vec{f}_{a_{A}}^\prime(V_a, \alpha_a, \beta_a, \vec{\omega}_B, ...,\vec{u}_a) \right) \notag \\ &+ \left( \vec{f}_{p_B}(M_\mathrm{a},T_S,...,\vec{x}_p) + \vec{f}_{p_{B}}^\prime(M_\mathrm{a},T_S,...,\vec{x}_p,\vec{u}_p) \right) \notag \\ &+ m \mat{R}_{\mathrm{\small BE}}(\phi, \theta, \psi) \underbrace{\left[0, 0, g\right]^T_E}_{\vec{g_E}} 
\end{align}

\subsection*{Attitude kinematics}
Attitude kinematics can be expressed relative to the velocity vector, or to the Earth surface by means of Euler angles:
\begin{equation} 
\begin{bmatrix}
	\dot{\mu} \cos\beta \\ \dot{\alpha} \cos\beta \\ \dot{\beta} 
\end{bmatrix}
=
\begin{bmatrix}
	{\cos\alpha} & 0 & {\sin\alpha} \\
	-\cos\alpha\sin\beta & \cos\beta & -\sin\alpha\sin\beta \\
	\sin\alpha & 0 & -\cos\alpha
\end{bmatrix}
\vec{\omega}_B
+
\begin{bmatrix}
	\cos\mu\sin\beta \\
	-\cos\mu \\
	-\sin\mu
\end{bmatrix}\,\dot{\gamma}
\;+\;
\begin{bmatrix}
	\sin\gamma \cos\beta + \sin\mu\cos\gamma\sin\beta \\
	-\sin\mu\cos\gamma \\
	\cos\mu\cos\gamma
\end{bmatrix}
\dot{\chi} \label{eq:muab}
\end{equation}
\begin{equation}\label{eq:wb2de}
\dot{\Theta} = 
\begin{bmatrix}
	\dot{\phi}  \\ \dot{\theta}  \\ \dot{\psi} 
\end{bmatrix}
=
\underbrace{
\begin{bmatrix}
	1 & \sin\phi\tan\theta & \cos\phi\tan\theta \\ 
	0 & \cos\phi & -\sin\phi \\
	0 & \sin\phi/\cos\theta & \cos\phi/\cos\theta
\end{bmatrix}
}_{\mat{R}_{\mathrm{\small \Theta B}}}
\vec{\omega}_B
\end{equation}
Depending on the selection of attitude states, the Euler angles are either obtained from integration or computation. The latter is based on:
\begin{align}
   \mat{R}_{\mathrm{\small BE}}(\phi,\theta,\psi) \mat{R}_{\mathrm{\small EF}}(\gamma,\chi) &= \mat{R}_{\mathrm{\small BF}}(\mu,\alpha,\beta)
\end{align}
One has to be careful, as $\alpha$, $\beta$ and $\mu$ are relative to the inertial velocity vector. The order of rotation from flight path to body axes is (1) \(\mu\), (2) \(-\beta\), and (3) \(\alpha\). This deviates from standard definition for inertial angles as used in for example~\cite{Brockhaus2011}. For computation of aerodynamic forces and moments, the air mass-referenced equivalents are needed. For those, the aforementioned order of rotation is standard. These angles are obtained by re-computing the angles after obtaining the airspeed vector:
\begin{subequations}\label{eq:mabtrans}
\begin{align}
	\vec{v_{a_B}} &= \mat{R}_{\mathrm{\small BE}}(\phi,\theta,\psi) \left(\left[\cos{\gamma}\cos{\chi},\cos{\gamma}\sin{\chi},-\sin{\gamma}\right]^T \vec{v} - \left[u_W, v_W, w_W \right]^T_E \right) \mathrm{\ \ or:} \\
	\vec{v_{a_B}} &= \mat{R}_{\mathrm{\small BF}}(\mu,\alpha,\beta) \left[V,0,0\right]^T  - \mat{R}_{\mathrm{\small BE}}(\phi,\theta,\psi) \left[u_W, v_W, w_W \right]^T_E \\
	\vec{v_{a_B}} &= \left[u_{a_B}, v_{a_B}, w_{a_B} \right]^T_B
\end{align}
\end{subequations}
Based on this, we obtain:
\begin{subequations}\label{eq:mabdef}
\begin{align}
	V_a &= \sqrt{\vec{v_{a_B}}^{\hskip -1.5ex T} \vec{v_{a_B}}}\\
	\tan\alpha_a &= \frac{w_{a_B}}{u_{a_B}} \\
	\sin\beta_a &= \frac{v_{a_B}}{V_a}
\end{align}
\end{subequations}
And relative to the local Earth surface:
\begin{subequations}\label{eq:attearth}
\begin{align}
	\vec{v_{a_E}} &= \left[\cos{\gamma}\cos{\chi},\cos{\gamma}\sin{\chi},-\sin{\gamma}\right]^T V - \left[u_W, v_W, w_W \right]^T_E \\
	\vec{v_{a_B}} &= \left[u_{a_E}, v_{a_E}, w_{a_E} \right]^T_E \\
	\sin\gamma_a &= \frac{-w_{a_E}}{V_a} \\
	\tan\chi_a &= \frac{v_{a_E}}{u_{a_E}}
\end{align}
\end{subequations}
Note that in the absence of wind and when using the same order of rotation as for aerodynamic angles, the rotation from aerodynamic into flight path coordinates becomes only dependent on \(\mu = \mu_a\).

\subsection*{Position kinematics}
Finally, position-related kinematics are defined by:
\begin{align}\label{eq:poseq}
	\dot{\vec{r}}_E = \left[\dot{X},\ \dot{Y},\ \dot{Z}\right]_E^T = \left[V_N,\ V_E,\ -\dot{h}\right]^T = \left[V \cos\gamma\cos\chi,\ V \cos\gamma\sin\chi,\ -V\sin\gamma\right]^T
\end{align}
These may be replaced with WGS84-standard longitude, latitude and elevation state equations, if needed.

\section*{A Personal Perspective: Acknowledgements}
The authors would like to thank Dale Enns, main author of the 1994 paper, \textit{``Dynamic inversion: an evolving methodology for flight control design,''} for his gracious permission to use a near-identical title and support to publish this article, which is reminiscent of his and his co-authors' seminal work. In addition, the valuable suggestions and remarks by all reviewers are highly appreciated.

\section*{Declaration of competing interest}
The authors declare that they have no known competing financial interests or personal relationships that could have appeared to influence the work reported in this article.

\section*{Funding}
The work was performed within the institutional research programme of the German Aerospace Center (DLR). %

\section*{Data availability}
The authors do not have permission to share the flight-test and simulation data underlying the results presented in this article. No other data were required for the work reported here, which otherwise draws on the published literature cited throughout.

\printcredits

\bibliographystyle{cas-model2-names}

\end{document}